\documentclass[a4paper,11pt]{article} 
\pdfoutput=1
\usepackage{jheppub}
\usepackage{slashed}
\usepackage[dvipsnames,table]{xcolor}
\usepackage{hyperref}
\usepackage{xspace}
\usepackage[tight]{subfigure}
\usepackage{amsmath}
\usepackage{amssymb}
\usepackage{amsfonts}
\usepackage{mathrsfs}
\usepackage{comment}
\usepackage{afterpage}
\usepackage{verbatim}
\usepackage{booktabs}
\usepackage{array}
\usepackage[title,titletoc]{appendix}
\usepackage{tabularx}
\newcolumntype{C}[1]{>{\centering\arraybackslash}p{#1}}
\usepackage{scalerel}
\newcommand{\loqcd}{\rm LO_{1}\xspace}
\newcommand{\loew}{\rm LO_{3}\xspace}
\newcommand{\loint}{\rm LO_{2}\xspace}
\newcommand{\nloone}{{\rm NLO}_1\xspace}
\newcommand{\nlotwo}{{\rm NLO}_2\xspace}
\newcommand{\nlothree}{{\rm NLO}_3\xspace}
\newcommand{\nlofour}{{\rm NLO}_4\xspace}
\newcommand{\lobqcd}{\rm LO_{\rm b,\,1}\xspace}
\newcommand{\lobew}{\rm LO_{\rm b,\,3}\xspace}
\newcommand{\lobint}{\rm LO_{\rm b,\,2}\xspace}
\newcommand{\nlobone}{{\rm NLO}_{\rm b,\,1}\xspace}
\newcommand{\nlobtwo}{{\rm NLO}_{\rm b,\,2}\xspace}
\newcommand{\nlobthree}{{\rm NLO}_{\rm b,\,3}\xspace}
\newcommand{\nlobfour}{{\rm NLO}_{\rm b,\,4}\xspace}
\def\refeq#1{\mbox{Eq.~(\ref{#1})}}
\def\refeqs#1{\mbox{Eqs.~(\ref{#1})}}
\def\refeqf#1{\mbox{(\ref{#1})}}
\def\reffi#1{\mbox{Fig.~\ref{#1}}}
\def\reffis#1{\mbox{Figs.~\ref{#1}}}
\def\refta#1{\mbox{Table~\ref{#1}}}
\def\reftas#1{\mbox{Tables~\ref{#1}}}
\def\refse#1{\mbox{Section~\ref{#1}}}
\def\refses#1{\mbox{Sections~\ref{#1}}}

\def\citere#1{\mbox{Ref.~\cite{#1}}}
\def\citeres#1{\mbox{Refs.~\cite{#1}}}

\newcommand{\newc}{\newcommand}
\newc{\beq}{\begin{equation}}
\newc{\eeq}{\end{equation}}
\newc{\bit}{\begin{itemize}}
\newc{\eit}{\end{itemize}}
\newc{\ben}{\begin{enumerate}}
\newc{\een}{\end{enumerate}}
\newc{\bce}{\begin{center}}
\newc{\ece}{\end{center}}
\newc{\bfi}{\begin{figure}}
\newc{\efi}{\end{figure}}

\newcommand{\ri}{\mathrm i}

\newcommand{\rT}{{\mathrm{T}}}

\newcommand{\rw}{{\mathrm{w}}}

\newcommand{\ie}{\emph{i.e.}\ }

\newcommand{\GeV}{\ensuremath{\,\text{GeV}}\xspace}
\newcommand{\TeV}{\ensuremath{\,\text{TeV}}\xspace}

\newcommand{\Pq}{\ensuremath{q}\xspace}
\newcommand{\PH}{\ensuremath{\text{H}}\xspace}
\newcommand{\Pj}{\ensuremath{\text{j}}\xspace}
\newcommand{\Pp}{\ensuremath{\text{p}}}
\newcommand{\Pe}{\ensuremath{\text{e}}\xspace}
\newcommand{\Pb}{\ensuremath{\text{b}}\xspace}
\newcommand{\Pt}{\ensuremath{\text{t}}\xspace}
\newcommand{\Pu}{\ensuremath{\text{u}}\xspace}
\newcommand{\Pd}{\ensuremath{\text{d}}\xspace}
\newcommand{\Ps}{\ensuremath{\text{s}}\xspace}
\newcommand{\Pc}{\ensuremath{\text{c}}\xspace}
\newcommand{\Pg}{\ensuremath{\text{g}}}

\newcommand{\PW}{\ensuremath{\text{W}}\xspace}
\newcommand{\PZ}{\ensuremath{\text{Z}}\xspace}
\newcommand{\Pl}{\ell}
                                    
\newcommand{\Mt}{\ensuremath{m_\Pt}\xspace}
\newcommand{\MH}{\ensuremath{M_\PH}\xspace}

\newcommand{\MZ}{\ensuremath{M_\PZ}\xspace}

\newcommand{\Gt}{\ensuremath{\Gamma_\Pt}\xspace}
\newcommand{\GH}{\ensuremath{\Gamma_\PH}\xspace}

\newcommand{\alphas}{\ensuremath{\alpha_\text{s}}\xspace}
\newcommand{\gs}{\ensuremath{g_\text{s}}\xspace}

\newcommand{\MSbar}{\ensuremath{\overline{{\text{MS}}\xspace}}}

\newcommand{\recola}{{\sc Recola}\xspace}

\newcommand{\mocanlo}{{\sc MoCaNLO}\xspace}
\newcommand{\collier}{{\sc Collier}\xspace}

\newcommand{\helacnlo}{{\sc Helac-NLO}\xspace}

\newcolumntype{.}{D{.}{.}{-1}}
\newcolumntype{d}[1]{D{.}{.}{#1}}
\colorlet{tableoverheadcolor}{gray!37.5}
\colorlet{tableheadcolor}{gray!25}
\colorlet{tablerowcolor}{gray!12.5}

\def\draftdate{\relax}
\def\mda{\relax}
\def\mua{\relax}
\def\mla{\relax}
\def\draft{
\def\thtystars{******************************}
\def\sixtystars{\thtystars\thtystars}
\typeout{}
\typeout{\sixtystars**}
\typeout{* Draft mode!
         For final version remove \protect\draft\space in source file *}
\typeout{\sixtystars**}
\typeout{}
\def\draftdate{\today}
\def\mua{\marginpar[\boldmath\hfil$\uparrow$]%
                   {\boldmath$\uparrow$\hfil}\color{black}%
                    \typeout{marginpar: $\uparrow$}\ignorespaces}
\def\mda{\color{red}\marginpar[\boldmath\hfil$\downarrow$]%
                   {\boldmath$\downarrow$\hfil}%
                    \typeout{marginpar: $\downarrow$}\ignorespaces}
\def\mla{\marginpar[\boldmath\hfil$\rightarrow$]%
                   {\boldmath$\leftarrow $\hfil}%
                    \typeout{marginpar: $\leftrightarrow$}\ignorespaces}
\def\Mua{\marginpar[\boldmath\hfil$\Uparrow$]%
                   {\boldmath$\Uparrow$\hfil}\color{black}%
                    \typeout{marginpar: $\uparrow$}\ignorespaces}
\def\Mda{\color{red}\marginpar[\boldmath\hfil$\Downarrow$]%
                   {\boldmath$\Downarrow$\hfil}%
                    \typeout{marginpar: $\downarrow$}\ignorespaces}
\def\Mla{\marginpar[\boldmath\hfil\textcolor{red}{$\Rightarrow$}]%
                   {\boldmath\textcolor{red}{$\Leftarrow $}\hfil}%
                    \typeout{marginpar: $\leftrightarrow$}\ignorespaces}
\overfullrule 5pt
\oddsidemargin 15mm
\marginparwidth 29mm
}

\newcommand{\mc}{\mathcal}
\let\nnb\notag

\let\Mw\MW

\let\Mz\MZ

\let\Mwo\MWOS
\let\Gwo\GWOS
\let\Mzo\MZOS
\let\Gzo\GZOS

\newcommand{\Mvo}{M^{\rm OS}_{V}}
\newcommand{\Mv} {M_{V}}
\newcommand{\Gvo}{\Gamma^{\rm OS}_{V}}
\newcommand{\Gv} {\Gamma_{V}}
\let\as\alphas
\newcommand{\pt}[1]{p_{\rT,{#1}}}
\newcommand{\tm}[1]{M_{\rT,{#1}}}

\usepackage{pifont}
\newcommand{\ttz}{{\rm t\bar{\rm t}Z}\xspace}
\newcommand{\ttw}{{\rm t\bar{\rm t}W^\pm}\xspace}

\newcommand{\csp}{\hspace{0.6em}}

\title{Complete NLO corrections to off-shell $\text{t}\overline{\text{t}}\text{Z}$ production at the LHC}
\author{Ansgar Denner,$^{a}$}
\author{Daniele Lombardi$^{a}$ and}
\author{Giovanni Pelliccioli$^{b}$}
\affiliation{$^{a}$Universit\"at W\"urzburg, Institut f\"ur Theoretische Physik und Astrophysik, 97074 W\"urzburg, Germany}
\affiliation[]{$^{b}$Max-Planck-Institut f\"ur Physik, F\"ohringer Ring 6,  80805 M\"unchen, Germany}
\emailAdd{ansgar.denner@physik.uni-wuerzburg.de}
\emailAdd{daniele.lombardi@physik.uni-wuerzburg.de}
\emailAdd{gpellicc@mpp.mpg.de}
\date{\draftdate}
\preprint{MPP-2023-132}
\abstract{
  Measuring precisely top-pair-associated processes at hadron colliders will become possible with the upcoming
  LHC running stages. The increased data statistics will especially enable differential measurements leading
  to an improved characterisation of such processes.
  Aiming at a consistent data--theory comparison, precise Standard-Model predictions are needed, including
  higher-order corrections and full off-shell effects.
  In this work we present NLO-accurate predictions for the production and decay of
  a top--antitop pair in association with a $\rm Z$ boson at the LHC, in the multi-lepton decay channel. 
  The complete set of LO contributions and NLO corrections of EW and QCD origin is included.
  The calculation is based on full matrix elements, computed with all resonant and non-resonant contributions,
  complete spin correlations and interference effects.
  Integrated and differential cross-sections are presented for a realistic fiducial setup.
}


\begin{document}
%%%%%%%%%%%%%%%%%%%%%%%%%%%%%%%%%
\maketitle
%%%%%%%%%%%%%%%%%%%%%
\section{Introduction}\label{sec:intro}
In view of the upcoming full Run-3 dataset and of the future high-luminosity stage of the Large Hadron Collider (LHC),
the increasing level of precision of LHC experiments requires to be
matched by theoretical predictions in the Standard Model (SM).
An improved interpretation of the data will then lead to cutting-edge results
either confirming the SM or highlighting new-physics effects.
The increased luminosity will allow for precise measurements of rare signatures that are expected to unveil
the interplay between the top-quark and electroweak (EW) sector of the SM.
At the LHC, a direct access to the top-quark coupling to EW bosons is given by the measurement of the production
of a top-quark pair in association with an EW boson, namely of the
$\ttz$ and $\ttw$ production processes.
Such signals are characterised by a large number of particles in the final state and rather intricate
resonance structures involving top quarks, $\PW$, $\PZ$, and Higgs bosons, such that a proper theoretical modelling
is required at production and decay level.

The $\ttz$ process admits access to the top-quark coupling to $\PZ$~bosons, which is still poorly constrained 
by experimental measurements, leaving room for new-physics effects \cite{Buckley:2015lku,BessidskaiaBylund:2016jvp}.
This freedom has already been exploited in different beyond-the-SM (BSM) scenarios. For instance,
anomalous values of the top--$\PZ$ coupling can be interpreted in terms of the presence of top-philic resonances
like $\PZ'$  \cite{Cox:2015afa,Kim:2016plm,Fox:2018ldq,Alvarez:2020ffi} and vector-like leptons \cite{Bissmann:2020lge}.
A number of studies on $\ttz$ production has been performed in the presence of higher-dimension effective operators
\cite{Baur:2004uw,Baur:2005wi,Berger:2009hi,Rontsch:2014cca,Rontsch:2015una,Buckley:2015lku,BessidskaiaBylund:2016jvp,
  Maltoni:2019aot,Durieux:2019rbz,Brivio:2019ius,Afik:2021xmi,Rahaman:2022dwp}.
It has been noticed  \cite{Dror:2015nkp,Maltoni:2019aot} that, similarly to the case of the $\ttw$ process,
the production of $\ttz$ in association with jets is enhanced by $\Pt\PZ\rightarrow \Pt\PZ$
and $\Pb\PW\rightarrow\Pt\PZ$ underlying scattering processes.
In the presence of anomalous couplings
between the top quark and EW bosons, this mechanism for $\ttz$ production
would lead to a unitarity-violating energy growth
of the amplitude that must be regularised by new-physics mechanisms.
Moreover, the $\ttz$ process with the $\PZ$~boson decaying into neutrinos represents a background to dark-matter LHC searches
\cite{Bevilacqua:2019cvp}.

The ATLAS and CMS collaborations have measured $\ttz$ production with Run-1 \cite{Aad:2015eua,Khachatryan:2015sha}
and Run-2 data \cite{Aaboud:2016xve,Sirunyan:2017uzs,Aaboud:2019njj,CMS:2019too,ATLAS:2021fzm}.
The mostly considered decay channels are those involving three or four charged leptons, with at least two opposite-sign,
same-flavour leptons. The four-lepton one, in spite of the lower decay branching fraction, has a very clean signature
and a low background yield, thanks to the optimised lepton identification in ATLAS and CMS detectors.
Like the $\rm t\bar{\rm t}\PW$ process, the $\ttz$ one  deserves high level of interest in the LHC community
also because it represents one of the largest backgrounds to $\rm t\bar{\rm t}H$ production in the multi-lepton decay
channel \cite{Maltoni:2015ena}. A proper modelling of $\ttz$ production
is therefore crucial for both ${\rm t}\bar{\rm t}V$ and $\rm t\bar{\rm t}H$ measurements.

In the context of SM modelling, the next-to-leading-order (NLO) QCD corrections to the $\ttz$ process are known since many years \cite{Lazopoulos:2008de,Kardos:2011na}
for on-shell top quarks and $\PZ$~bosons, while the subleading NLO corrections are more recent \cite{Frixione:2015zaa,Frederix:2018nkq}.
\looseness -1
The matching of NLO QCD corrections to parton shower (PS) has been performed with leading-order (LO) decays simulated in the narrow-width approximation 
for top quarks and $\PZ$~bosons \cite{Garzelli:2011is,Garzelli:2012bn,Maltoni:2015ena} and more recently including full off-shell effects related
to the lepton pair from the $\PZ$-boson decay \cite{Ghezzi:2021rpc}.
The NLO QCD corrections to the top-quark decays in $\ttz$ production
have also been computed in the narrow-width approximation \cite{Rontsch:2014cca}.
Resummed calculations are available in the literature up to NLO+NNLL accuracy
\cite{Broggio:2017kzi,Kulesza:2018tqz,Broggio:2019ewu,Kulesza:2020nfh} in the case of on-shell resonances.
For a realistic comparison with experimental data in fiducial setups, it is necessary to properly model the decay effects,
with the possible inclusion of higher orders in perturbation theory.
The need for precise off-shell predictions is urgent, given the recent measurement of differential cross-sections
in angular and transverse-momentum observables \cite{CMS:2019too,ATLAS:2021fzm} with Run-2 data. The increased luminosity
of the upcoming data will allow for an even more precise differential description of the process.
The first off-shell calculation for the $\ttz$ reaction at NLO QCD was performed 
for the $\PZ$ boson decaying into neutrinos, with
the purpose of modelling the sizeable QCD background in dark-matter searches \cite{Bevilacqua:2019cvp}. An analogous
calculation (at NLO QCD) has been recently presented in the four-charged-lepton decay channel \cite{Bevilacqua:2022nrm}.

In this work, we present SM predictions for the $\ttz$ process that go
one step forward in accuracy and modelling. 
In addition to the NLO QCD corrections to the leading tree-level process that were first computed
in \citere{Bevilacqua:2022nrm}, we calculate all subleading
LO contributions and NLO corrections of EW and QCD type.
All kinds of non-resonant contributions, spin correlations, and interferences are accounted for at LO and at NLO. 
Photon-induced and bottom-induced partonic channels are consistently included, giving a complete NLO calculation in
the five-flavour scheme. At variance with \citere{Frixione:2015zaa}, we do not include heavy-boson radiation at NLO EW.
Although these contributions typically account for a percent in inclusive setups, they are usually
regarded as reducible backgrounds owing to a different final-state signature given by the additional heavy-boson decay products.

This article is structured as follows.
In \refse{sec:calcdetails} we detail our fixed-order calculation of the $\ttz$ process, highlighting
the most relevant theoretical aspects of EW and QCD corrections in the full off-shell description, the
partonic channels that contribute in Born-like and real topologies, and the interplay between different perturbative orders.
A comparison with existing results in the literature is presented in \refse{sec:valid}.
Our numerical results obtained in a realistic fiducial setup are shown in \refse{sec:numresults}, including
integrated cross-sections and differential distributions in typical LHC observables. In \refse{sec:conclusions} we summarise
the main achievements of the calculation and present our conclusions.

\section{Details of the calculation}\label{sec:calcdetails}

In this paper we present the complete set of NLO EW and NLO QCD corrections to the process
\beq\label{eq:procdef}
\Pp\Pp\to\Pe^+\nu_\Pe\,\mu^-\overline{\nu}_\mu\,{\Pb}\,\overline{\Pb}\,{\tau}^+{\tau}^-\,.
\eeq
The calculation has been performed with the in-house program \mocanlo,
a multichannel Monte Carlo generator that has already proven suitable for the
evaluation of processes with high-multiplicity final states and involving 
 top quarks at NLO QCD and NLO EW accuracy
\cite{Denner:2020hgg,Denner:2021hqi,Denner:2015yca,Denner:2016jyo,Denner:2017kzu,Denner:2016wet}.
It is interfaced with \recola \cite{Actis:2012qn, Actis:2016mpe}, which 
provides the tree-level SM matrix elements together with the spin-correlated and colour-correlated
ones, required for the definition of the unintegrated subtraction counterterms.
\recola also computes all the required one-loop amplitudes using the \collier
library \cite{Denner:2016kdg} to perform the reduction and numerical
evaluation of one-loop integrals
\cite{Denner:2002ii,Denner:2005nn,Denner:2010tr}.%
\footnote{While \collier proved capable in calculating all required
  one-loop integrals for the practically relevant NLO corrections to
  the process \refeqf{eq:procdef}, some improvements in the
  reduction of tensor integrals were required, in particular, for the reduction of
  rank-6 tensor integrals in 10-point functions appearing in the (very small)
  NLO corrections to the $\gamma\Pg$ and $\gamma\gamma$ partonic processes.}

We explicitly calculate the process in \refeq{eq:procdef}, whose final state involves three different
lepton flavours. Despite that, predictions for leptons of the same flavour can be roughly recovered
by multiplying our results by appropriate factors, which account for the number of identical
particles in the final state. This procedure is expected to deliver the correct cross-section
up to interference effects, which are known to have a marginal impact. Indeed,
it has been shown in \citere{Bevilacqua:2022nrm} for the same hadronic process discussed here that differences
in the LO QCD cross-section induced by the presence of same-flavour leptons are at the per-mille level for
sufficiently inclusive cuts.

It is worth mentioning from the very beginning that in our setup for all diagrams involving a resonant
$\PZ$~boson the corresponding diagrams with photon-mediated
contributions, as well as $\PZ/\gamma^*$ interference effects, are properly taken into account.
In \citere{Bevilacqua:2022nrm} it was also explicitly checked with a LO calculation that
diagrams involving a Higgs boson, despite being quite numerous, contribute at the per-mille level and can be
neglected to a first approximation. In our work all Higgs contributions are  instead retained both in LO and
in NLO terms.
We also point out that throughout our calculation a diagonal quark-mixing matrix with unit entries is assumed.

\subsection{Leading-order contributions}
\label{sec:leadingorder}

%%%%%%%%%%%%%%%%%%%%%%%%%%%%%%%%%%%%%%%%%%%%%%%%%%%%%%
\begin{figure*}
  \centering
  \includegraphics[scale=0.35]{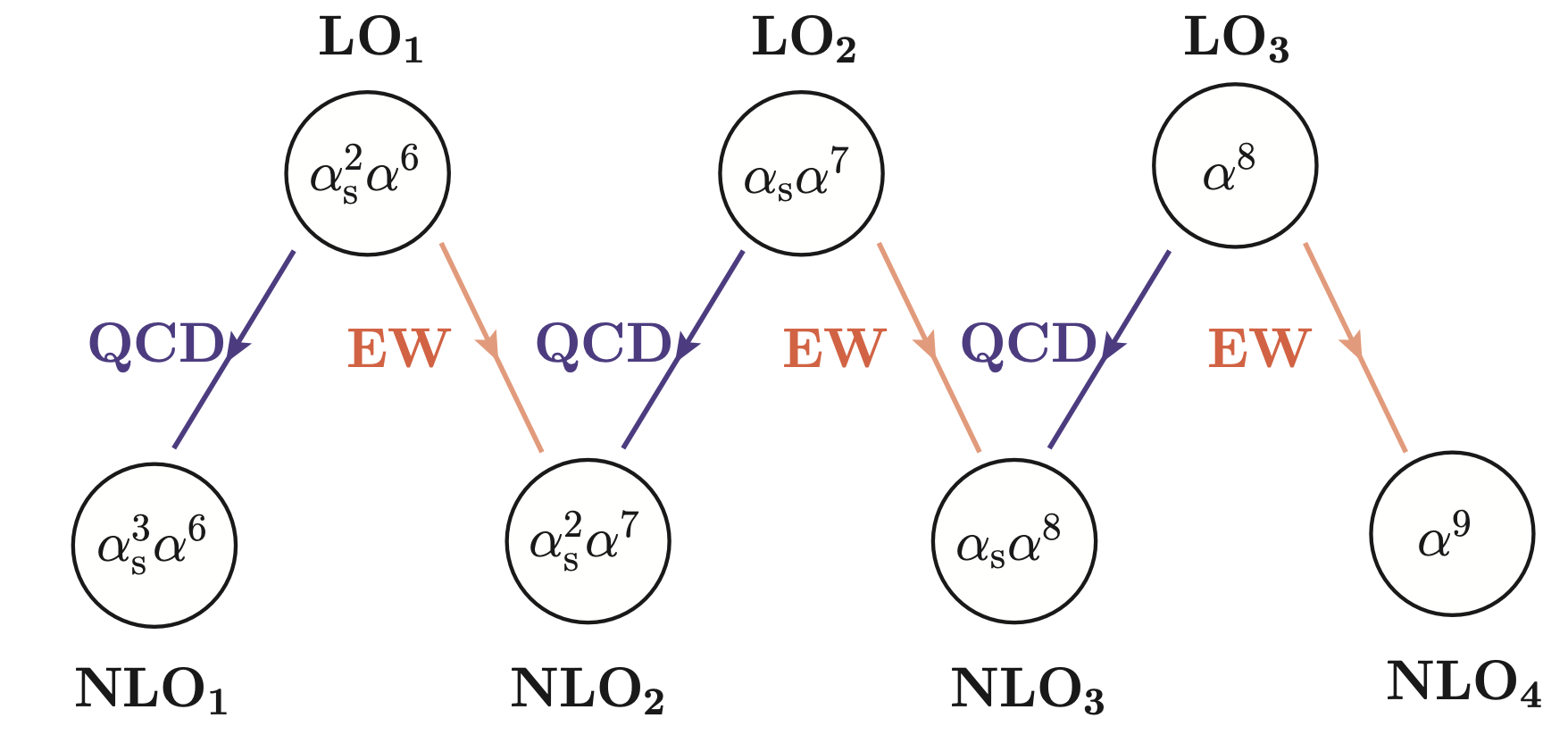}
  \caption{
    Perturbative orders contributing at LO and NLO
    for $\Pt\overline{\Pt}\PZ$ hadro-production in the
    four-charged-lepton channel.
  }\label{fig:orders}
\end{figure*}
%%%%%%%%%%%%%%%%%%%%%%%%%%%%%%%%%%%%%%%%%%%%%%%%%%%%%%
For the process \refeqf{eq:procdef} three LO contributions with different powers of
$\as$ and $\alpha$ are allowed, if the possible partons initiating the reaction are
not only selected from gluons ($\Pg$) and light quarks ($\Pq\in\{\Pu,\,\Pd,\,\Ps,\,\Pc\}$)
but also from photons ($\gamma$) and bottom quarks ($\Pb$), which are considered massless here.
This is illustrated in \reffi{fig:orders}, where
the splitting and naming of the different contributions both for the LO and
the NLO case make use of the notation proposed in \citeres{Frederix:2017wme, Frederix:2018nkq}.

\begin{figure}
  \centering
  \subfigure[\label{fig:loqcd2}]{   \includegraphics[scale=0.16]{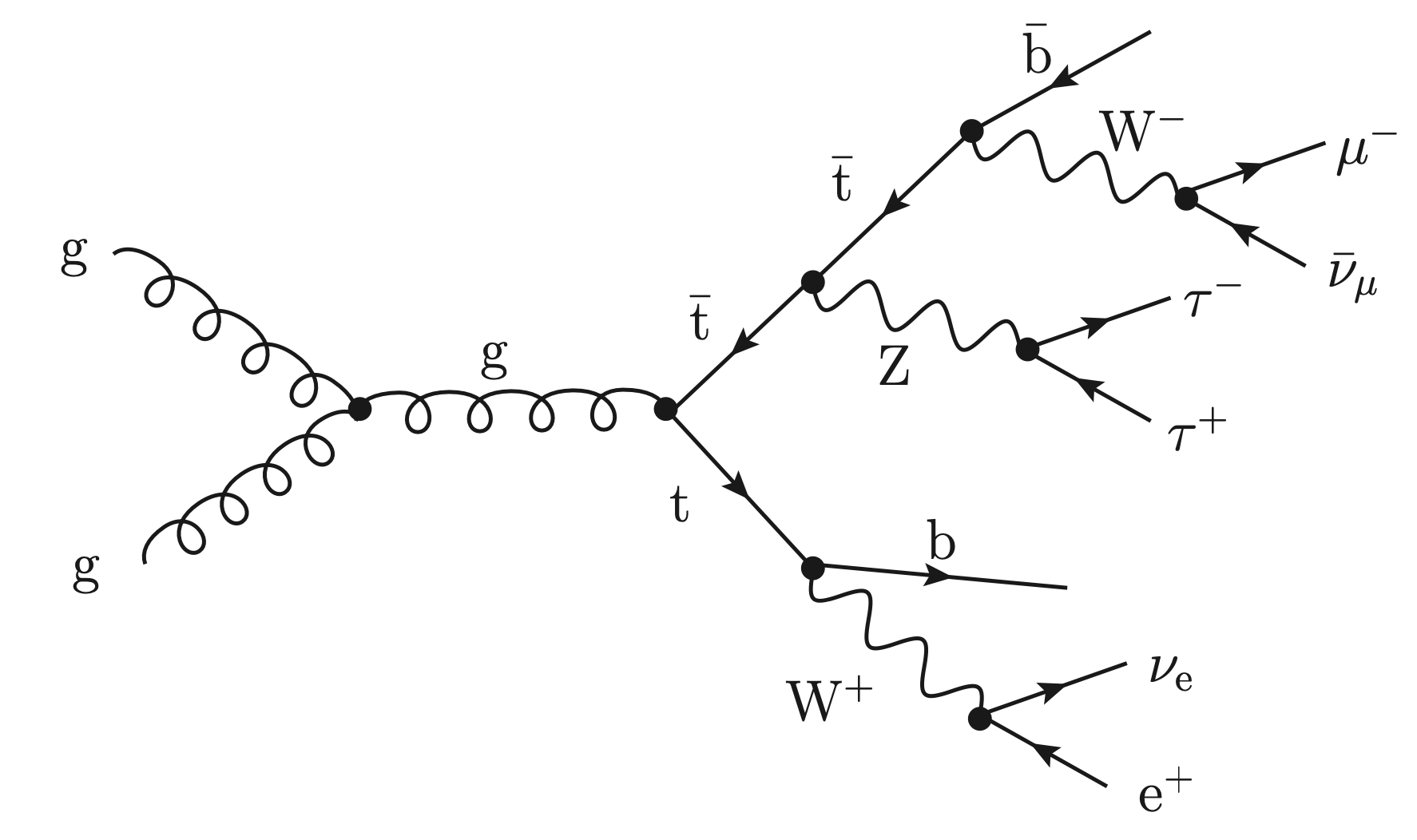}}
\subfigure[\label{fig:loqcd1}]{ \includegraphics[scale=0.16]{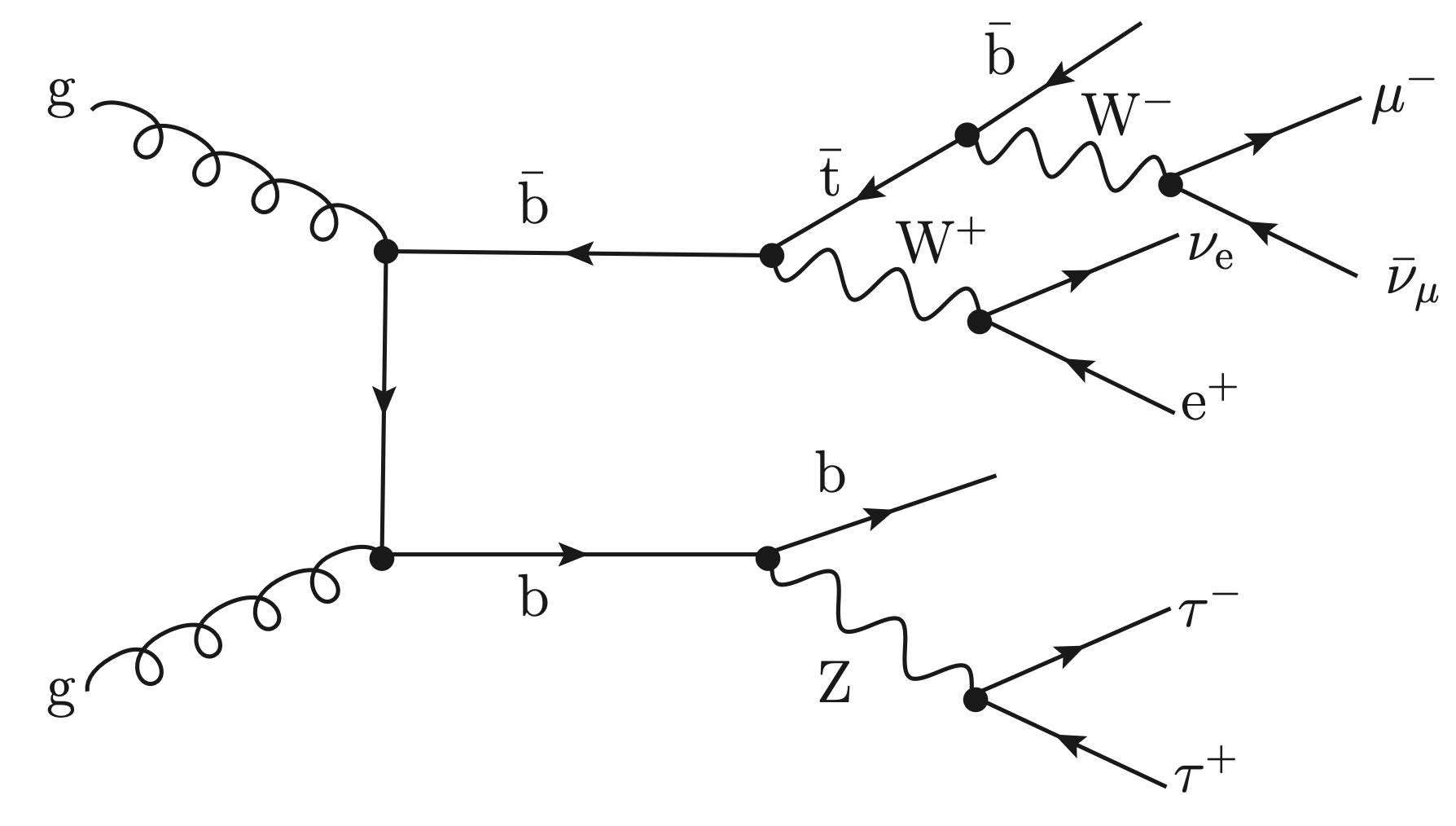}}
\subfigure[\label{fig:loqcd3}]{   \includegraphics[scale=0.16]{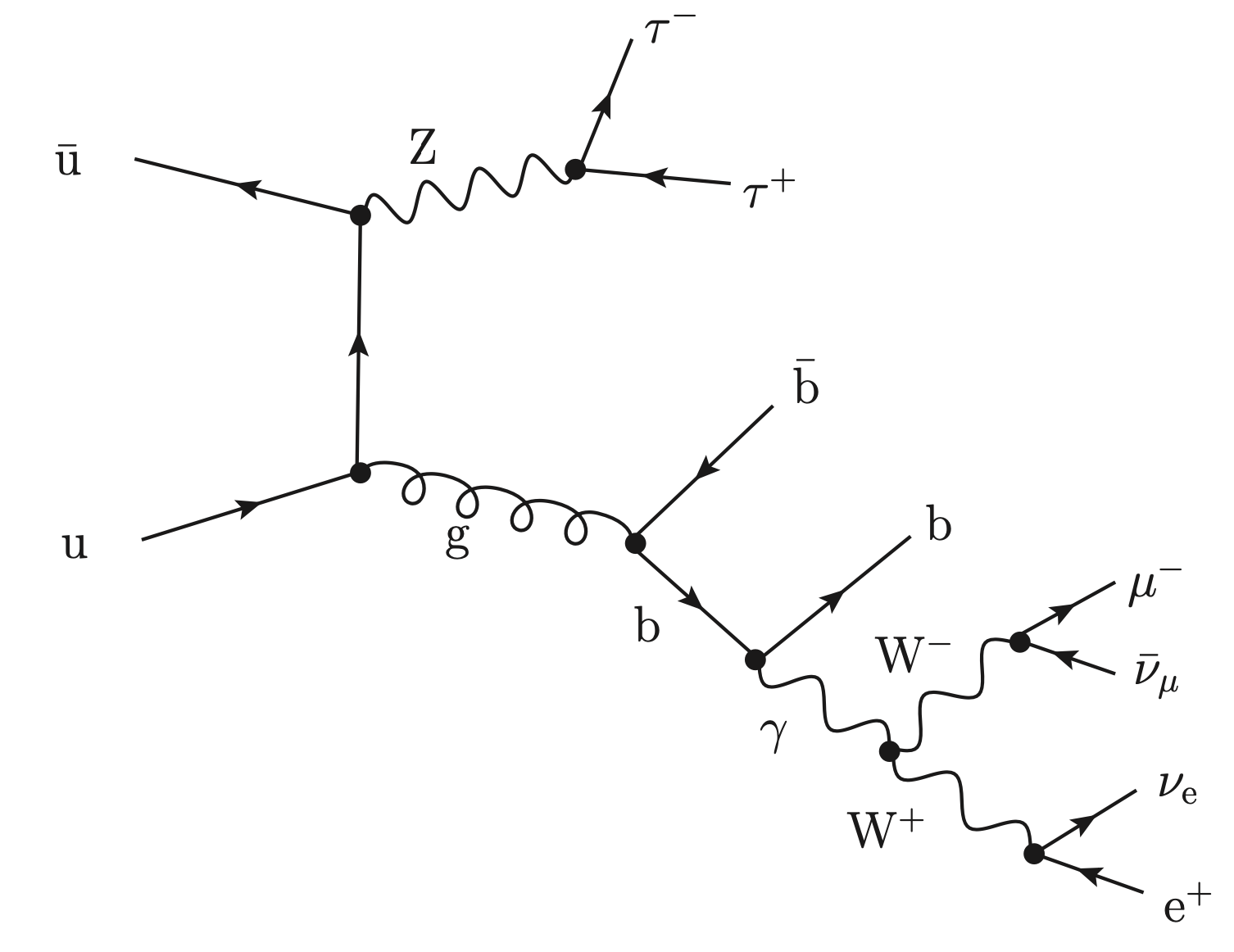}}
\caption{Sample $\loqcd$ diagrams both in the $\Pg\Pg$ and the $\Pq\bar\Pq$ channel with two~[\ref{fig:loqcd2}], one~[\ref{fig:loqcd1}], and
  zero~[\ref{fig:loqcd3}] top-quark resonances.}\label{fig:loone}
\end{figure}

The largest
LO contribution originates from QCD-mediated partonic processes of order $\mc O(\as^2\alpha^6)$,
which we dub $\loqcd$. In this case, the colliding partons can be two gluons $\Pg\Pg$ or a
light quark--antiquark pair $\Pq\bar{\Pq}$, namely:
\begin{align}\label{lo1channels}
\Pg\Pg \to{}& \Pe^+\nu_\Pe\,\mu^-\bar{\nu}_\mu\,{\Pb}\,\bar{\Pb}\,{\tau}^+{\tau}^-\,,
\quad\quad\quad\Pq\bar{\Pq} \to{} \Pe^+\nu_\Pe\,\mu^-\bar{\nu}_\mu\,{\Pb}\,\bar{\Pb}\,{\tau}^+{\tau}^-\,,
\end{align}
where the $\Pg\Pg$ partonic reaction alone requires the computation of roughly $2000$ Feynman
diagrams\footnote{The approximate number of Feynman diagrams contributing at a given perturbative
order is obtained from the one of integration channels provided by \mocanlo: even though the two numbers are
not identical, their orders of magnitude match.}, while each of the
four\footnote{Our counting is defined as follows: Each channel with two different initial-state partons has to be computed
separately for the two contributions differing by the interchange of
the initial states. On the other hand,
\mocanlo allows to compute together partonic reactions that differ by
the exchange of the first and second generation of all initial-state light quarks,
since the corresponding partonic amplitudes are identical
and just need to be reweighted by appropriate PDF
factors.\label{foot:channels}
Thus, in~\refeq{lo1channels} the four different $\Pq\bar{\Pq}$ channels to be calculated have
the initial states $\Pu\bar{\Pu}$,  $\bar\Pu\Pu$ $\Pd\bar{\Pd}$,  $\bar\Pd\Pd$.}
$\Pq\bar{\Pq}$ channels involves almost $1000$ Feynman diagrams. Indeed, gluon-initiated reactions can proceed
both via $s$-channel diagrams involving a triple-gluon coupling or $t$-channel diagrams, where the two
incoming gluons are attached to a top- or bottom-quark line, as illustrated by sample diagrams in \reffi{fig:loqcd2} and
\reffi{fig:loqcd1}, respectively. In $\Pq\bar{\Pq}$-initiated
reactions, an internal gluon always propagates in the $s$ channel,
like in the diagram shown in \reffi{fig:loqcd3}. Clearly, the gluon-induced channel represents
the dominant contribution, due to the large gluon luminosity in the protons. Despite the large number of diagrams, in both
cases more than the $90\%$ of the cross-section%
\footnote{The numerical impact of resonant and non-resonant contributions has been
estimated by means of a survey of phase-space integration channels.}
originates from diagrams involving two resonant
top quarks, either arising from an $s$-channel gluon splitting into a top--antitop pair or
from a top-quark line directly attached to the incoming gluons (for the $\Pg\Pg$-initiated processes).
Single-resonant and especially non-resonant top-quark diagrams have a much smaller
impact if compared to the doubly-resonant ones.
%All diagrams can include up to two $\PZ$ and/or two $\PW^\pm$~boson lines. 
\begin{figure}
  \centering
  \subfigure[\label{fig:loint1}]{   \includegraphics[scale=0.19]{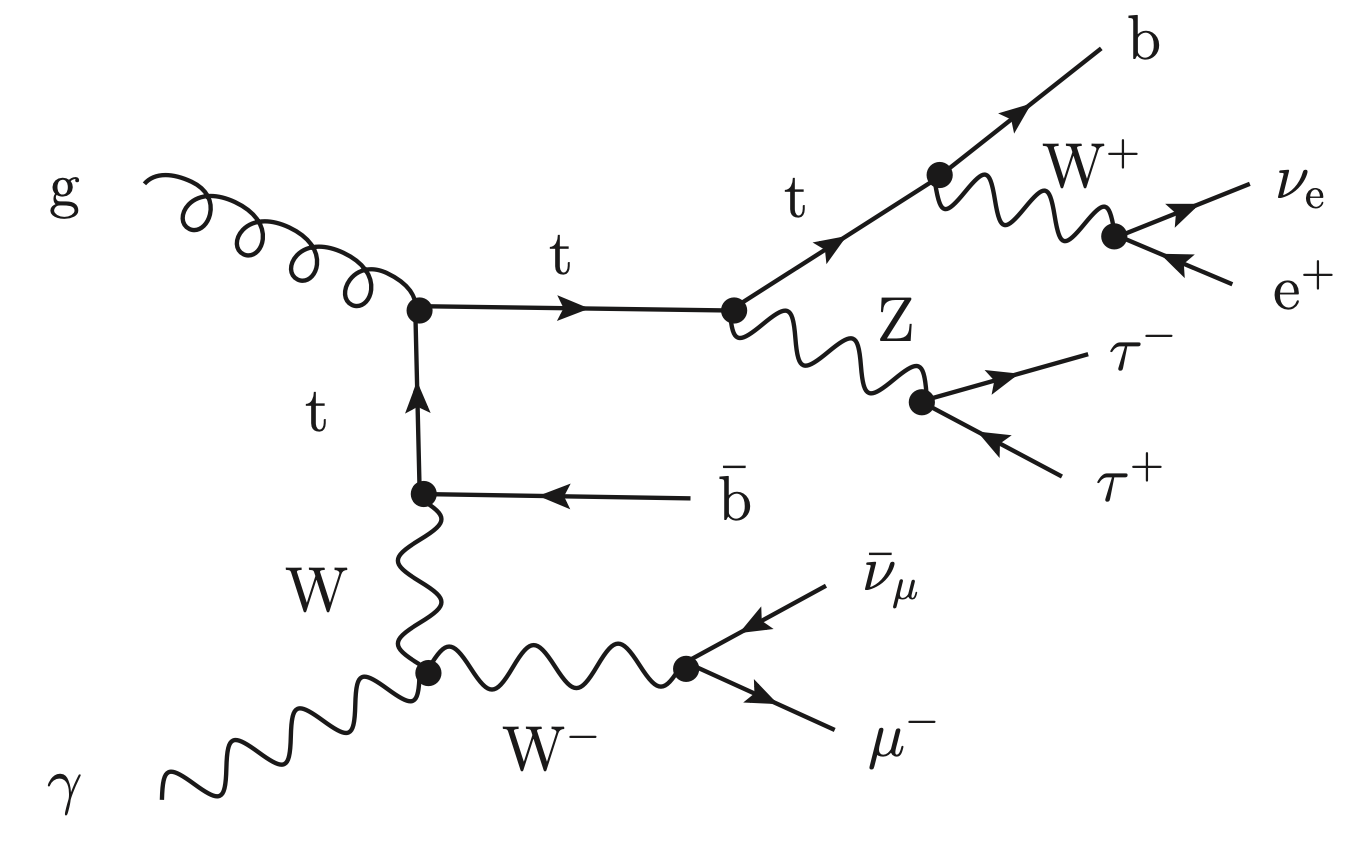}}
    \hspace{0.4cm}%
\subfigure[\label{fig:loew1}]{ \includegraphics[scale=0.16]{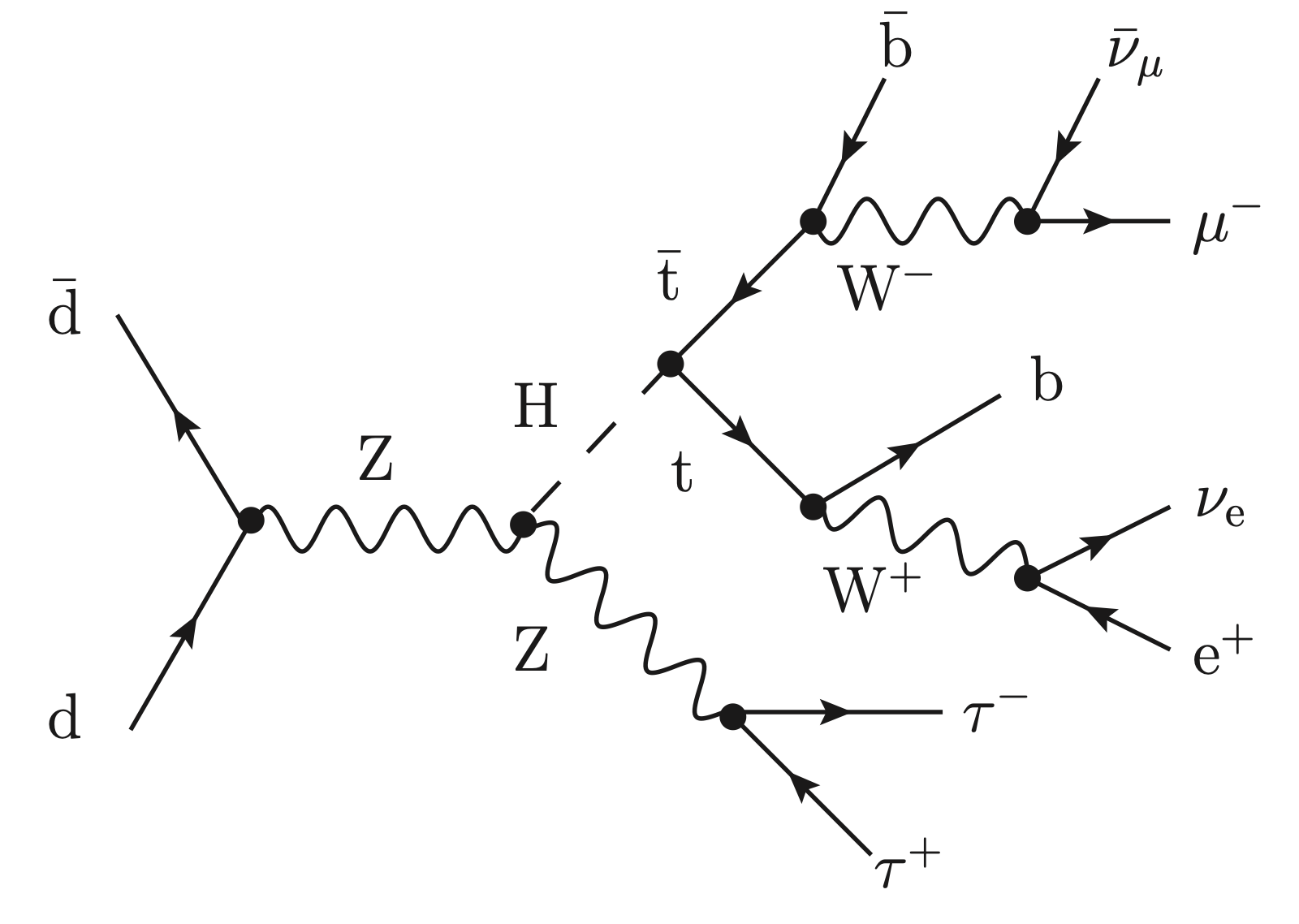}}
  \hspace{0.4cm}%
\subfigure[\label{fig:loew2}]{   \includegraphics[scale=0.17]{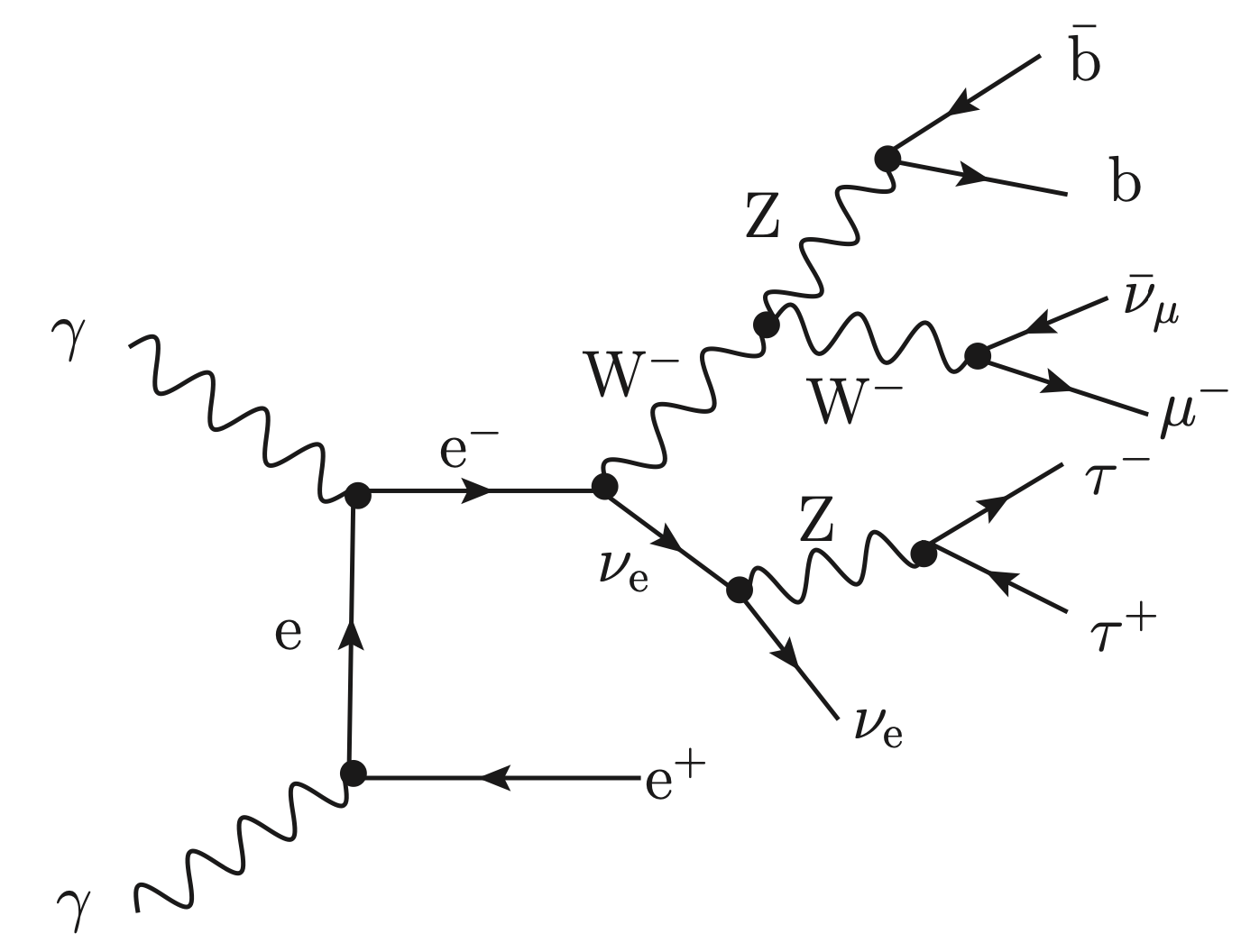}}
\caption{Sample $\loint$ diagram for the $\gamma\Pg$ channel with one top-quark resonance~[\ref{fig:loint1}] and $\loew$ diagrams
  in the $\Pq\bar\Pq$~[\ref{fig:loew1}] and $\gamma\gamma$ channel~[\ref{fig:loew2}] involving two and zero
  top-quark resonances, respectively.}\label{fig:losub}%
\end{figure}%

At the order $\mc O(\as \alpha^7)$, labelled $\loint$, as long as initial-state bottom quarks are excluded,
the only non-zero contribution arises from the photon--gluon-induced partonic channel
\begin{align}\label{lo2channels}
  \gamma\Pg \to{}& \Pe^+\nu_\Pe\,\mu^-\bar{\nu}_\mu\,{\Pb}\,\bar{\Pb}\,{\tau}^+{\tau}^-\,.
\end{align}
All other contributions at this perturbative order come from the interference between EW- and QCD-mediated quark-induced
diagrams, which vanish owing to colour algebra. Despite the fact that the LO
contribution from these channels is expected to be just a small fraction of the $\loqcd$,
due to the small photon luminosity, their calculation involves roughly $4000$ 
Feynman diagrams.
Moreover, the role of doubly-resonant top-quark topologies is not as dominant as for the $\loqcd$
case: an equally important fraction of the result is ruled by single-resonant diagrams, which
involve an incoming photon connected to the rest of the process by a $\gamma\PW^+\PW^-$ triple-gauge-boson
vertex [like the one reported in \reffi{fig:loint1}].

Finally, an even smaller contribution originates from the fully EW-mediated processes, entering at order
$\mc O(\alpha^7)$, that we refer to as $\loew$ term. Due to the suppression caused by the small ratio of the  $\alpha$ and the
$\as$ couplings, this contribution is expected to amount to roughly 
$1\%$ of the $\loqcd$. In addition to the $\Pq\bar{\Pq}$-initiated partonic reactions,
sharing the same external particles as its $\loqcd$ counterpart in \refeq{lo1channels}
(but involving now only EW interaction vertices), the $\gamma\gamma$-induced channel also participates
in the process through the reaction
\begin{align}\label{lo3channels}
  \gamma\gamma \to{}& \Pe^+\nu_\Pe\,\mu^-\bar{\nu}_\mu\,{\Pb}\,\bar{\Pb}\,{\tau}^+{\tau}^-\,.
\end{align}
In comparison with the $\Pq\bar{\Pq}$ channels, the $\gamma\gamma$ one represents just a tiny fraction of the overall $\loew$ result,
as a consequence of the additional suppression from the photon
parton-distribution functions (PDFs).
In spite of a negligible numerical impact, the EW nature of these processes allows for a very high number
of topologies, involving approximately $6000$ Feynman diagrams for $\Pq\bar{\Pq}$ channels and $17000$ Feynman diagrams
for the photon-initiated ones, where resonant and non-resonant configurations play an equally important
role. Sample Feynman diagrams in the two configurations contributing at $\loew$ are shown in \reffis{fig:loew1} and \ref{fig:loew2}.

Throughout our calculation, we made use of a five-flavour scheme, where the bottom quarks are treated as
massless. Within this scheme, a consistent calculation should also include bottom-induced processes.
The small value of the bottom PDFs is expected to highly suppress these contributions, as explicitly
confirmed by a computation carried out  in \citere{Stremmer:2021bnk} for $\Pt\bar{\Pt}\PH$ production.
Nevertheless, we decided to carry out their calculation for the $\Pt\bar{\Pt}\PZ$ process consistently at all LO and NLO orders.
Having a quantitative control on bottom-induced contributions is especially relevant in the calculation we present,
because the size of the bottom-quark channels may become comparable to some of the subleading NLO corrections
to the corresponding light-quark ones.

\begin{figure}
  \centering
  \subfigure[\label{fig:lob1}]{   \includegraphics[scale=0.16]{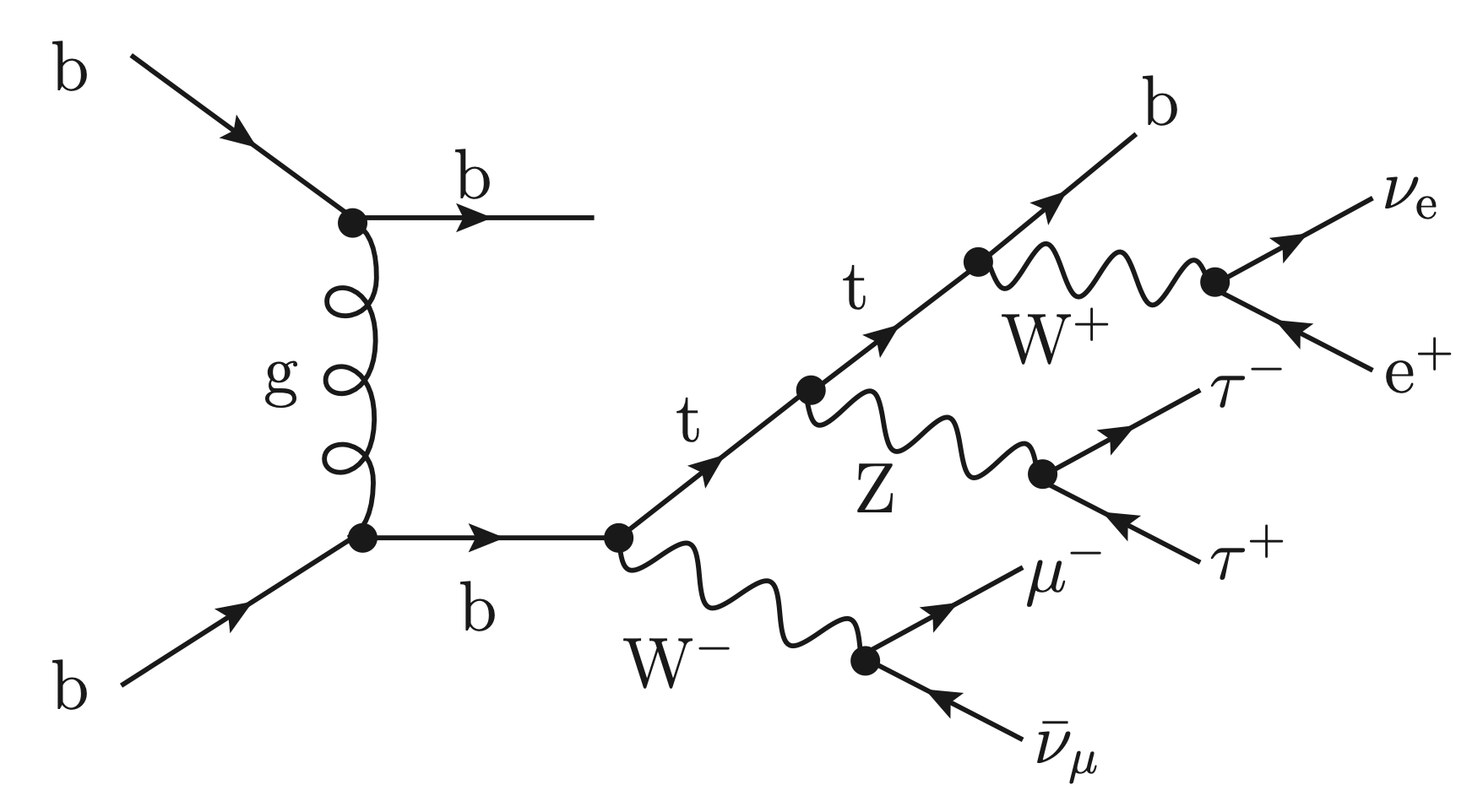}}
    \hspace{0.4cm}
\subfigure[\label{fig:lob2}]{ \includegraphics[scale=0.15]{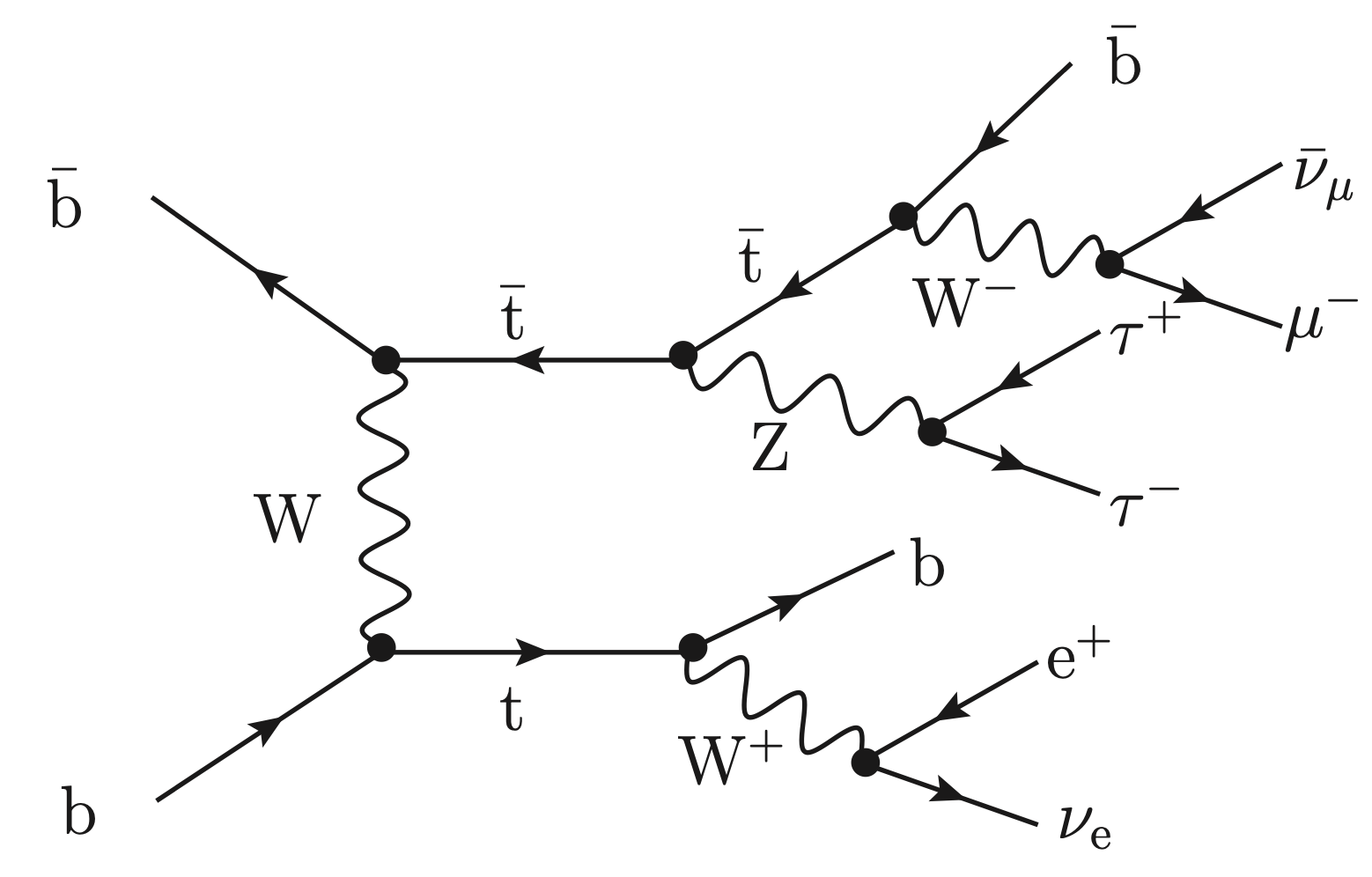}}
  \hspace{0.4cm}
\subfigure[\label{fig:lob3}]{   \includegraphics[scale=0.15]{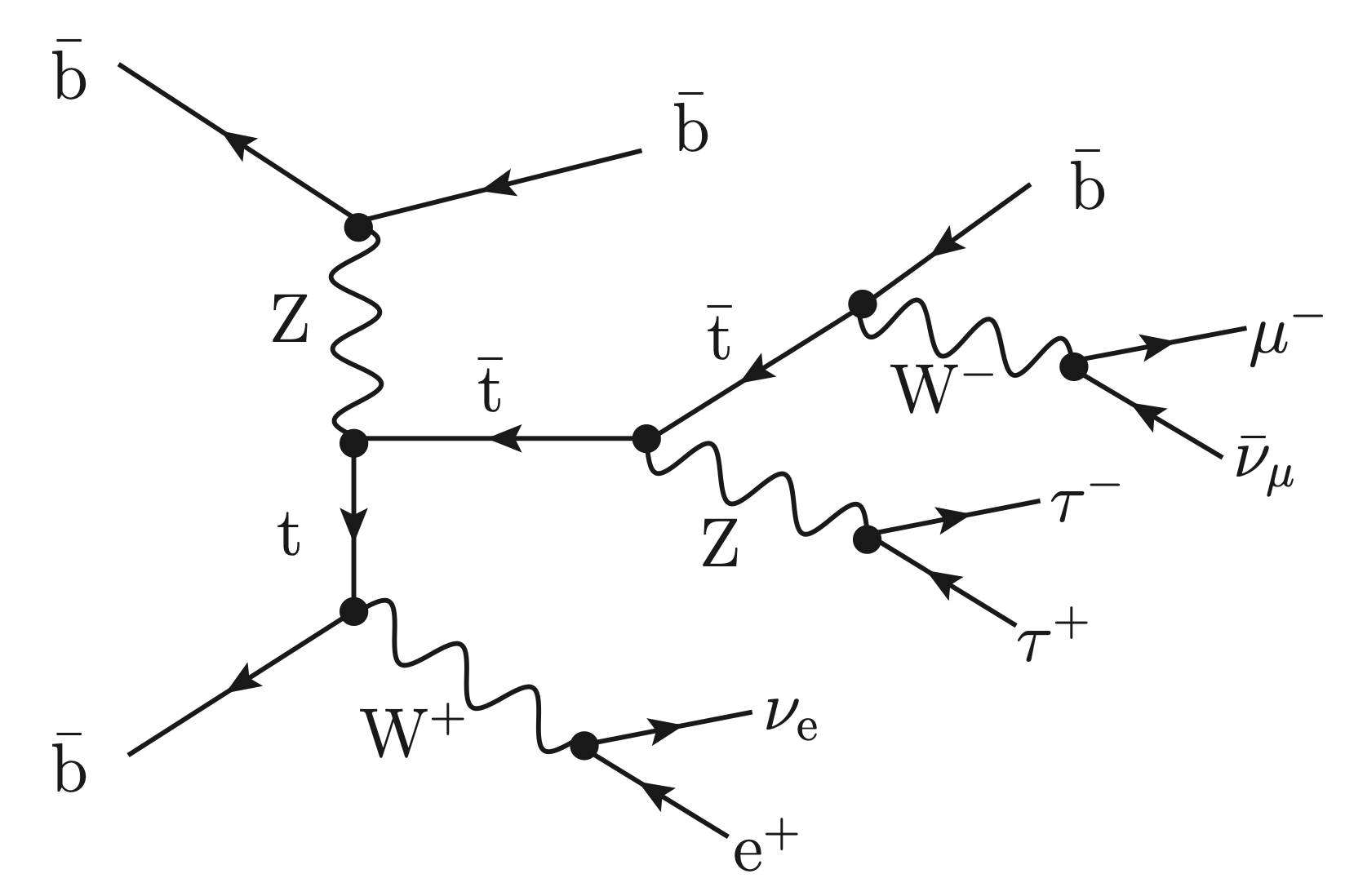}}
\caption{Sample $\loqcd$ diagram for the $\Pb\Pb$ bottom channel [\ref{fig:lob1}] and $\loew$ diagrams
  in the $\Pb\bar{\Pb}$ [\ref{fig:lob2}] and $\bar{\Pb}\bar{\Pb}$ [\ref{fig:lob3}] bottom channels.}\label{fig:lobottoms}
\end{figure}

Customary $\Pb$-jet taggings are charge blind, which means that bottom- and antibottom-initiated jets are not distinguished
and have to be treated on the same footing. Therefore, already at LO additional partonic channels are present on top of
the $\Pb\bar{\Pb}$ ones, namely the $\Pb\Pb$ and $\bar{\Pb}\bar{\Pb}$ reactions,
\begin{align}\label{lob1channels}
&\quad\quad\quad\quad\Pb\bar{\Pb} \to{} \Pe^+\nu_\Pe\,\mu^-\bar{\nu}_\mu\,{\Pb}\,\bar{\Pb}\,{\tau}^+{\tau}^-\,,\nonumber\\
\Pb\Pb \to{}& \Pe^+\nu_\Pe\,\mu^-\bar{\nu}_\mu\,{\Pb}\,{\Pb}\,{\tau}^+{\tau}^-\,,
\quad\quad\quad\bar{\Pb}\bar{\Pb} \to{} \Pe^+\nu_\Pe\,\mu^-\bar{\nu}_\mu\,\bar{\Pb}\,\bar{\Pb}\,{\tau}^+{\tau}^-\,.
\end{align}
The LO terms for the bottom channels contributing at $\mc O(\as^2\alpha^6)$ (from now on $\lobqcd$)
require a computation of roughly $2000$  Feynman diagrams for each partonic
process. The $\Pb\bar{\Pb}$~channels are expected to dominate the
$\loqcd$ cross-section, since they mainly
receive contributions from doubly-resonant diagrams;  the $\Pb\Pb$/$\bar{\Pb}\bar{\Pb}$ bottom channels
are instead resonance-suppressed, since only single or non-resonant
topologies with a $t$-channel gluon exchange can contribute, as shown in \reffi{fig:lob1}.

At the order $\mc O(\as \alpha^7)$, named $\lobint$, differently from the light-quark mediated case,
non-vanishing contributions result from bottom-initiated reactions,
both for the $\Pb\bar{\Pb}$ and the $\Pb\Pb$/$\bar{\Pb}\bar{\Pb}$ ones.
For instance, for the $\Pb\bar{\Pb}$ case
non-vanishing contributions arise by interfering  
$\rm EW$ tree-level diagrams with a $t$-channel $W$~boson [as the one in~\reffi{fig:lob2}] and
 $\rm QCD$ ones with an $s$-channel gluon, and vice versa $\rm EW$ tree-level diagrams
with an $s$-channel photon or $\PZ$~boson and $\rm QCD$ ones with a $t$-channel gluon.
For the $\Pb\Pb$/$\bar{\Pb}\bar{\Pb}$ case no $s$-channel topologies are allowed both
for $\rm EW$ and $\rm QCD$ diagrams. Nevertheless, due to the presence of identical bottom
quarks in the final state non-vanishing terms still emerge by interfering
$t$-channel $\rm QCD$ diagrams [as the one in~\reffi{fig:lob1}] and
$u$-channel $\rm EW$ ones, and vice versa.

Finally, we have also included the bottom channels at the order $\mc
O(\alpha^8)$ or $\lobew$.  This computation requires to deal with
approximately $11000$ Feynman diagrams for each contribution
in~\refeq{lob1channels}, where the same channels now involve only EW
interaction vertices.  Besides configurations obtained by replacing a
gluon with a neutral EW boson in a given $\lobqcd$ diagram, a new
class of reactions appears at this order. This contains Feynman
diagrams involving a $t$-channel EW boson exchange like those reported
in \reffi{fig:lob2}, where a $\PW$-mediated scattering between a
$\Pb$-quark and a $\bar{\Pb}$-quark takes place, and in
\reffi{fig:lob3}, which illustrates a $\bar{\Pb}$-quark and an
antitop-quark scattering via a $\PZ$~boson exchange.

\subsection{Next-to-leading-order corrections}
\label{sec:nexttoleadingorder}
As summarised in \reffi{fig:orders}, NLO corrections to the $\Pt\bar{\Pt}\PZ$
process can be classified in four different categories, according to the perturbative order
they contribute to. Pure QCD corrections enter at order $\mc O(\as^3 \alpha^6)$ and are
denoted by $\nloone$ contributions, followed by $\nlotwo$ and $\nlothree$ corrections entering at the orders
$\mc O(\as^2 \alpha^7)$ and $\mc O(\as \alpha^8)$, respectively, and finally pure EW corrections, dubbed $\nlofour$,
which contribute at order $\mc O(\alpha^9)$. Perturbative corrections to the bottom-induced LO contributions
will be occasionally referred to as $\nlobone$, $\nlobtwo$, $\nlobthree$, and $\nlobfour$, respectively.

Both QCD and QED singularities of infrared and collinear origin that plague the real
contributions are treated using the dipole subtraction formalism
\cite{Catani:1996vz,Dittmaier:1999mb,Catani:2002hc,Dittmaier:2008md}.
The initial-state collinear singularities are absorbed in the PDFs in the $\MSbar$ factorisation scheme. 

Throughout our calculation, the complex-mass scheme for all unstable particles is used
\cite{Denner:1999gp,Denner:2005fg,Denner:2006ic,Denner:2019vbn},
resulting in complex input values for the EW boson masses,
the top-quark mass, and the EW mixing angle,
\beq
\mu_V^2 = \Mv^2-\ri\Gv \Mv\quad (V={\PW,\PZ})\,,\qquad 
\mu_\Pt^2 = \Mt^2-\ri\Gt \Mt\,,\qquad 
\cos^2\theta_{\rw} = \frac{\mu_{\PW}^2}{\mu_{\PZ}^2}\,.
\eeq

\subsubsection{Contributions of order $\mc O(\as^3 \alpha^6)$}
\label{sec:nlo1}

\begin{figure}
  \centering
  \subfigure[$\nloone$ loop\label{fig:loop1}]{ \includegraphics[scale=0.18]{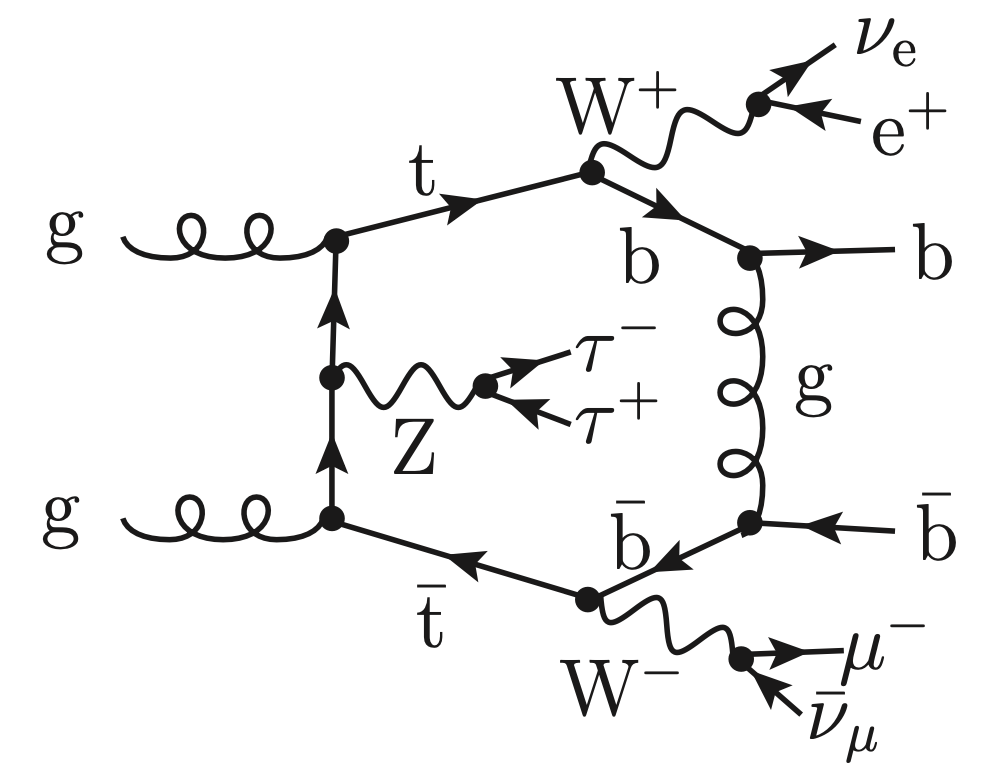}}
  \subfigure[$\nlotwo$ loop\label{fig:loop2}]{   \includegraphics[scale=0.17]{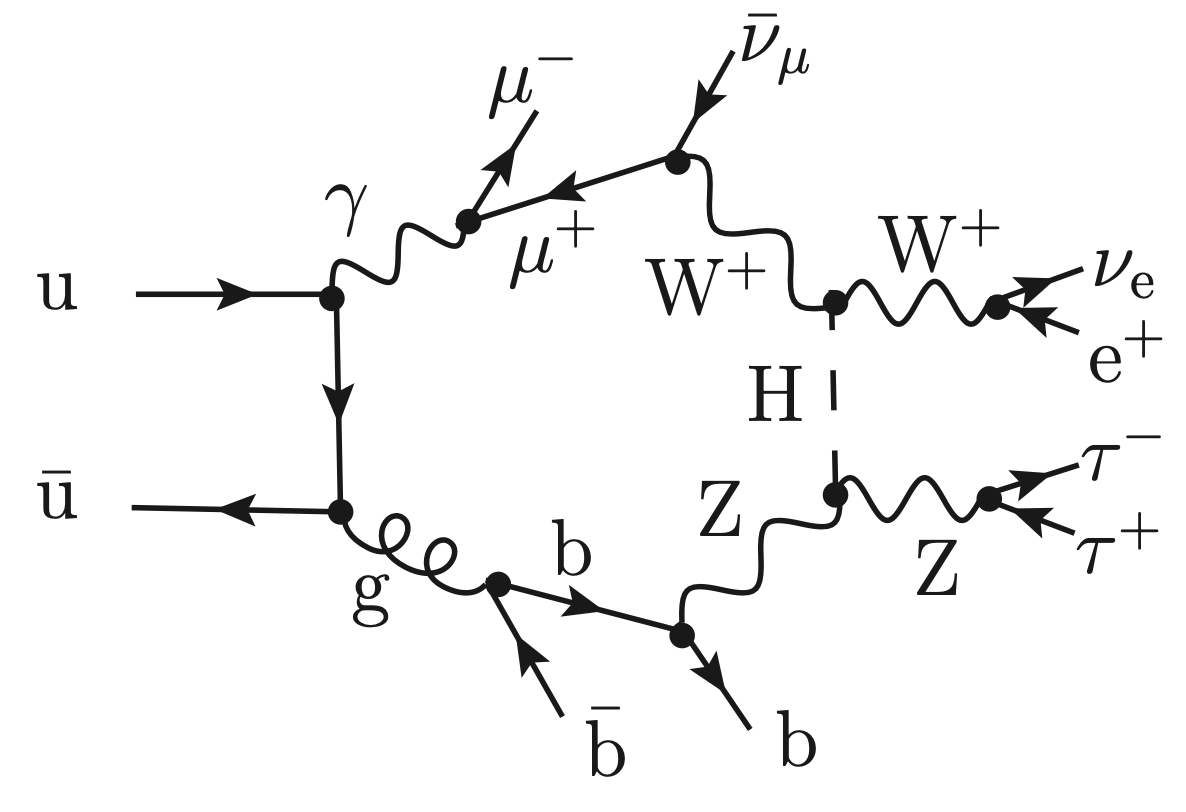}}
  \subfigure[$\nlothree$ loop\label{fig:loop3}]{   \includegraphics[scale=0.16]{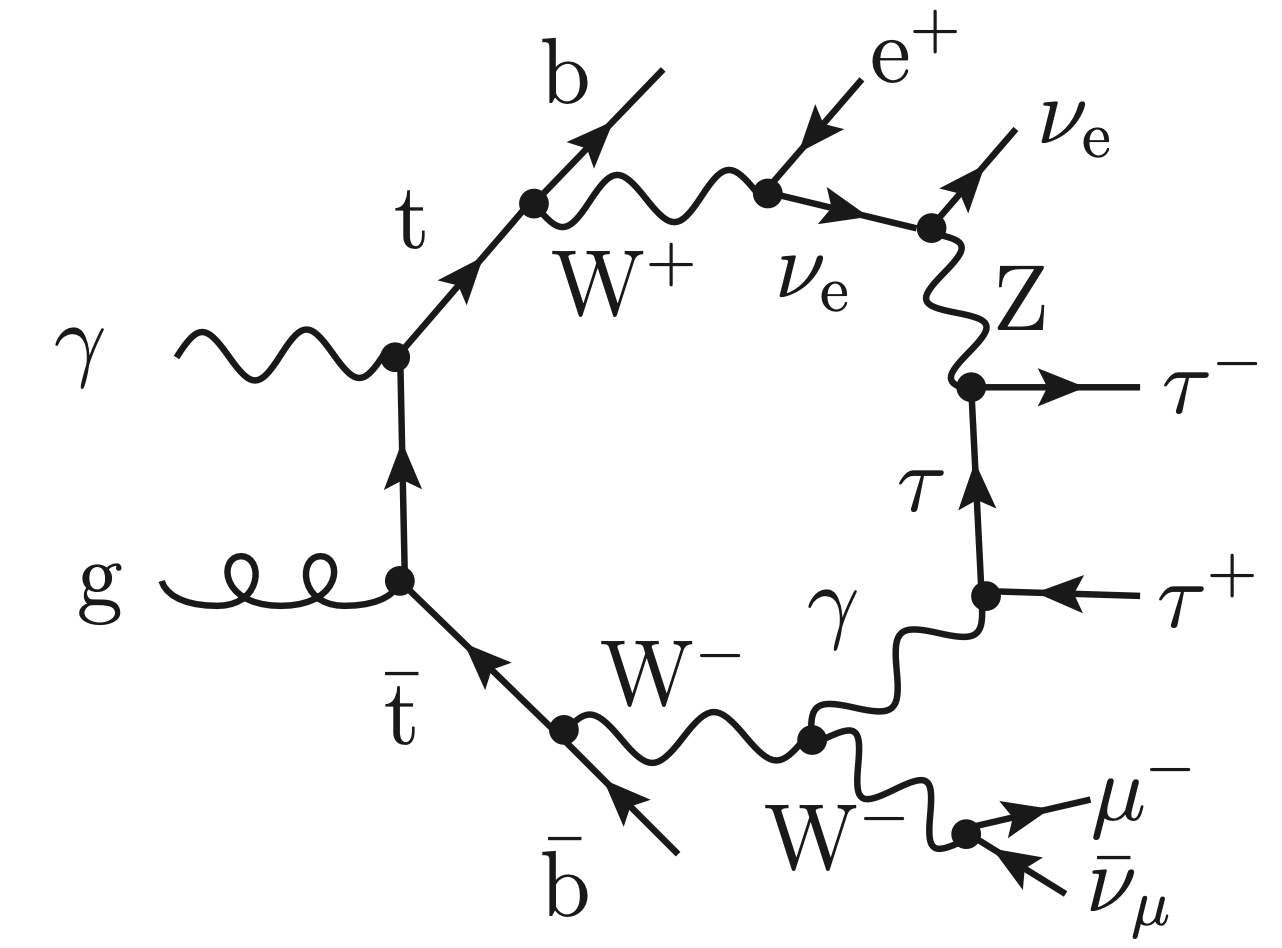}}
    \subfigure[$\nlofour$ loop\label{fig:loop4}]{   \includegraphics[scale=0.15]{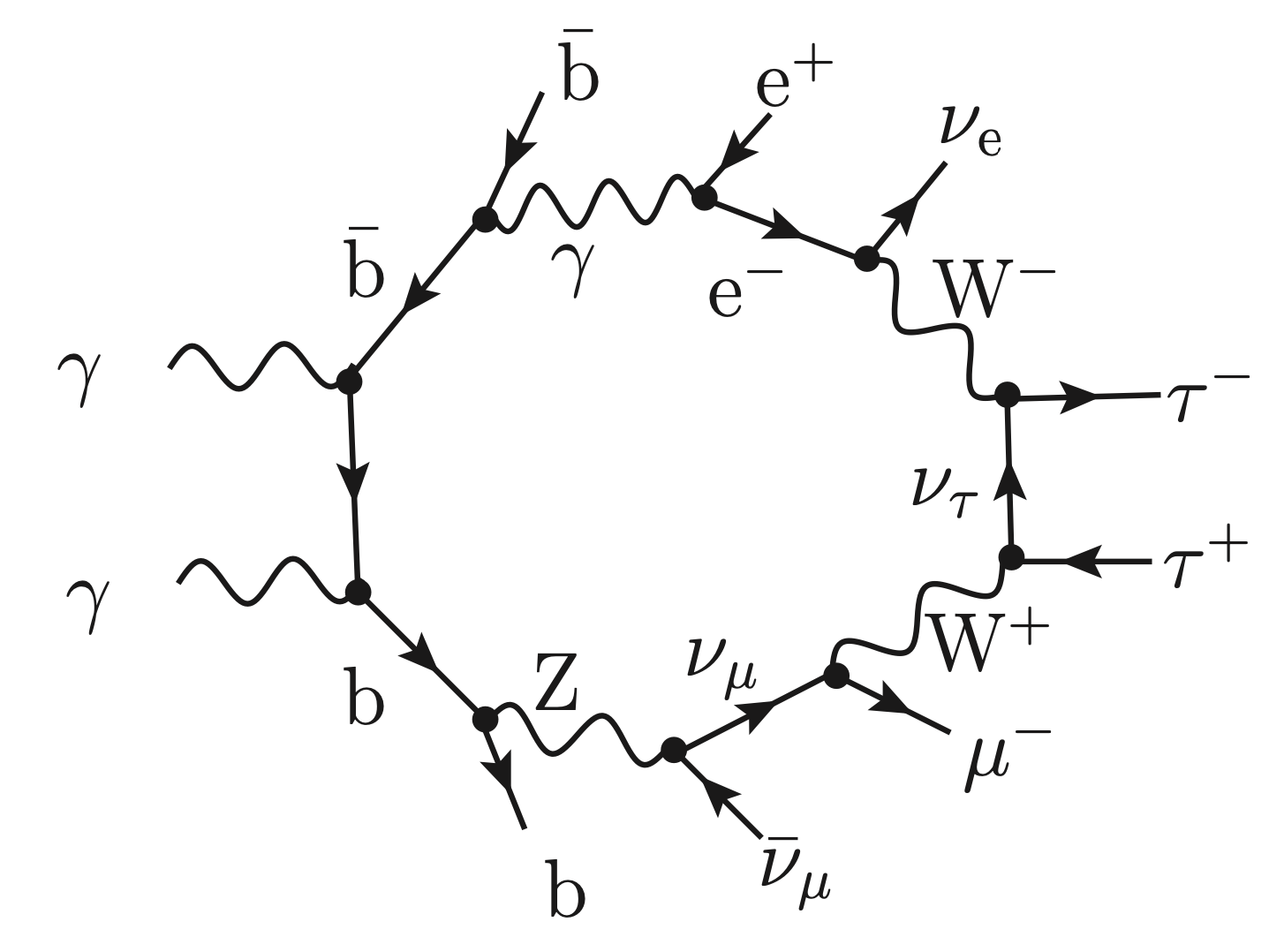}}
    \caption{Sample one-loop diagrams involving loop functions of different complexity:
      $7$-point, rank-$6$ loop functions for the $\Pg\Pg$ channel entering at $\nloone$ [\ref{fig:loop1}];
      $8$-point, rank-$3$ loop functions for the $\Pq\bar\Pq$ channel entering at $\nlotwo$ [\ref{fig:loop2}];
      $9$-point, rank-$5$ loop functions for the $\gamma\Pg$ channel entering at $\nlothree$ [\ref{fig:loop3}];
     $10$-point, rank-$6$ loop functions for the $\gamma\gamma$ channel entering at $\nlofour$ [\ref{fig:loop4}].
}\label{fig:loop}
\end{figure}

The QCD corrections to $\loqcd$ give by far
the largest NLO contribution. The real corrections originate from
 $13$ different partonic processes (according to the counting explained in footnote~\ref{foot:channels}),
which can be summarised in the following reactions,
\begin{align}\label{nlo1realchannels}
\Pg\Pg \to{}& \Pe^+\nu_\Pe\,\mu^-\bar{\nu}_\mu\,{\Pb}\,\bar{\Pb}\,{\tau}^+{\tau}^-\Pg\,,
\quad\quad\quad\Pq\bar{\Pq} \to{} \Pe^+\nu_\Pe\,\mu^-\bar{\nu}_\mu\,{\Pb}\,\bar{\Pb}\,{\tau}^+{\tau}^-\Pg\,,\nonumber\\
\Pg\bar{\Pq} \to{}& \Pe^+\nu_\Pe\,\mu^-\bar{\nu}_\mu\,{\Pb}\,\bar{\Pb}\,{\tau}^+{\tau}^-\bar{\Pq}\,,
\quad\quad\quad\Pg\Pq \to{} \Pe^+\nu_\Pe\,\mu^-\bar{\nu}_\mu\,{\Pb}\,\bar{\Pb}\,{\tau}^+{\tau}^-\Pq\,.
\end{align}
Owing to the large gluon luminosity, a considerable fraction of real
corrections is due to the first partonic channel, which,
together with the $\Pq\bar{\Pq}$ channel, is the
one characterised by the largest number of IR singularities. The evaluation of the $\Pg\Pg$-initiated real channel
is particularly challenging from a computational viewpoint, both for its large numerical contribution and for
the higher number of integration channels compared to the $\Pq\bar{\Pq}$ real
one.
The calculation of the virtual corrections
requires the evaluation of up to $7$-point loop functions, having at most rank-$4$ and rank-$6$ for the $\Pq\bar{\Pq}$ and
the $\Pg\Pg$ case, respectively. Owing to the presence of the $\Pg\Pg$ channel, loop diagrams involving $7$-point functions
can also occur for doubly-resonant topologies [see for instance \reffi{fig:loop1}], as opposed to the $\Pt\bar\Pt\PW$ case \cite{Denner:2021hqi}.
Since the $\nloone$ contribution to the $\Pt\bar\Pt\PZ$ process was already computed
in \citere{Bevilacqua:2022nrm} within the \helacnlo framework \cite{Bevilacqua:2011xh}, we
have benefited from this calculation and, as a first step,
we reproduced these corrections with \mocanlo. We compared our results with the ones
of \citere{Bevilacqua:2022nrm} finding very good agreement, as shown and further discussed in
\refse{sec:valid}.

At this perturbative order we have also included for the first time the QCD corrections to the $\lobqcd$
contribution, whose real part comprises $8$ partonic contributions to be calculated with the following channels,
\begin{align}\label{nlob1realchannels}
&\quad\quad\quad\quad\Pb\bar{\Pb} \to{} \Pe^+\nu_\Pe\,\mu^-\bar{\nu}_\mu\,{\Pb}\,\bar{\Pb}\,{\tau}^+{\tau}^-\Pg\,,\nonumber\\
\Pb\Pb \to{}& \Pe^+\nu_\Pe\,\mu^-\bar{\nu}_\mu\,{\Pb}\,{\Pb}\,{\tau}^+{\tau}^-\Pg\,,
\quad\quad\quad\bar{\Pb}\bar{\Pb} \to{} \Pe^+\nu_\Pe\,\mu^-\bar{\nu}_\mu\,\bar{\Pb}\,\bar{\Pb}\,{\tau}^+{\tau}^-\Pg\,,\nonumber\\
\Pg\bar{\Pb} \to{}& \Pe^+\nu_\Pe\,\mu^-\bar{\nu}_\mu\,{\Pb}\,\bar{\Pb}\,{\tau}^+{\tau}^-\bar{\Pb}\,,
\quad\quad\quad\Pg\Pb \to{} \Pe^+\nu_\Pe\,\mu^-\bar{\nu}_\mu\,{\Pb}\,\bar{\Pb}\,{\tau}^+{\tau}^-\Pb\,,
\end{align}
where the $\Pg\Pb$/$\Pg\bar{\Pb}$-initiated ones involve three bottom quarks in the final state.
The related diagrams include additional
collinear singularities as compared to the ones in \refeq{nlo1realchannels}, due to a
gluon splitting into a $\Pb\bar{\Pb}$ pair. Nevertheless, these singularities do not need to be
subtracted by introducing a dedicated subtraction term, since a $\Pt\bar\Pt\PZ$ signature
requires at least two resolved $\Pb$ jets in the final state that survive the fiducial cuts.
Therefore, configurations involving two collinear $\Pb$ quarks are simply cut away if recombination
rules that cluster two $\Pb$ jets into a light one are enforced as part of the jet-clustering algorithm,
as further described in \refse{sec:input} in \refeq{eq:recrules}.

\subsubsection{Contributions of order $\mc O(\as^2 \alpha^7)$}
\label{sec:nlo2}

The $\nlotwo$ corrections to the process \refeqf{eq:procdef} are the result of two
different types of contributions, as shown pictorially in \reffi{fig:orders}:
EW corrections to $\loqcd$ and QCD corrections to $\loint$.

This distinction does not lead to any ambiguity for the real part of the calculation,
where all squared diagrams can be attributed to one of the two contributions. Real EW
corrections to the $\loqcd$ term are characterised by the presence of an additional photon,
either as an initial-state parton or as a final-state particle, as summarised in the following
reactions,
\begin{align}\label{nlo2realchannels}
  \Pg\Pg \to{}& \Pe^+\nu_\Pe\,\mu^-\bar{\nu}_\mu\,{\Pb}\,\bar{\Pb}\,{\tau}^+{\tau}^-\gamma\,,
  \quad\quad\quad\Pq\bar{\Pq} \to{} \Pe^+\nu_\Pe\,\mu^-\bar{\nu}_\mu\,{\Pb}\,\bar{\Pb}\,{\tau}^+{\tau}^-\gamma\,,\nonumber\\
\gamma\bar{\Pq} \to{}& \Pe^+\nu_\Pe\,\mu^-\bar{\nu}_\mu\,{\Pb}\,\bar{\Pb}\,{\tau}^+{\tau}^-\bar{\Pq}\,,
\quad\quad\quad\gamma\Pq \to{} \Pe^+\nu_\Pe\,\mu^-\bar{\nu}_\mu\,{\Pb}\,\bar{\Pb}\,{\tau}^+{\tau}^-\Pq\,.
\end{align}
The $\Pg\Pg$ and $\Pq\bar\Pq$ channels are again the most CPU-intensive ones to evaluate,
since they represent the most significant fraction of the $\nlotwo$ corrections and comprise many
more integration channels as the corresponding real QCD counterparts in \refeq{nlo1realchannels}.
%roughly $60\%$ and $80\%$ diagrams more than the corresponding real QCD reactions, respectively.
%gg -> g 12428  uux -> g 6440
% gg -> a 20222 uux -> a 11964
\begin{figure}%[t]
  \centering
  \subfigure[$\nlotwo$ real diagram \label{fig:nlo41}]{   \includegraphics[scale=0.20]{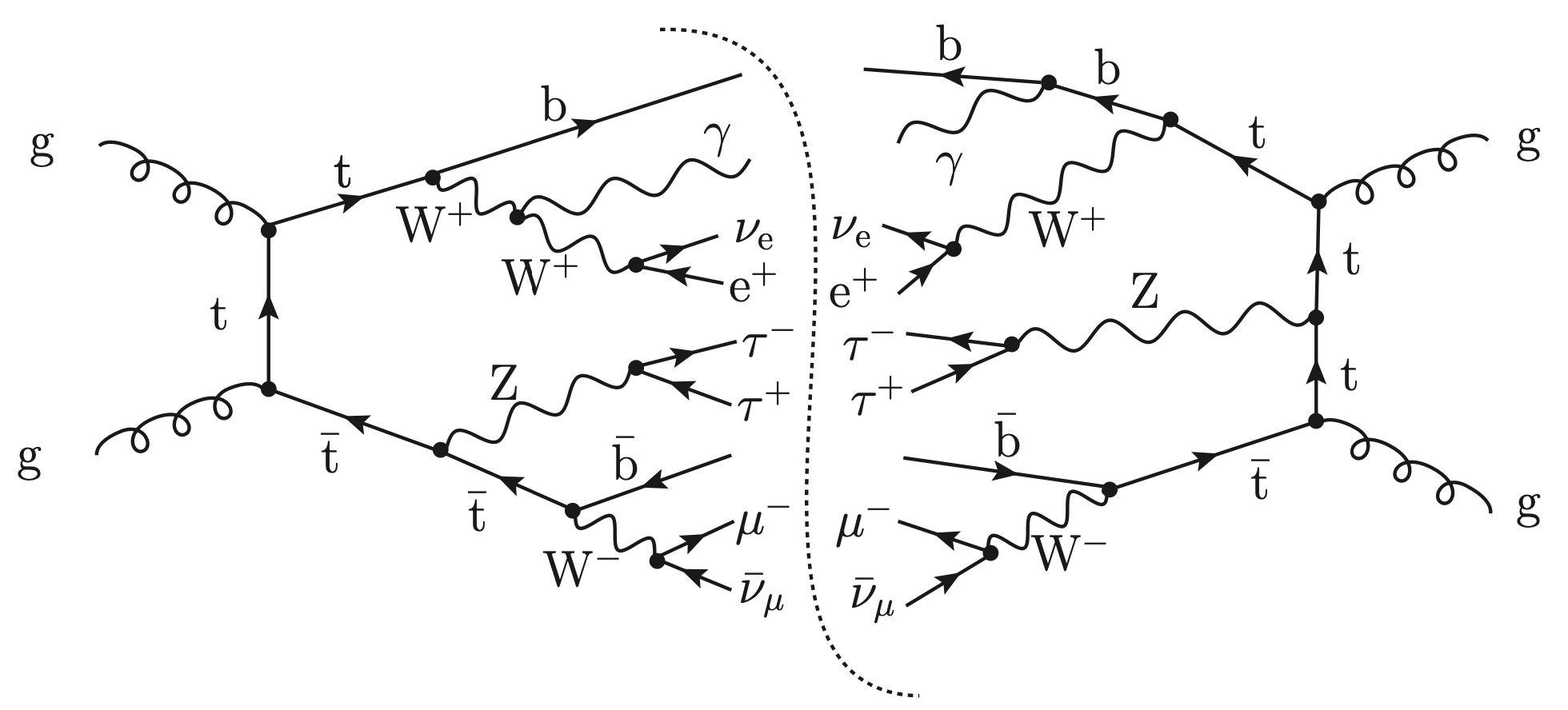}}
  \hspace{0.00cm}
  \subfigure[$\nlothree$ real diagram \label{fig:nlo42}]{ \includegraphics[scale=0.21]{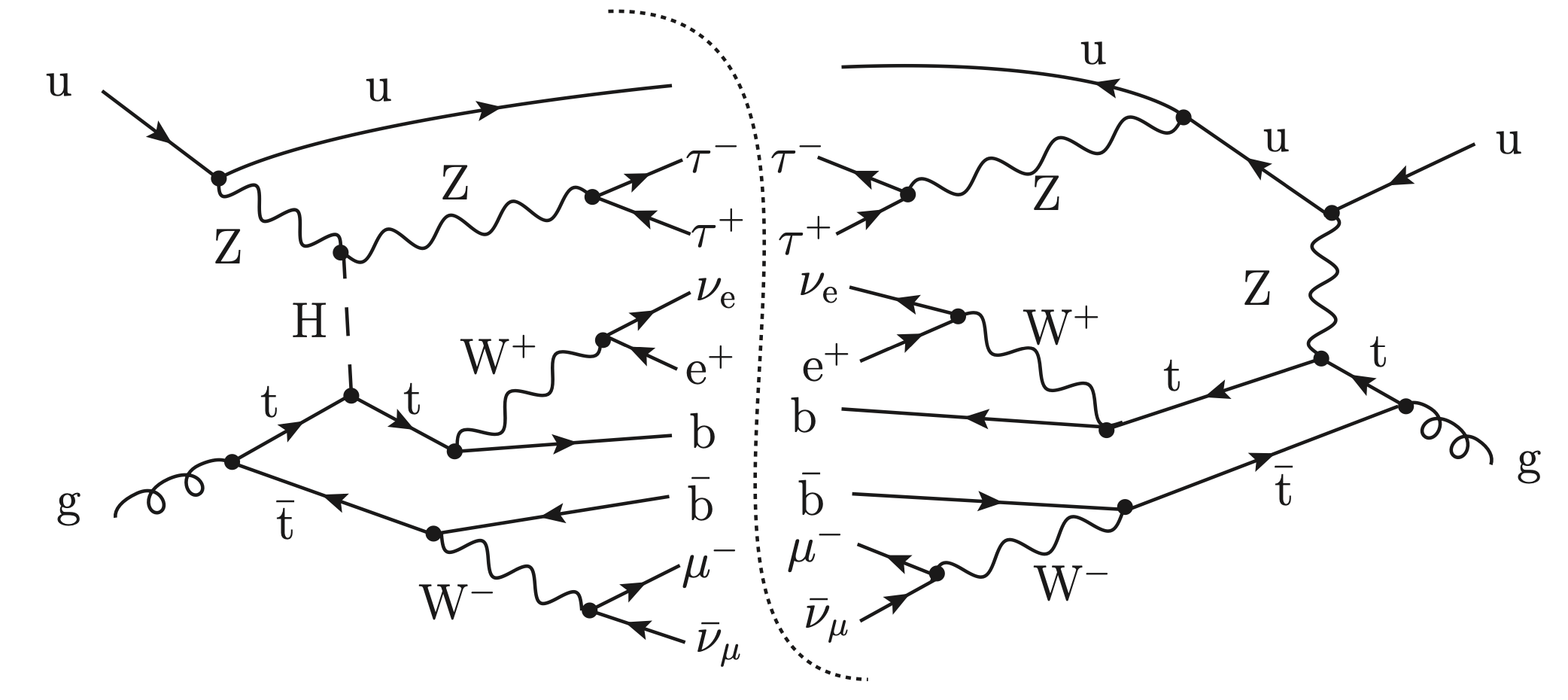}}
  \caption{A sample real squared diagram contributing at $\nlotwo$~[\ref{fig:nlo41}], obtained as an EW correction
    to $\loqcd$, and a sample squared diagram contributing at $\nlothree$~[\ref{fig:nlo42}], showing a $\Pt\PZ$-scattering
    topology, representing a QCD correction to $\loew$.}\label{fig:nlo23}
\end{figure}
Indeed, the final-state photon can be radiated both by quark and lepton lines, resulting in an even
richer singularity structure. A sample $\Pg\Pg$ squared diagram belonging to this class of corrections is
shown in \reffi{fig:nlo41}. Channels involving one initial-state photon, although they contribute to the
same perturbative order, are highly suppressed by the photon PDF. A similar discussion applies to
the real EW corrections to the $\lobqcd$ contributions, which are characterised by the following reactions:
\begin{align}\label{nlob2realchannels}
&\quad\quad\quad\quad\Pb\bar{\Pb} \to{} \Pe^+\nu_\Pe\,\mu^-\bar{\nu}_\mu\,{\Pb}\,\bar{\Pb}\,{\tau}^+{\tau}^-\gamma\,,\nonumber\\
\Pb\Pb \to{}& \Pe^+\nu_\Pe\,\mu^-\bar{\nu}_\mu\,{\Pb}\,{\Pb}\,{\tau}^+{\tau}^-\gamma\,,
\quad\quad\quad\bar{\Pb}\bar{\Pb} \to{} \Pe^+\nu_\Pe\,\mu^-\bar{\nu}_\mu\,\bar{\Pb}\,\bar{\Pb}\,{\tau}^+{\tau}^-\gamma\,,\nonumber\\
\gamma\bar{\Pb} \to{}& \Pe^+\nu_\Pe\,\mu^-\bar{\nu}_\mu\,{\Pb}\,\bar{\Pb}\,{\tau}^+{\tau}^-\bar{\Pb}\,,
\quad\quad\quad\gamma\Pb \to{} \Pe^+\nu_\Pe\,\mu^-\bar{\nu}_\mu\,{\Pb}\,\bar{\Pb}\,{\tau}^+{\tau}^-\Pb\,.
\end{align}
The second class of real contributions arises from the QCD corrections to $\loint$,
obtained by interfering a tree-level $\loqcd$ with a tree-level $\loew$ diagram, with an additional
QCD radiation exchange amongst the two. We notice that, even if the $\loint$ is zero for the $\Pg\Pg$ and $\Pq\bar\Pq$ channels
due to colour algebra, the interference of the corresponding QCD-real amplitudes is non-vanishing
if the gluon is exchanged between an initial quark or gluon and a final $\Pb$-quark line of the two interfering diagrams,
as in \reffi{fig:nlo21}. Moreover,
\begin{figure}
  \centering
  \subfigure[$\nlotwo$ real diagram\label{fig:nlo21}]{   \includegraphics[scale=0.21]{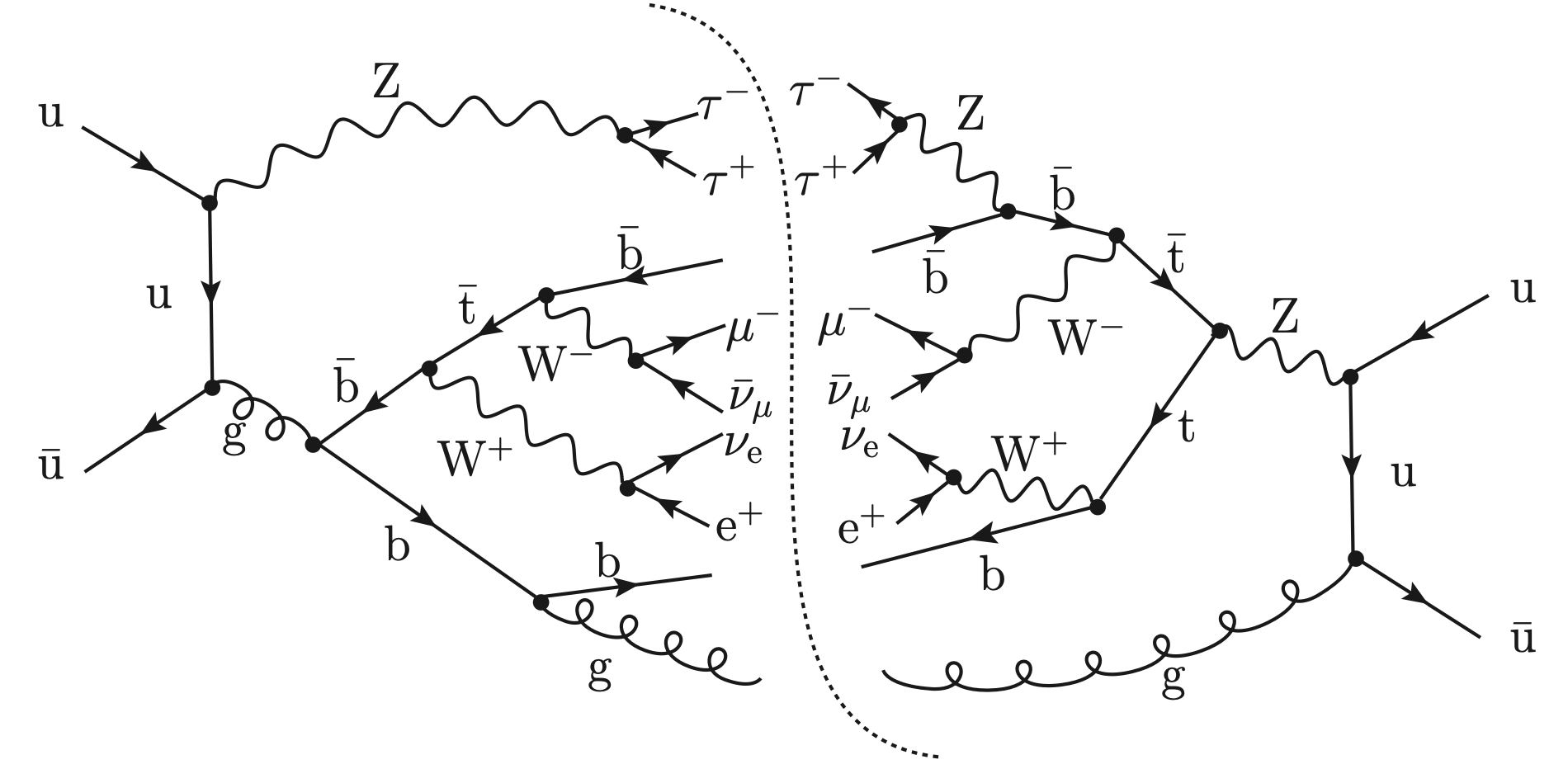}}
  \hspace{0.00cm}
  \subfigure[$\nlotwo$ real diagram \label{fig:nlo22}]{ \includegraphics[scale=0.21]{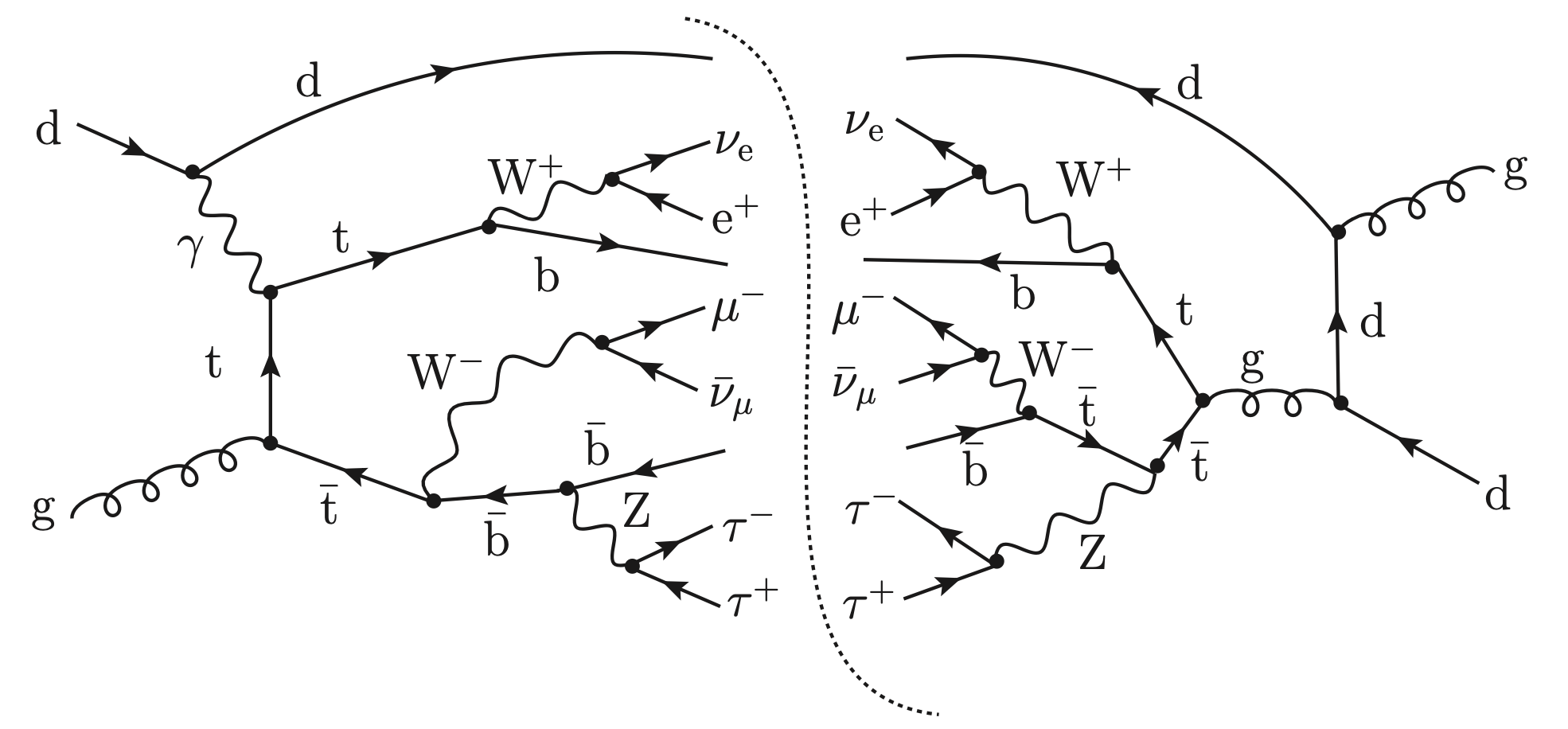}}
  \caption{Sample real interference diagrams contributing at $\nlotwo$ representing QCD corrections
    to $\loint$ interference terms.}\label{fig:nlo2int}
\end{figure}
an additional interference contribution is permitted for  $\Pg\Pq$ and $\Pg\bar\Pq$ channels,
where a quark line connects a genuinely QCD diagram with its EW counterpart, like in \reffi{fig:nlo22}. Finally, real QCD corrections
to $\loint$ contributions, which are already non-zero at leading order, \ie
$\lobint$ and $\gamma\Pg$-initiated
ones, are taken properly into account. For the latter the following real reaction
has to be considered:
\begin{align}\label{agrealchannels}
\gamma\Pg \to{}& \Pe^+\nu_\Pe\,\mu^-\bar{\nu}_\mu\,{\Pb}\,\bar{\Pb}\,{\tau}^+{\tau}^-\Pg\,.
\end{align}

\begin{figure}
  \centering
  \subfigure[$\nlotwo$ virtual diagram \label{fig:virt2}]{ \includegraphics[scale=0.23]{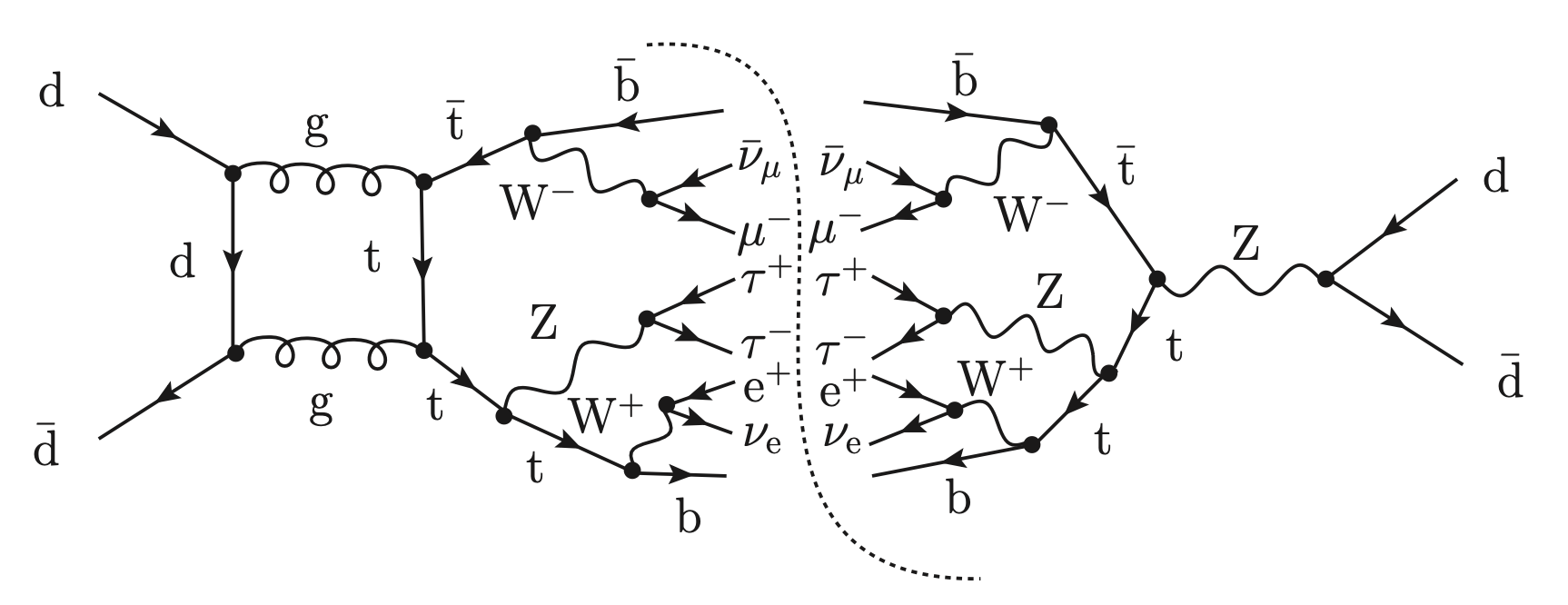}}
  \hspace{0.00cm}
  \subfigure[$\nlotwo$ virtual diagram\label{fig:virt1}]{   \includegraphics[scale=0.23]{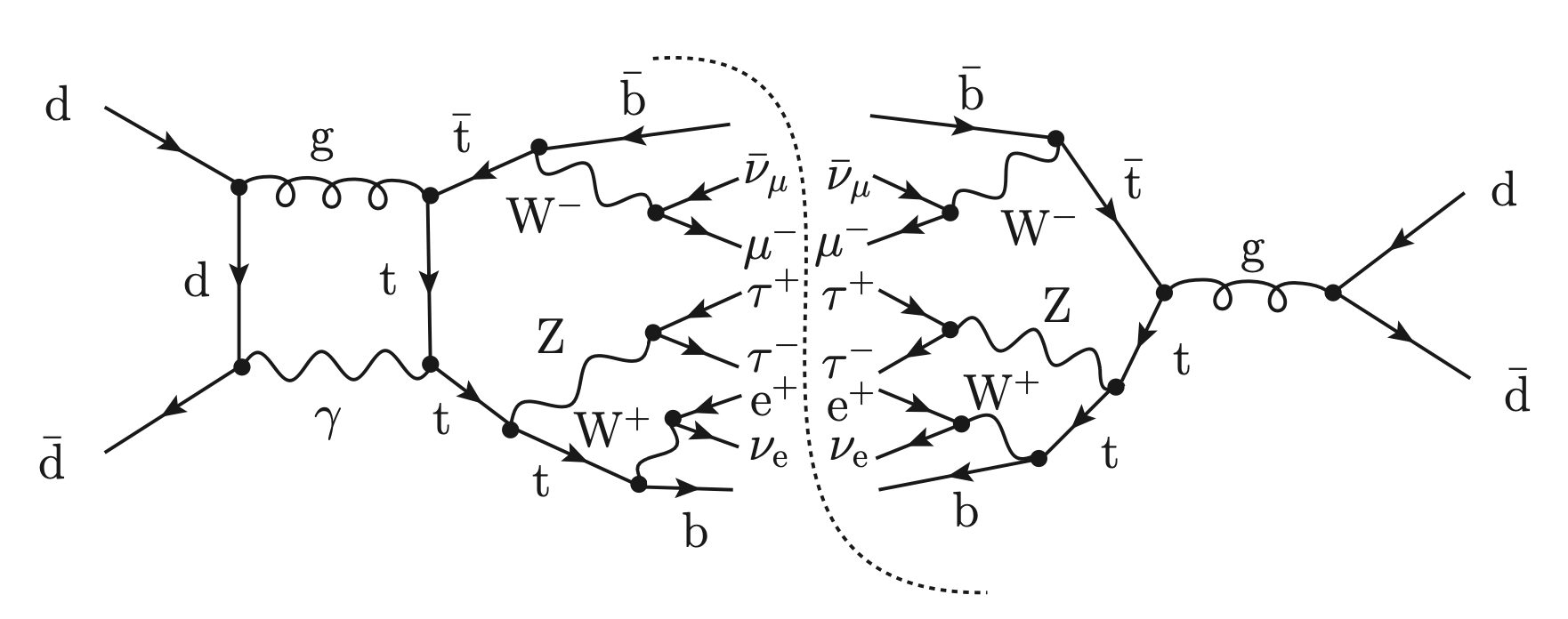}}
  \caption{Sample virtual interference diagrams contributing at $\nlotwo$:  the virtual diagram~[\ref{fig:virt2}]
    is uniquely identified as a QCD correction to an interference term; the virtual diagram~[\ref{fig:virt1}] can be considered both as an
    EW correction to a $\loqcd$ amplitude and as a QCD correction to an interference contribution.
}\label{fig:virt}
\end{figure}
The neat distinction between the two classes of $\nlotwo$
contributions is not possible anymore when
considering the virtual corrections, as already pointed out in \citeres{Denner:2021hqi,Denner:2016jyo}. Indeed,
interference of one-loop amplitudes of order $\mc O(\gs^4 g^6)$, obtained from the insertion of a gluon
propagator in a QCD-mediated diagram, with tree-level EW diagrams of
order $\mc O(g^8)$ are unambiguously classified as QCD corrections to $\loint$, as shown in \reffi{fig:virt2}. But, for instance, the
interference of one-loop amplitudes of order $\mc O(\gs^2 g^8)$ and tree-level QCD ones of order $\mc O(\gs^2 g^6)$
can either be considered as an EW correction to $\loqcd$ or a QCD correction to $\loint$. These terms are obtained
by inserting an EW-particle propagator anywhere in a $\loqcd$ squared amplitude [see for instance \reffi{fig:virt1}]. This ambiguity
is also reflected in the IR structure of this class of loop diagrams, whose singularities are fully cancelled
only once contributions from both classes of real corrections are taken into account.
Already for $\nlotwo$ virtual contributions, the evaluation of loop integrals involving EW particles
becomes extremely challenging due to the higher number of loop diagrams to account for and to the
complexity of the loop integrals, which can involve up to $10$-point functions with a maximal rank of
$6$, as in the $\Pg\Pg$ channel, where two external vector bosons are attached to the loop. In \reffi{fig:loop2}
a sample $\nlotwo$ diagram involving the computation of $8$-point, rank-$3$ loop functions illustrates how the number of
loop diagrams to be evaluated grows when EW particles are allowed to run in the loop, because of the larger
number of topologies permitted.

\subsubsection{Contributions of order $\mc O(\as \alpha^8)$}
\label{sec:nlo3}

The largest part of the $\nlothree$ contribution arises from the QCD corrections to the $\loew$ terms.
The real-emission diagrams are obtained by adding an external gluon (either as initial or final state) to $\loew$ amplitudes:
\begin{align}\label{nlo3realchannels}
\gamma\gamma \to{}& \Pe^+\nu_\Pe\,\mu^-\bar{\nu}_\mu\,{\Pb}\,\bar{\Pb}\,{\tau}^+{\tau}^-\Pg\,,
\quad\quad\quad\Pq\bar{\Pq} \to{} \Pe^+\nu_\Pe\,\mu^-\bar{\nu}_\mu\,{\Pb}\,\bar{\Pb}\,{\tau}^+{\tau}^-\Pg\,,\nonumber\\
\Pg\bar{\Pq} \to{}& \Pe^+\nu_\Pe\,\mu^-\bar{\nu}_\mu\,{\Pb}\,\bar{\Pb}\,{\tau}^+{\tau}^-\bar{\Pq}\,,
\quad\quad\quad\Pg\Pq \to{} \Pe^+\nu_\Pe\,\mu^-\bar{\nu}_\mu\,{\Pb}\,\bar{\Pb}\,{\tau}^+{\tau}^-\Pq\,,
\end{align}
and similarly for the bottom-initiated contributions, where reactions with the same external particles like the ones in \refeq{nlob1realchannels} are found.
Clearly, all channels are EW mediated: the only QCD vertex is the one at which the external
gluon is attached.
In line with the observation in \citere{Denner:2021hqi} for the $\ttw$ process,
we also expect $\nlothree$ corrections to $\ttz$ not to be negligible with respect to the $\nlotwo$ ones.
Indeed, even though these contributions are suppressed by an $\alpha/\as$ factor, the real $\Pg\bar{\Pq}$/$\Pg\Pq$ channels
embed a top-quark scattering against a $\PZ$~boson [as shown in the
left sub-diagram in \reffi{fig:nlo42}]. Their contribution is larger
than the corresponding $\loew$ and the largest of all $\nlothree$ corrections,
as explicitly shown and further discussed in~\refse{sec:integrated}.
Conversely, we do not expect similar effects for the $\Pg\bar{\Pb}$/$\Pg\Pb$ reactions.
Apart from the enhancement by the gluon PDFs, which still renders these partonic channels sizeable
 real  $\nlobthree$ corrections, no additional enhancement with respect to $\lobew$ due to the opening up of new topologies
  occurs in this case. Indeed, diagrams including a $t$-channel scattering of bottom quarks mediated by a $\PW$~boson,
  which are the dominant topologies at $\nlothree$, are already present at $\lobew$, as exemplified
  by~\reffi{fig:lob2}.
Despite the larger number of diagrams, which is due to the EW nature of the process, the complexity of the loop functions to be evaluated for
the QCD corrections to $\loew$ is comparable to the one encountered for the $\nloone$ case.

\begin{figure} 
  \centering
  \subfigure[$\nlothree$ real diagram for the $\gamma\Pg$ channel \label{fig:nlo31}]{   \includegraphics[scale=0.21]{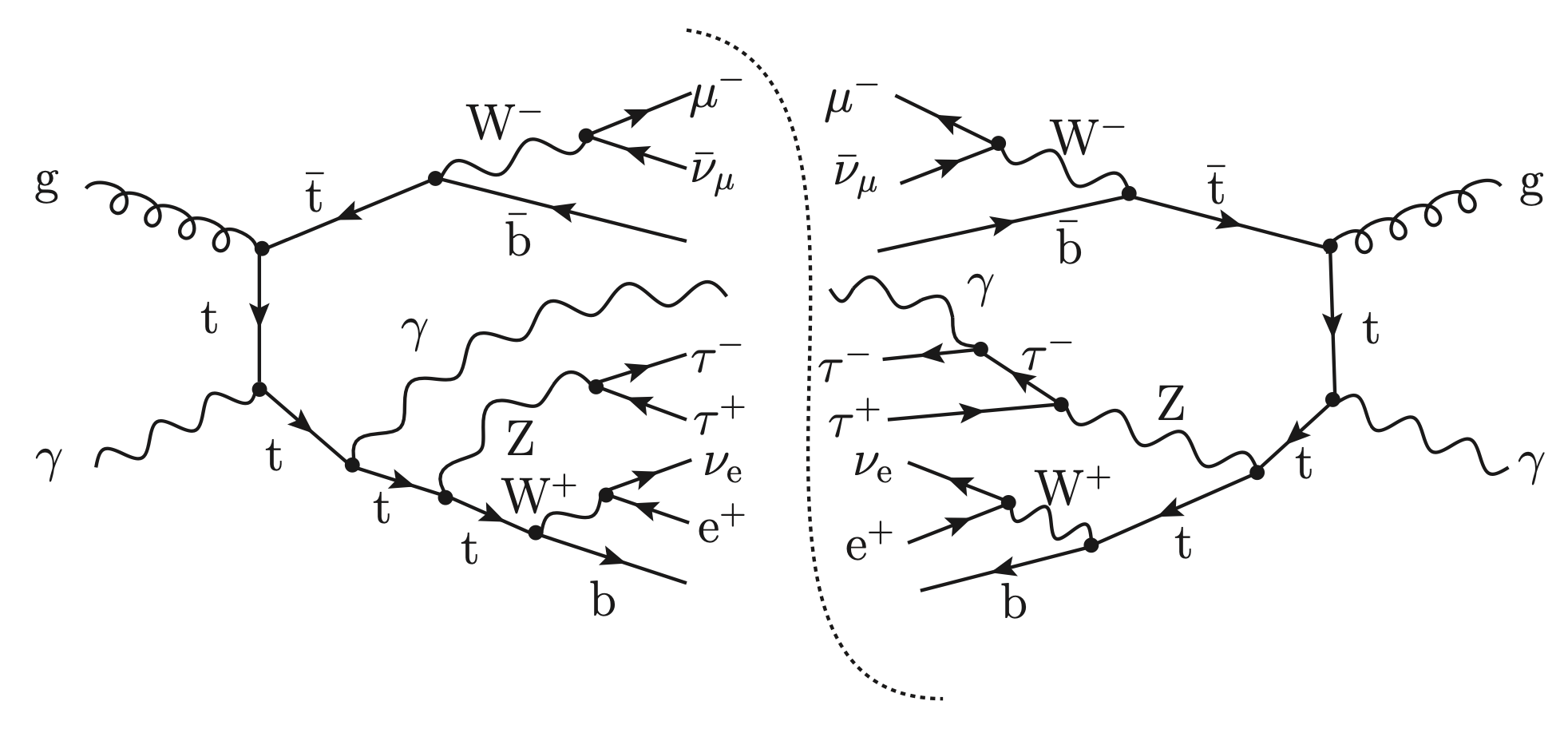}}
  \hspace{0.00cm}
  \subfigure[$\nlothree$ real diagram for the bottom contribution\label{fig:nlo32}]{ \includegraphics[scale=0.19]{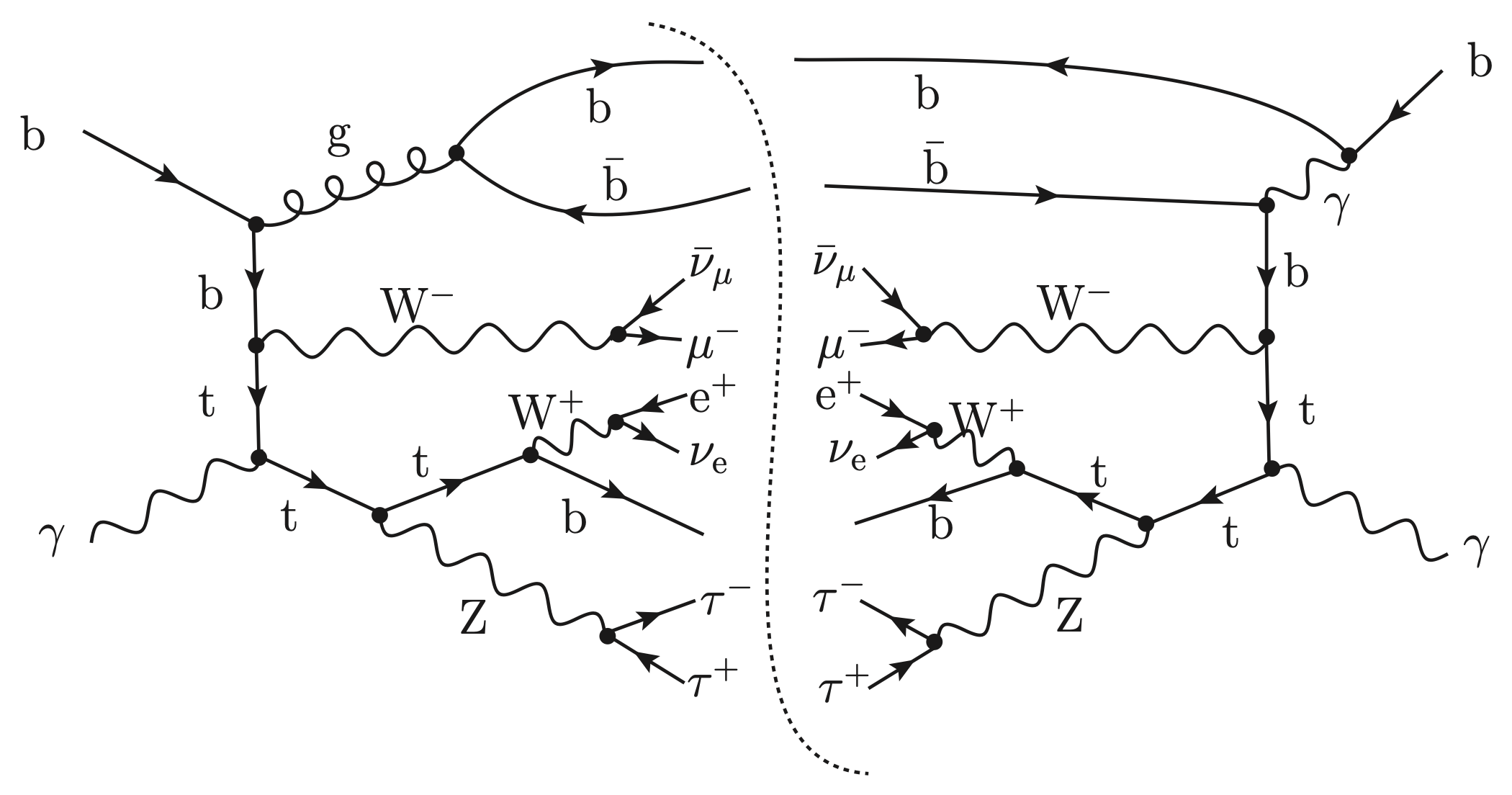}}
  \caption{Sample real squared diagrams contributing at $\nlothree$ representing EW corrections
    to $\loint$ for the $\gamma\Pg$ channel~[\ref{fig:nlo31}] and the bottom interference
    terms~[\ref{fig:nlo32}].}\label{fig:nlo3int}
\end{figure}
\begin{figure} 
  \centering
  \subfigure[$\nlothree$ virtual diagram \label{fig:virt2b}]{ \includegraphics[scale=0.24]{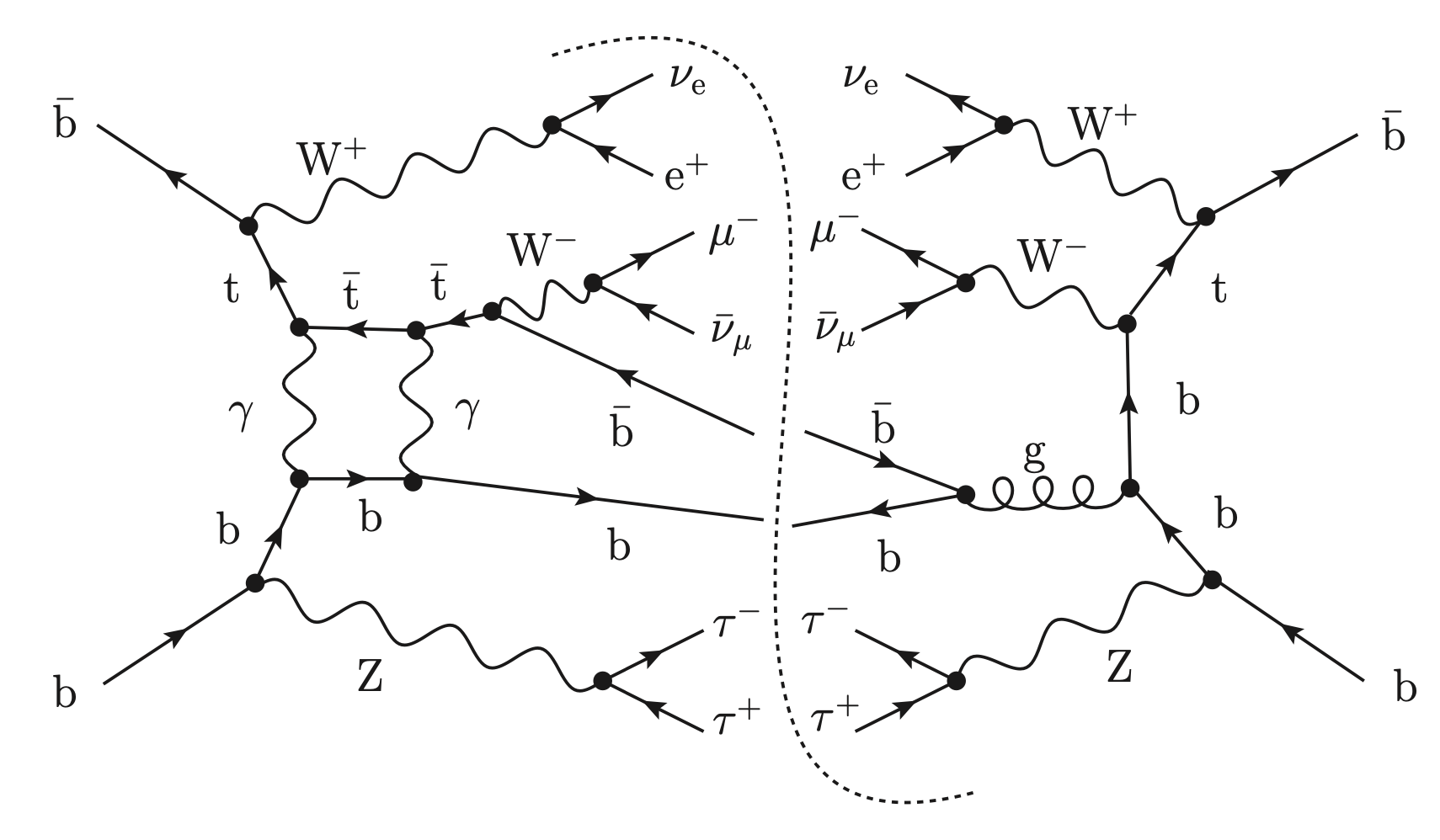}}
  \hspace{0.00cm}  
  \subfigure[$\nlothree$ virtual diagram\label{fig:virt1b}]{   \includegraphics[scale=0.24]{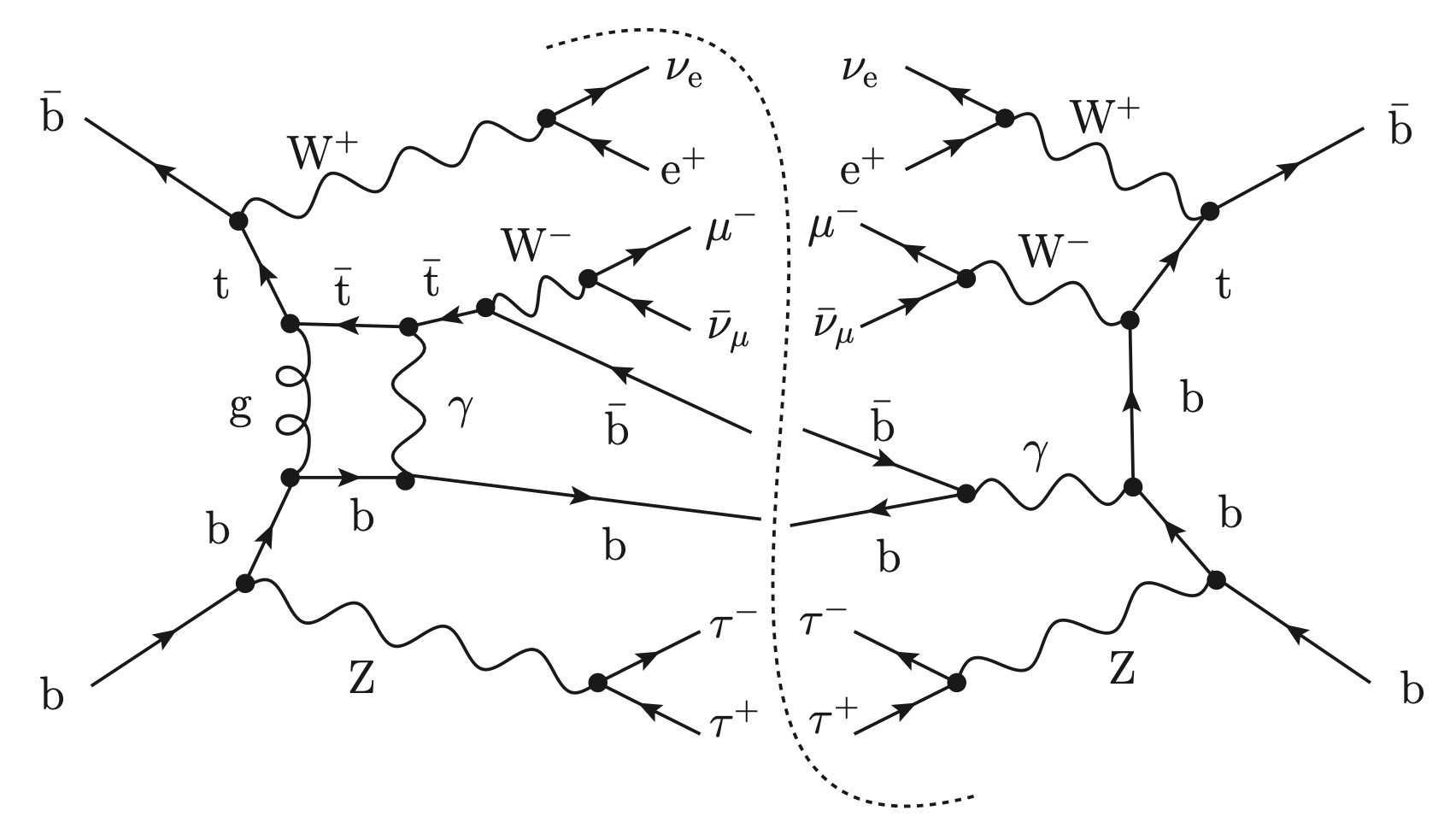}}
  \caption{Sample virtual interference diagrams contributing at $\nlothree$ for the bottom channels:  the virtual diagram~[\ref{fig:virt2b}]
    is uniquely identified as an EW correction to an interference term;
    the virtual diagram~[\ref{fig:virt1b}]
    can be considered both as a QCD correction to a $\loew$ amplitude and as an EW correction to an interference contribution.
}\label{fig:virtb}
\end{figure}

An additional class of corrections contributing at order $\mc O(\as
\alpha^8)$ results from
EW corrections to $\loint$, which are non-vanishing only for those terms whose $\loint$ is already different
from zero, since EW corrections do not modify the colour structure of the LO amplitude.
That means that only EW corrections to the $\gamma\Pg$-initiated channel and to $\lobint$ have to be
taken into account. For the $\gamma\Pg$ channel, the real reaction
[diagrammatically shown in \reffi{fig:nlo31}],
\begin{align}\label{ag3realchannels}
\gamma\Pg \to{}& \Pe^+\nu_\Pe\,\mu^-\bar{\nu}_\mu\,{\Pb}\,\bar{\Pb}\,{\tau}^+{\tau}^-\gamma\,,
\end{align}
and the corresponding virtual term have to be computed. The latter turned out to be one of the most challenging virtual contributions
to be evaluated within $\nloone$, $\nlotwo$, and $\nlothree$. Indeed, it requires the evaluation of $10$-point functions
up to rank $6$, already encountered at $\nlotwo$, but with a much larger number of topologies induced by the presence of
a photon as initial-state parton. An exemplary $\nlothree$ loop diagram for this channel, involving up to $9$-point, rank-$5$
loop functions is shown in~\reffi{fig:loop3}. As far as the bottom contributions are concerned, real EW corrections are simply
obtained by the emission of a photon off $\Pb\bar\Pb$-, $\Pb\Pb$-, or $\bar\Pb\bar\Pb$-initiated QCD diagrams, which is then absorbed
by an appropriate EW counterpart or by connecting a $\gamma\Pb$/$\bar\gamma\Pb$-induced QCD diagram and its EW counterpart
via a bottom-quark line [see for instance \reffi{fig:nlo32}]. As already found at $\nlotwo$, the $\nlothree$ virtual contributions for
the bottoms cannot be unambiguously separated into EW corrections to $\lobint$ and QCD corrections to $\lobew$. 
The interference of one-loop amplitudes of order $\mc O(g^{10})$ %, obtained from the insertion of an EW particle in an EW-mediated diagram,
with tree-level diagrams of order $\mc O(\gs^2 g^6)$ are uniquely identified as EW corrections to $\lobint$,
as also clarified by \reffi{fig:virt2b}.
On the contrary, the interference of one-loop amplitudes of order $\mc O(\gs^2 g^8)$ and tree-level ones of order $\mc O(g^8)$
%,obtained inserting a QCD particle in a $\lobew$ squared amplitude,
can be attributed to both classes of corrections [see \reffi{fig:virt1b}].

\subsubsection{Contributions of order $\mc O(\alpha^9)$}
\label{sec:nlo4}
The $\nlofour$ contribution represents the last missing ingredient to furnish
the full set of NLO corrections to $\ttz$ production. As suggested by
power counting and as explicitly shown in on-shell calculations \cite{Frederix:2018nkq}, these corrections affect the overall result
at the sub-per-mille level and are out of reach in any realistic measurement.
The evaluation of these contributions requires to consider the set of real reactions
\begin{align}\label{nlo4realchannels}
\gamma\gamma \to{}& \Pe^+\nu_\Pe\,\mu^-\bar{\nu}_\mu\,{\Pb}\,\bar{\Pb}\,{\tau}^+{\tau}^-\gamma\,,
\quad\quad\quad\Pq\bar{\Pq} \to{} \Pe^+\nu_\Pe\,\mu^-\bar{\nu}_\mu\,{\Pb}\,\bar{\Pb}\,{\tau}^+{\tau}^-\gamma\,,\nonumber\\
\gamma\bar{\Pq} \to{}& \Pe^+\nu_\Pe\,\mu^-\bar{\nu}_\mu\,{\Pb}\,\bar{\Pb}\,{\tau}^+{\tau}^-\bar{\Pq}\,,
\quad\quad\quad\gamma\Pq \to{} \Pe^+\nu_\Pe\,\mu^-\bar{\nu}_\mu\,{\Pb}\,\bar{\Pb}\,{\tau}^+{\tau}^-\Pq\,,
\end{align}
together with the bottom-induced ones as reported in \refeq{nlob2realchannels} but fully EW mediated. 
The bottleneck here is the numerical evaluation of virtual corrections.
The most difficult term to be computed, which turns out to be also the least sizeable,
is represented by the virtual EW corrections to the $\gamma\gamma$ channel: the presence
of two external photons renders the number of $10$-point, rank-$6$ loop functions to be evaluated
even larger. A loop diagram involving loop functions of such a level of complexity
is shown in \reffi{fig:loop4} for the $\gamma\gamma$ initial state.

We present results for the $\nlofour$ contribution in a fully off-shell calculation
for $\ttz$ production at the integrated level for fiducial cross-sections in \refse{sec:integrated},
but we refrain from including them in any differential result.

\subsection{Validation}\label{sec:valid}

Our calculation has been validated by reproducing NLO QCD results ($\loqcd+\nloone$ in our notation) for fully off-shell
$\ttz$ production recently published in \citere{Bevilacqua:2022nrm} and obtained
within the \helacnlo framework. Since those results have also been computed for a final state involving three different
leptonic flavours, we just had to adapt our SM input parameters, presented in \refse{sec:input}, to the ones reported
in \citere{Bevilacqua:2022nrm}.
The most important features of the setup of
\citere{Bevilacqua:2022nrm} that differ from those of our default
setup and, thus, had to be readjusted
for validation purposes are the following:
\begin{itemize}
\item the NLO top-quark width does not include NLO EW corrections;
\item the factorisation and renormalisation scales are set to $\mu^{(c)}_0$, defined in \refeq{eq:scaleA};
\item different PDF sets are used for $\loqcd$ and $\loqcd+\nloone$ results (\texttt{NNPDF31\_lo\_as\_0118} and \texttt{NNPDF31\_nlo\_as\_0118}, respectively);
\item no bottom-induced contributions are included.
\end{itemize}
Since the selection cuts used to generate our main results in \refse{sec:numresults} (and described
in \refse{sec:input}) match  the ones in \citere{Bevilacqua:2022nrm}, no additional adjustment to our calculation
has been required.

In this setup we have obtained the following integrated cross-sections:
\begin{align}
&  \sigma_{\loqcd\,\rm nob} = 80.39(1)^{+25.54\,(32\%)}_{-18.03\,(22\%)}\,\text{ab}\quad 
\left(\text{\citere{Bevilacqua:2022nrm}}:  80.32^{+25.51 \,(32\%)}_{-18.02\,(22\%)}\,\text{ab}\right)\,, \nonumber\\
&  \sigma_{\nloone\,\rm nob} = 99.3(2)^{+1.25\,(1\%)}_{-5.83\,(6\%)}\,\text{ab}\quad \quad\,
\left(\text{\citere{Bevilacqua:2022nrm}}: 98.88^{+1.22\,(1\%)}_{-5.68\,(6\%)}\,\text{ab}\right)\,,
\end{align}
to be compared with Eq.~(4.3) of \citere{Bevilacqua:2022nrm} (reported
here in parentheses as a reference).
%The digits in parentheses indicate the integration errors and the percentages
%in super-/sub-scripts the relative 7-point scale variations.
The digits in parentheses indicate the integration errors, while the absolute
  7-point scale variations are reported as super-/sub-scripts with the respective
  relative variations as percentages in parenthesis.
The agreement between the two $\loqcd$ results is extremely good. The two central
values differ at the $0.1\%$ level, which confirms the size of Higgs-diagram contributions at $\loqcd$ stated in \citere{Bevilacqua:2022nrm}.
Indeed, the presence of Higgs contributions, included in our results, is an unavoidable difference between the two calculations.
The theory uncertainty bands obtained via scale variations are also perfectly reproduced. A good agreement is found after including QCD corrections.
In this case, although larger Monte Carlo uncertainties partially limit the comparison, the two results agree within $1.9\,\sigma$ (computed with the Monte Carlo error estimate of our result),  which is more than satisfactory,
especially considering the different treatment of Higgs diagrams that still persists at NLO.

\begin{figure} 
  \centering
\subfigure[Azimuthal-angle separation between the positron and the muon\label{fig:vphi}]
{\includegraphics[scale=0.30]{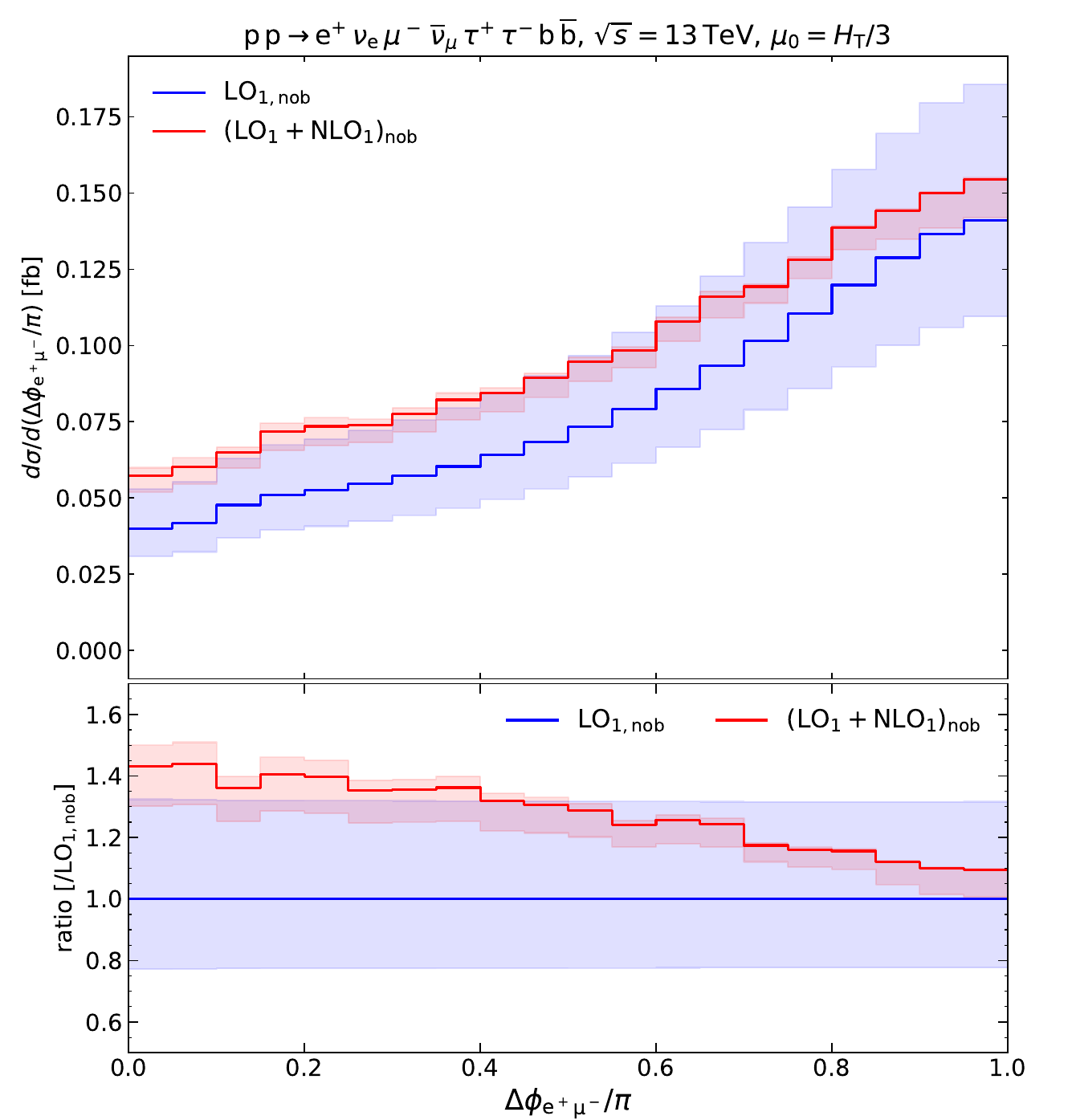}}
 \subfigure[Transverse momentum of the $\tau^+\tau^-$ pair\label{fig:vpttt}]{   \includegraphics[scale=0.30]{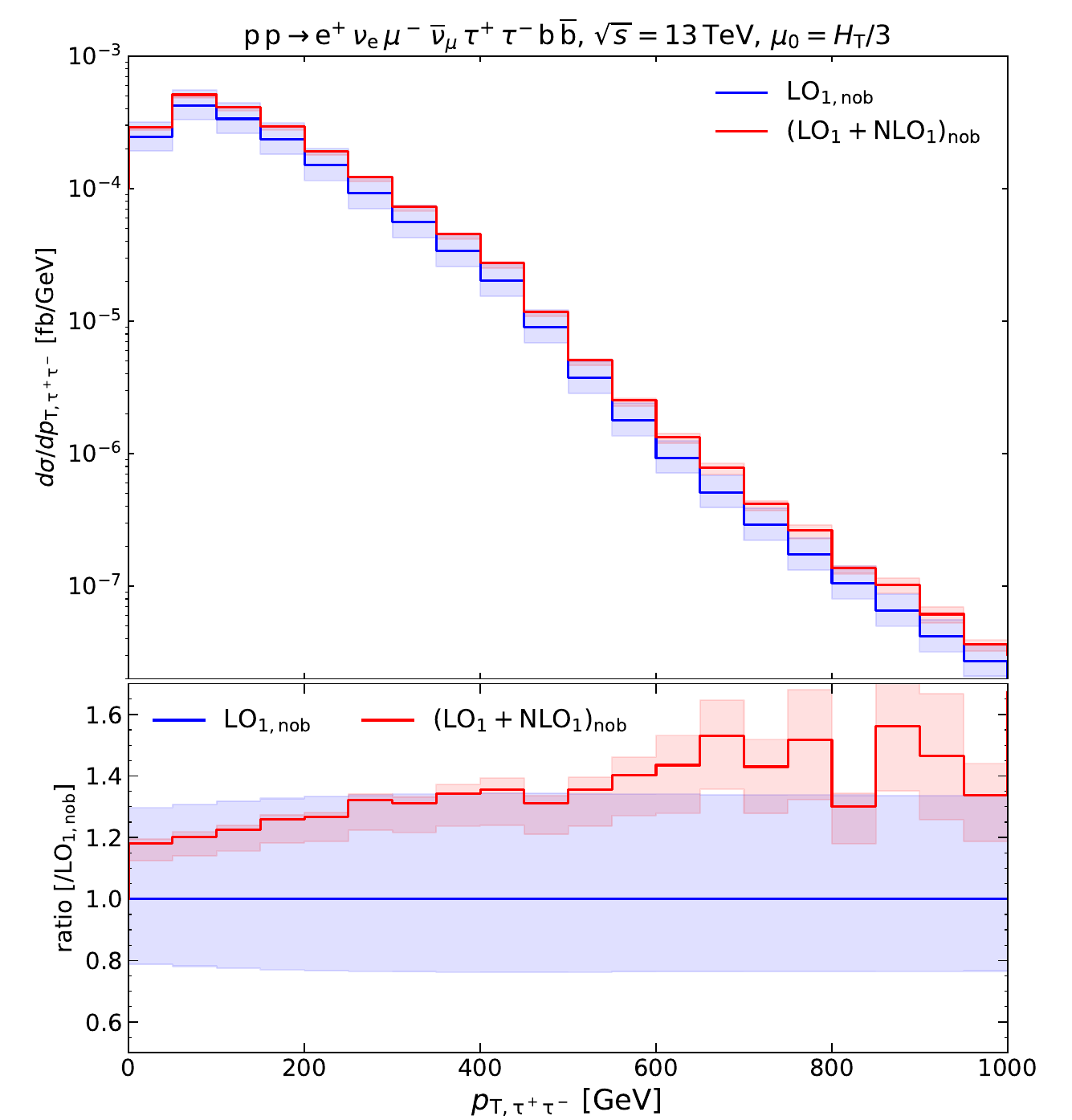}}
  %\hspace{0.cm}
\subfigure[Transverse momentum of the subleading $\Pb$ jet \label{fig:vptb2}]{   \includegraphics[scale=0.30]{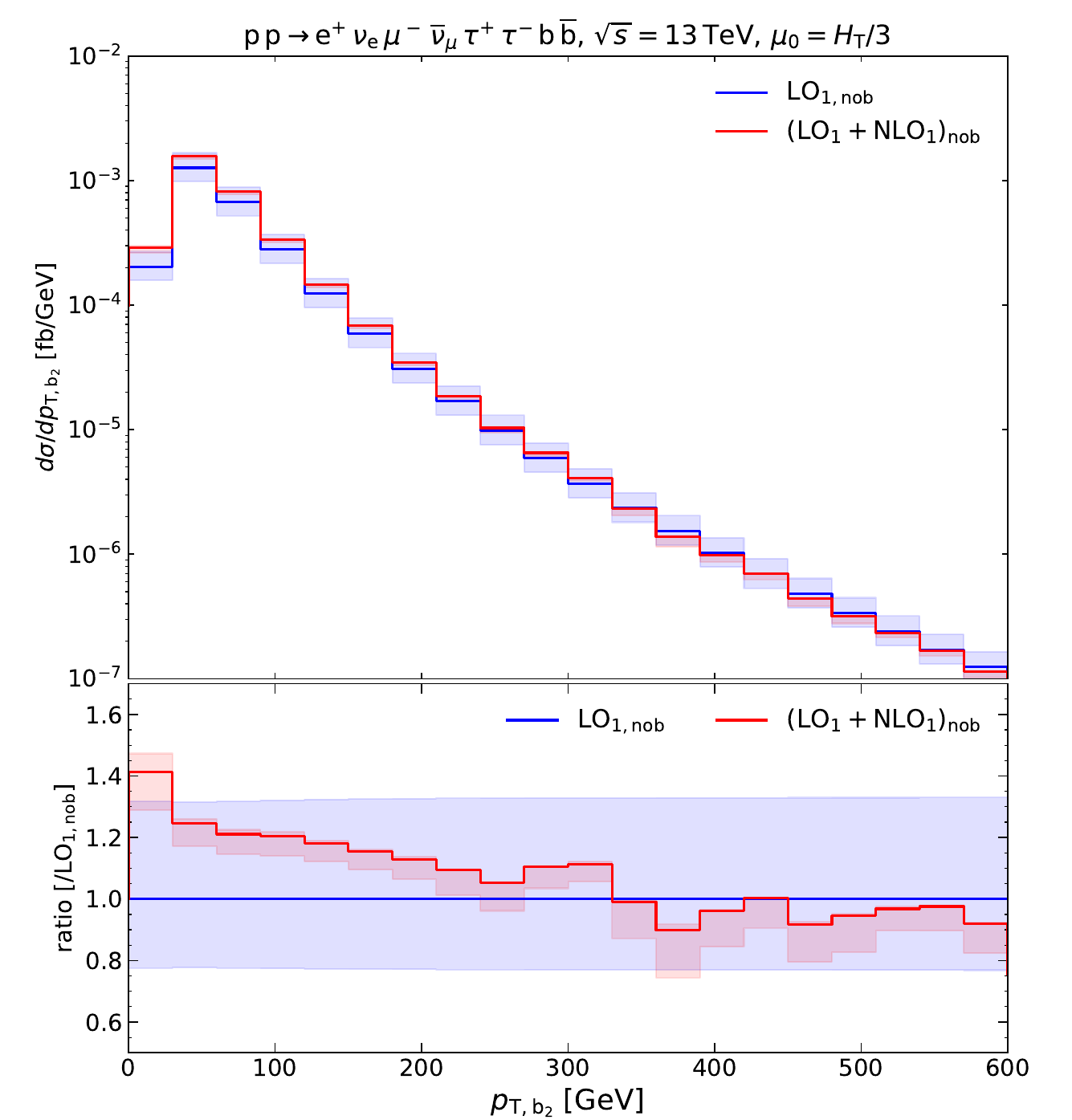}}
\subfigure[Transverse momentum of the positron \label{fig:vptep}]{   \includegraphics[scale=0.30]{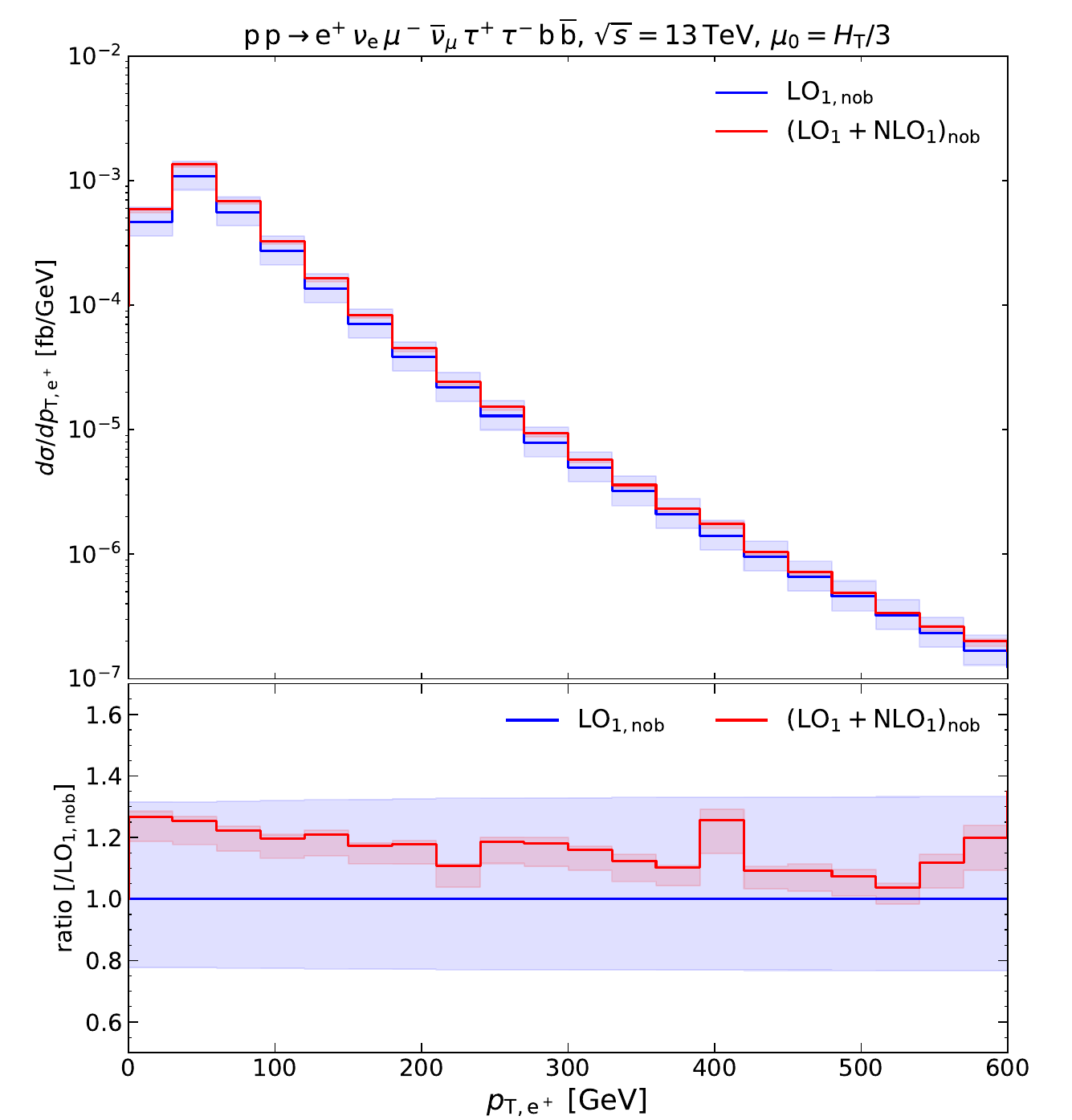}}
  \caption{Selected distributions used for validating our results against the ones
  obtained within the \helacnlo framework in \citere{Bevilacqua:2022nrm}.}\label{fig:Aachen_distributions}
\end{figure}

The comparison with the calculation of \citere{Bevilacqua:2022nrm} has been carried out also at the differential level,
by reproducing distributions in all kinematic variables reported therein. In all cases a good agreement was found,
as confirmed by the selection of distributions in \reffi{fig:Aachen_distributions}.
We show the distribution in the azimuthal angle between the positron
and the muon in \reffi{fig:vphi} as well as the distributions in
the transverse momentum of the $\tau^+\tau^-$ pair [\reffi{fig:vpttt}], of the subleading $\Pb$ jet [\reffi{fig:vptb2}], and of the positron [\reffi{fig:vptep}].
In the main panels the $\loqcd$ and $\loqcd+\nloone$ results (with no bottom contributions included) are presented with a blue and red curve,
respectively. In the ratio panels the same results normalised to the $\loqcd$ one are reported. The corresponding distributions %of \reffi{fig:Aachen_distributions}
can be found in Figs.~6, 7, and 8 of \citere{Bevilacqua:2022nrm}. A
direct comparison reveals that the relative $\nloone$ corrections as well as the
size of the QCD-uncertainty bands obtained with $7$-point scale variations
[see~\refeq{eq:scaleset}] are correctly reproduced by our calculation.

\section{Numerical results}\label{sec:numresults}
\subsection{Input parameters}\label{sec:input}
In the following, we present results for the LHC at a centre-of-mass energy of 13\TeV.  We consider
the process \refeqf{eq:procdef} with four different charged leptons in the final
state. All leptons are assumed to be massless, \ie also $m_\tau = 0$.
%% We note that our results can be used as well 
%% for the case of identical leptons in the final state after applying
%% appropriate symmetry factor $1/2$ up to interference effects. These
%% interference effects are not doubly resonant and suppressed by factors
%% $\Gt/\Mt$. 
We work in the five-flavour scheme, therefore we treat all light and bottom quarks as massless.
A unit Cabibbo--Kobayashi--Maskawa matrix is understood.

The on-shell values for the masses and widths of the EW bosons
are chosen according to \cite{ParticleDataGroup:2020ssz},
\begin{align}
\Mwo ={}& 80.379 \GeV,\qquad\,\,\, \Gwo = 2.085\GeV\,, \nnb\\
\Mzo ={}& 91.1876 \GeV,\qquad \Gzo \,= 2.4952\GeV\,, \nnb\\
\MH  ={}& 125 \GeV, \qquad\,\,\, \GH\, \,\,= 0.00407\,\GeV\,.
\end{align}
The masses of the vector bosons are converted to their pole values by
means of the following relations \cite{Bardin:1988xt}:
\beq
\Mv = \frac{\Mvo}{\sqrt{1+({\Gvo}/{\Mvo})^2}}\,,\qquad\, 
\Gv = \frac{\Gvo}{\sqrt{1+({\Gvo}/{\Mvo})^2}}\,.
\eeq
The top-quark mass and width are fixed as
\beq
\label{eq:masswidthtop}
\Mt = 173.0\GeV\,, \qquad \Gt^{\rm LO}   = 1.4437\GeV\,,
\qquad \Gt^{\rm NLO}   = 1.3636\GeV\, .
\eeq
The top-quark width at LO has been computed with the
formulas of \citere{Jezabek:1988iv} and using the pole mass and width
for the $\PW$~boson as input. In order to meet the perturbative accuracy addressed
in this work, the NLO width has been
obtained upon applying QCD- and EW-correction factors from~\citere{Basso:2015gca}
to the LO width. All LO and NLO results in Sections~\ref{sec:integrated} and~\ref{sec:differential}
are obtained by using the NLO top-quark width in \refeq{eq:masswidthtop}.

The EW coupling is extracted from the Fermi constant $G_\mu$ by
means of \cite{Denner:2000bj}
\beq
\alpha = \frac{\sqrt{2}}{\pi}\,G_\mu\Mw^2\biggl(1-\frac{\Mw^2}{\Mz^2}\biggr)\,,
\eeq
where $G_\mu = 1.16638\cdot10^{-5} \GeV^{-2}$.

The masses of unstable particles, \ie the EW vector bosons
and the top quark, are treated in the complex-mass scheme
\cite{Denner:1999gp,Denner:2005fg,Denner:2006ic,Denner:2019vbn} in all
parts of the computation. As a consequence, the EW mixing
angle and the related couplings are complex valued.

For both the LO and the NLO calculation, we use \texttt{NNPDF31\_nlo\_as\_0118\_luxqed} 
PDFs \cite{Ball:2014uwa,Bertone:2017bme}
extracted at NLO with $\as(\Mz)=0.118$.  The usage of this PDF set allows us
to properly account for the photon PDF. The strong coupling constant
$\as$ used in the calculation of the amplitudes matches the one used
in the evolution of PDFs.  The PDFs and the running of $\as$
are obtained by interfacing \mocanlo with {\scshape LHAPDF6}
\cite{Buckley:2014ana}. 

%%%%%%%%%%%%%%%%%%%%%%%%%%%%%%%%%%%%%%%%%%%%%%%%%%%%%%%%%%%%%%%%%%%%%
The QCD partons with pseudorapidity $|\eta|<5$ are
promoted to candidate jets and then used as inputs for the jet clustering performed with
%clustered into jets by means of
the anti-$k_t$ algorithm \cite{Cacciari:2008gp} with resolution
radius $R=0.4$. As part of the jet clustering, we recombine a $\Pb$~jet and a light jet ($\Pj$) into a
$\Pb$~jet and two $\Pb$~jets into a light jet using the
following recombination rules:
\beq \label{eq:recrules}
\Pj+\Pj\to\Pj, \qquad
\Pj_{\rm b}+\Pj\to\Pj_{\rm b}, \qquad
\Pj_{\rm b}+\Pj_{\rm b}\to\Pj,
\eeq
where the last rule is crucial for bottom-induced contributions, which can include up to three
bottoms in the final state, and where simply requiring two $\Pb$ jets does not remove all
IR divergences (as discussed in \refse{sec:leadingorder}).

To facilitate the comparison of our results with the ones of~\citere{Bevilacqua:2022nrm},
we made use of the selection cuts of that paper, whose choice is motivated 
by a recent ATLAS analysis \cite{ATLAS:2021fzm} (see Table 2 therein) and a
CMS one \cite{CMS:2019too}.  In
particular, we ask for at least two $\Pb$~jets in the final state, 
assuming a perfect $\Pb$-tagging efficiency, 
%(estimated to be $77\%$ from ${\Pt\overline{\Pt}}$ measurements). 
which are required to fulfil
\beq\label{eq:setup1}
\pt{\Pb} > 25 \GeV\,, \quad |\eta_{\Pb} |<2.5\,, \quad \Delta R_{\Pb \Pb}>0.4\,.
\eeq
Charged leptons are dressed with anti-$k_t$ clustering algorithm,
but with $R=0.1$. For all charged leptons, we require 
\beq\label{eq:setup2}
\pt{\Pl_i} > 20 \GeV\,, \quad |\eta_{\Pl_i} |<2.5\,, \quad \Delta R_{\Pl_i \Pl_j}>0.4\,,
\eeq
where $\Pl_i\in\{\Pe^+, \mu^-, \tau^+, \tau^-\}$. While these cuts are applied
to the light leptons $\Pl=\Pe,\,\mu$ by
ATLAS, we apply them also to $\tau$~leptons.
Finally, we apply a cut on the missing transverse momentum arising from the undetected neutrinos
\beq\label{eq:setup3}
\pt{\text{miss}}>40\GeV\,,
\eeq
where for the specific case at hand  $\pt{\text{miss}}$ is computed as the transverse component of the
sum of the momenta of the two neutrinos at Monte Carlo-truth level.
No specific veto is imposed on
the additional light- or b-jet activity that may arise as part of real QCD radiation at NLO.

%%%%%%%%%%%%%%%%%%%%%%%%%%%%%%%%%%%%%%%%%%%%%%%%%%%%%%%%%%%%%%%%%%%%%
%%% Setup closer to ATLAS analysis
%%%%%%%%%%%%%%%%%%%%%%%%%%%%%%%%%%%%%%%%%%%%%%%%%%%%%%%%%%%%%%%%%%%%%
%% Our choice of selection cuts reflects those applied by ATLAS in a
%% recent analysis \cite{ATLAS:2021fzm} (see Table 2 therein).  In
%% particular, we ask for exactly two $\Pb$~jets in the final state,
%% assuming a perfect $\Pb$-tagging efficiency, 
%% which are required to fulfil
%% \beq
%% \pt{\Pb} > 25 \GeV\,, \quad |\eta_{\Pb} |<2.5\,.
%% \eeq
%% We apply a rapidity cut of $|\eta_\Pl |<2.5$ to all charged leptons,
%% while we set transverse-momentum cuts of decreasing magnitude to the
%% leading $\Pl_1$, subleading $\Pl_2$, third $\Pl_3$ and fourth $\Pl_4$ leptons:
%% \beq
%% \pt{\Pl_1} > 27 \GeV\,, \quad \pt{\Pl_2} > 20 \GeV\,, \quad \pt{\Pl_3} > 10 \GeV\,, \quad \pt{\Pl_4} > 5 \GeV\,.
%% \eeq
%% For the OSSF lepton pair associated to the $Z$ boson we also apply the invariant mass cut
%% \beq
%% |m_{\Pl^+\Pl^-}-\Mz|<10\GeV\,.
%% \eeq
%% We also require visible final state particles defining the process to be
%% well separated by imposing:
%% \beq
%% \Delta R_{\Pl_i \Pl_j}>0.4\, \quad \Delta R_{\Pl_i \Pb}>0.4\,, \quad \Delta R_{\Pb \Pb}>0.4\,.
%% \eeq
%% with $\Pl_i=\{\Pe^+, \Pe^-, \mu^-, \tau^+\}$.
%%%%%%%%%%%%%%%%%%%%%%%%%%%%%%%%%%%%%%%%%%%%%%%%%%%%%%%%%%%%%%%%%%%%%

As usual, we set  the factorisation and renormalisation scales to the same central scale $\mu_0$,
\ie $\mu_{\rm R}=\mu_{\rm F}=\mu_0$. For $\mu_0$
we made use of two different choices.
In order to directly compare our results with the ones of~\citere{Bevilacqua:2022nrm},
we adopted the following dynamical choice:
\beq\label{eq:scaleA}
\mu^{(c)}_0=\frac{H_{\rm T}}{3}\quad\text{with}\quad H_{\rm T}=\sum_{i=1}^{2}\pt{\Pb_i}+\pt{\tau^+}+\pt{\tau^-}+\pt{ \mu^-}+\pt{\Pe^+}+\pt{\text{miss}}\,,
\eeq
where the sum over the transverse momenta of the final states in $H_{\rm T}$ does not run over
additional light jets. This scale has been
adopted in \refse{sec:valid} to validate our calculation at $\nloone$ level: since no bottom-induced
channels are included for that comparison, the two $\Pb$~jets entering the definition of $H_{\rm T}$
are the only ones present in the partonic reaction. We notice that
the definition in~\refeq{eq:scaleA} is insensitive to specific resonant
topologies which lead to the final state of interest. 

Our main results, discussed in \refses{sec:integrated} and \ref{sec:differential}, rely on
a second scale choice, whose definition was first introduced in~\citere{Denner:2020hgg}
for $\ttw$ production and which was in turn constructed in line with the choice of
\citeres{Denner:2016wet, Denner:2017kzu} for $\Pt\overline{\Pt}$ and
$\Pt\overline{\Pt}\PH$ production:
\beq\label{eq:scaleB}
\mu^{(d)}_0 =
\frac{1}{2}{\left(\tm{\Pt}\,\tm{\overline{\Pt}}\right)}^{1/2}=
\frac{1}{2}\left(\sqrt{\Mt^2+{p_{\rT,\Pt}^2}}\,\sqrt{\Mt^2+{p_{\rT,\overline{\Pt}}^2}}\right)^{1/2}  \,,
\eeq
where the top and antitop transverse momenta are reconstructed from
their decay products based on the Monte Carlo truth. As pointed out in~\citere{Denner:2020hgg}, since the
determination of the top-quark momentum is subject to an ambiguity, we use
the lepton--neutrino pair that, combined with the $\Pb$ quark, gives rise to an
invariant mass which is the closest to $\Mt$. This scale definition is physically
motivated by the expectation that top--antitop resonant structures dominate the
cross-section, even when considering off-shell effects. Moreover, the overall
factor of one half in~\refeq{eq:scaleB} has been shown in~\citere{Denner:2020hgg} to give
a smaller scale sensitivity of the results when using the conventional $7$-point scale
variation and to reduce the size of QCD corrections (of $\nloone$ type).

The uncertainties in our results are estimated by computing the $7$-point scale envelope, \ie
by considering the maximum and minimum values of the cross-section evaluated on the set
of scales $(\mu_{\rm R},\,\mu_{\rm F})$ defined as
\beq\label{eq:scaleset}
\biggl(\frac{\mu_{\rm R}}{\mu_0},\,\frac{\mu_{\rm F}}{\mu_0}\biggr)\in\{(0.5,0.5)\,,(0.5,1)\,,(1,0.5)\,,(1,1)\,,(2,1)\,,(1,2)\,,(2,2)\}\,.
\eeq
Note that the $\as$ coupling entering the calculation of the NLO top-quark width is kept fixed
for the evaluation of the scale envelope.

In the following two sections, we present results for the fiducial cross-sections and
differential distributions where all contributions described in \refse{sec:calcdetails}
are combined in an \emph{additive} scheme, \ie
\begin{align}
\label{eq:addscheme}
  \sigma_{\rm LO+NLO} ={}& \sigma_{\rm LO_1}\,+\,\sigma_{\rm NLO_1} 
    \,+\,\sigma_{\rm LO_2}\,+\,\sigma_{\rm NLO_2}
    \,+\,\sigma_{\rm LO_3}\,+\,\sigma_{\rm NLO_3}+\,\sigma_{\rm NLO_4}\,,
\end{align}
which provides an exact result at the order of truncation of the perturbative expansion.

\subsection{Fiducial cross-sections}\label{sec:integrated}

In this section we present results for the integrated cross-section at different levels of perturbative accuracy in the
fiducial region discussed in \refse{sec:input}.
%%%%%%%%%%%%%%%%%%%%%%%%

\newcolumntype{R}[1]{>{\raggedright\let\newline\\\arraybackslash\hspace{0pt}}m{#1}}

\begin{table*}\small
  \centerline{
    \renewcommand{\arraystretch}{1.3}
  \begin{tabular}{C{1.8cm}|C{2.5cm}C{2.8cm}C{2.5cm}}%
  \hline
\textrm{Channel}  & $\loqcd$   &  $\loint$  & $\loew$   \\%[0.5ex] 
\hline
$\Pg\Pg$ &74.760(4)&-&- \\%[0.5ex]
$\Pq\bar\Pq$ &32.486(3)&-&0.2848(1)\\%[0.5ex]
$\Pb\bar\Pb$ &\csp\csp\csp0.29208(9)&$-$0.6330(2)&0.7821(2)\\%[0.5ex]
$\bar\Pb\bar\Pb$/$\Pb\Pb$ &\csp\csp\csp0.02171(2)&\csp\csp\csp0.002516(9)&\csp\csp0.005817(9) \\%[0.5ex]
$\gamma\Pg$ &-&\csp\:0.7522(2)&-\\%[0.5ex]
$\gamma\gamma$ &-&-&\csp\csp0.001431(6)\\%[0.5ex]
\hline
\textrm{sum}  & 107.560(5)& \csp\:0.1217(3)&\,1.0742(3)\\%[0.5ex]
\hline
  \end{tabular}
  }
\caption{
  LO cross-sections (in ab) in the fiducial setup defined by \refeqs{eq:setup1}--\refeqf{eq:setup3}
  for the different sets of partonic channels contributing to the reaction
  $\Pp\Pp\to\Pe^+\nu_\Pe\,\mu^-\overline{\nu}_\mu\,{\Pb}\,\overline{\Pb}\,{\tau}^+{\tau}^-$.
    In the last line the sum of all partonic channels contributing at that specific order
  is reported as a reference. Integration errors are given in parentheses.
}\label{table:sigmainclLO_channels}
\end{table*}

In \refta{table:sigmainclLO_channels} we report the different LO contributions 
separately for the various partonic processes. Their sum at a given order is shown
in the last line as a reference.
As expected, the LO result is dominated by the contribution of the $\Pg\Pg$~channel, 
which just enters at order $\mc O(\as^2 \alpha^6)$, due to the high luminosity of the gluon PDF.
A significant fraction of the LO result is also represented by the $\Pq\bar\Pq$ channel, whose $\loqcd$
is roughly $30\%$ of the full $\loqcd$ result. 
%is roughly $43\%$ of the $\Pg\Pg$ $\loqcd$ result. 
The $\Pq\bar\Pq$ partonic process enters also at $\loew$,
which only amounts to roughly  $1\%$ of the corresponding $\loqcd$, as expected by the $\alpha/\as$ power suppression.
Its $\loint$ (interference) contributions exactly vanish owing to colour algebra.

The $\Pb\bar\Pb$ and $\Pb\Pb$/$\bar{\Pb}\bar{\Pb}$ channels participate in all three LO terms, even though they
contribute overall at the sub-percent level. The bottom contribution is dominated by the 
$\Pb\bar\Pb$~partonic channels, due to the enhancement of  doubly-resonant top-quark topologies, which are absent
for the $\bar\Pb\bar\Pb$/$\Pb\Pb$ cases. All LO contributions are comparable in size,
contrary to the expectation from a naive power counting, with the interference terms in
$\lobint$ having negative sign. Indeed,
at $\mc O(\alpha^8)$ the $\Pb\bar\Pb$ partonic channel embeds doubly-resonant topologies with a $t$-channel $\PW$-boson exchange 
[like the one shown in~\reffi{fig:lob2}], which is absent
in the corresponding tree-level QCD diagrams.
Since the $\Pb\Pb$/$\bar{\Pb}\bar{\Pb}$ channels allow only for a
single top resonance their integrated cross-sections are small fractions of the $\Pb\bar\Pb$
ones at each perturbative order:
the largest $\Pb\Pb$/$\bar{\Pb}\bar{\Pb}$ contribution is provided by the $\lobqcd$ term,
which is already only $7\%$ of the $\lobqcd$ one for the $\Pb\bar\Pb$ channels.

Two additional partonic channels are allowed at LO,
namely the $\gamma \Pg$ and $\gamma \gamma$ ones. The former enters at $\mc O(\as \alpha^7)$ and
is comparable in size with the contribution of the $\Pb\bar\Pb$ reactions: indeed the
photon PDF suppression is partially compensated by the gluon PDF and by
the appearence of topologies involving
a $t$-channel $\PW$-boson exchange [as in~\reffi{fig:loint1}]. Overall, it amounts to roughly $1\%$ of the contribution from the dominant $\Pg\Pg$ channel.
The purely photon-induced contributions are even further suppressed by the photon luminosity
and by the EW nature of the reaction, accounting for a few milli-percent of the $\Pg\Pg$ channel and for $0.1\%$ of the full $\loew$ result.

%%%%%%%%%%%%%%%%%%%%%%%%
\begin{table*}\small
  \centerline{
    \renewcommand{\arraystretch}{1.3}
  \begin{tabular}{C{1.8cm}|C{2.5cm}C{2.7cm}C{2.5cm}C{2.5cm}}%
  \hline
\textrm{Channel}  &   $\nloone$  & $\nlotwo$   &  $\nlothree$  & $\nlofour$ \\%[0.5ex] 
\hline
$\Pg\Pg$ &\hspace{-0.5cm}$-$14.9(1)&$-$0.107(9)&-&- \\%[0.5ex]
$\Pq\bar\Pq$ &\hspace{-0.27cm}$-$12.35(7)& \hspace{0.1cm}$-$1.177(6)  &\hspace{0.1cm}0.013(4)&$-$0.0380(9) \\%[0.5ex]
%$\Pb\bar\Pb$ &\hspace{0.07cm}$-$0.106(2)&\hspace{0.2cm}0.195(2)  &\hspace{-0.1cm}$-$0.324(4)& $-$0.0194(9)\\%[0.5ex]
$\Pb\bar\Pb$ &\hspace{0.07cm}$-$0.106(2)&\hspace{0.2cm} 0.253(2)  &\hspace{-0.1cm}$-$0.324(4)& $-$0.0194(9)\\%[0.5ex]
%$\bar\Pb\bar\Pb$/$\Pb\Pb$ &\hspace{0.735cm}0.00031(7)&\hspace{0.1cm}$-$0.0016(1)&$-$0.0022(2)&\hspace{0.2cm}$-$0.00059(2) \\%[0.5ex]
$\bar\Pb\bar\Pb$/$\Pb\Pb$ &\hspace{0.735cm}0.00031(7)&\hspace{0.1cm}$-$0.0017(1)&$-$0.0022(2)&\hspace{0.2cm}$-$0.00059(2) \\%[0.5ex]
$\gamma\Pg$ &-&$-$0.136(2)&\hspace{0.3cm}0.0101(8)& -\\%[0.5ex]
$\gamma\gamma$ &-&-&\hspace{0.2cm}$-$0.00020(3)& \hspace{0.2cm}$-$0.00010(2)\\%[0.5ex]
\hline
$\Pg\Pq$/$\Pg\bar\Pq$ &15.77(3)&\hspace{0.5cm}0.0570(5)&\hspace{0.1cm}1.102(1)& -\\%[0.5ex]
$\Pg\Pb$/$\Pg\bar\Pb$ &\hspace{0.3cm}0.624(2)&$-$0.146(2)&\hspace{0.1cm}0.237(2)& -\\%[0.5ex]
$\gamma\Pq$/$\gamma\bar\Pq$ &-&\hspace{0.5cm}0.4774(8)&-& \hspace{0.5cm}0.00403(2)\\%[0.5ex]
$\gamma\Pb$/$\gamma\bar\Pb$ &-&\hspace{0.6cm}0.00347(9)&\hspace{0.1cm}$-$0.00026(1)&\hspace{0.5cm}0.00194(1) \\%[0.5ex]
\hline
%\textrm{sum} &\hspace{-0.5cm}$-$10.9(1)&\hspace{-0.1cm}$-$0.83(1)&\hspace{0.1cm}1.037(6)&\hspace{-0.1cm}$-$0.052(1) \\%[0.5ex]
\textrm{sum} &\hspace{-0.5cm}$-$10.9(1)&\hspace{-0.1cm}$-$0.78(1)&\hspace{0.1cm}1.037(6)&\hspace{-0.1cm}$-$0.052(1) \\%[0.5ex]
\hline
  \end{tabular}
  }
\caption{
  NLO corrections (in ab) to the LO cross-section in the fiducial setup defined by
  \refeqs{eq:setup1}--\refeqf{eq:setup3}
  shown for the different sets of partonic channels contributing to the reaction
  $\Pp\Pp\to\Pe^+\nu_\Pe\,\mu^-\overline{\nu}_\mu\,{\Pb}\,\overline{\Pb}\,{\tau}^+{\tau}^-$.
  In the last line the sum of all partonic channels contributing at that specific order
  is reported as a reference.  Integration errors are given in parentheses.
}\label{table:sigmainclNLO_channels}
\end{table*}
%%%%%%%%%%%%%%%%%%%%%%%%

In~\refta{table:sigmainclNLO_channels} we illustrate separately the contributions of the
various partonic channels to the different NLO corrections. The sum of the channels at
each perturbative order is reported in the last line as a reference.
At $\nloone$ the new gluon-induced partonic channels $\Pg\Pq$ open up. In our fiducial
setup they provide the largest correction in absolute value, which is of positive sign,
since the $\Pg\Pq$ channels just comprise real tree-level contributions (together with
  corresponding subtraction counterterms cancelling initial-state collinear singularities).
The corresponding bottom counterparts, namely $\Pg\Pb$, are also present,
but just amount to roughly $4\%$ of the $\Pg\Pq$ ones.
A sizeable and negative correction at this order arises from the
virtual and real corrections to the $\Pg\Pg$ and $\Pq\bar\Pq$ channels, roughly $-20\%$ and $-38\%$
of the corresponding $\loqcd$, respectively. Similarly, for the $\Pb\bar\Pb$ process
the $\lobqcd$ corrections amount to roughly $-36\%$ of the corresponding
LO, but their contribution to the overall $\nloone$ corrections is only about $1\%$.
The QCD corrections to the $\Pb\Pb$/$\bar{\Pb}\bar{\Pb}$ bottom
channels are instead fully negligible.

All partonic channels contributing to $\nloone$ also receive corrections at the order \sloppy $\mc O(\as^2 \alpha^7)$
together with the $\gamma\Pg$ one, whose $\nlotwo$ contribution can be unambiguously identified as
a QCD correction to its $\loint$. In our fiducial region the $\nlotwo$ corrections to $\gamma\Pg$ are negative and amount
to $-18\%$ of their LO. 
At $\mc O(\as^2 \alpha^7)$  also the $\gamma\Pq$ and $\gamma\Pb$ channels open up, which are clearly
EW real corrections to $\loqcd$. Actually, the $\gamma\Pq$ cross-section is the second-largest contribution (in absolute value)
to $\nlotwo$ after the one of the  $\Pq\bar\Pq$ reaction.
We remind the reader that the $\Pq\bar\Pq$  partonic channel,
together with $\Pb\bar\Pb$, $\Pb\Pb$, and $\bar\Pb\bar\Pb$,
receives both EW corrections from $\loqcd$ and QCD corrections from $\loint$,
whose contributions cannot be unambiguously separated.
The EW corrections to $\Pg\Pg$ (also contributing at $\nlotwo$ order) %, which are the last class of NLO corrections to this partonic channel,
 are quite small if compared to the corresponding $\nloone$ ones (just a $0.8\%$ of them).
The real channels $\Pg\Pq$ and $\Pg\Pb$ enter at order $\mc O(\as^2 \alpha^7)$ as genuine QCD corrections to the LO interference.
The $\nlotwo$ result for the $\Pg\Pq$ channel is $0.4\%$ of its corresponding contribution at $\nloone$, while for the $\Pg\Pb$ one
it amounts to $23\%$ of the corresponding $\nloone$ counterpart but of negative relative sign.
More strikingly, the impact of the  $\Pg\Pb$ partonic channels is more sizeable than the $\Pg\Pq$ one.
This can be traced back to the different topologies contributing in
the two cases. For $\Pg\Pq$ initial states, the colour structure is only
non-vanishing if the additional gluon connects the light-quark and the
bottom-quark lines [see \reffi{fig:nlo21}]. In
the $\Pg\Pb$ case, the presence of $t$-channel gluon exchange in the LO QCD amplitude allows for more possibilities for the additional
gluon to connect the two b-quark lines, opening diagram topologies  with a non-vanishing colour factor
that are not kinematically suppressed.

At $\nlothree$ the largest positive contribution comes from the $\Pg\Pq$ channel, which plays the role of a QCD correction to
$\loew$, but is actually almost $4$ times larger than the corresponding $\loew$ cross-section in the $\Pq\bar\Pq$
channel. As anticipated in \refse{sec:nlo3}, this result can be explained by the presence of $\Pt\PZ$-scattering
topologies that open up at this order [see left sub-diagram in~\reffi{fig:nlo42}]. For the $\Pg\Pb$ reaction, the $\nlothree$ contribution is a non-negligible fraction
of its $\nloone$ one, but 
roughly a factor of $3$ smaller than the corresponding $\loew$ cross-section in the $\Pb\bar\Pb$ channel.
Indeed, differently from the $\Pg\Pq$ case, all relevant  scattering topologies contributing to the $\Pg\Pb$ reaction
are already present at $\lobew$. Therefore, the $\nlothree$ corrections
to the $\Pb\bar\Pb$ channels are comparable in size to the $\Pg\Pb$ ones, and, being negative, they largely cancel them.
Moreover, since the $\lobew$ contribution is enhanced by EW-boson scattering topologies [see~\reffis{fig:lob2} and~\ref{fig:lob3}], its QCD corrections
to the $\Pb\bar\Pb$ and the $\Pb\Pb$/$\bar{\Pb}\bar{\Pb}$ channels are not suppressed with respect to the
$\nloone$ and $\nlotwo$ corrections to the same channels.
On the contrary,  the $\mc O(\as\alpha^8)$ corrections to $\Pq\bar\Pq$
are $\alpha$-suppres\-sed, as expected: they amount to roughly a per-mille of the corresponding $\nloone$ term.
The $\nlothree$ QCD corrections to the $\gamma\gamma$-induced process are fully negligible.
In the $\gamma\Pg$ partonic channel, the $\nlothree$ corrections are of purely EW nature and
amount to $7\%$ of the QCD corrections of order $\nlotwo$. 
An even stronger suppression is visible in the $\gamma\Pb$ channels, which enter this order as
real EW corrections to the LO interference.

All $\nlofour$ corrections for each partonic channel are consistently found to be
$\alpha$-suppres\-sed with respect to the corresponding $\loew$ and
overall negligible compared to the other corrections.
This last statement holds true for all channels receiving an $\mc O(\alpha^9)$ contribution,
with the exception of the $\Pb\Pb$/$\bar{\Pb}\bar{\Pb}$ bottom ones
(whose $\nloone$ corrections were already $3\times 10^{-4}\,$ab)
and the $\Pq\bar\Pq$ ones, with a negative $\nlofour$ term of the same order of magnitude
as the $\nlothree$ one.
The overall $\nlofour$ result is dominated by the negative contribution of
$\Pq\bar\Pq$/$\Pb\bar\Pb$ channels, with a positive $\gamma\Pq$/$\gamma\Pb$ contribution
roughly giving a $10\%$ of the former.
The EW corrections to the di-photon channel, whose virtual contribution
has been the most CPU-expensive part of the calculation to  evaluate, is
the smallest correction amongst the ones in \refta{table:sigmainclNLO_channels} (less than
half of the QCD corrections to this channel) and therefore fully negligible.

\begin{table*}\small
\renewcommand{\arraystretch}{1.3}
\renewcommand{\tabcolsep}{5pt}
  \centerline{
%\begin{tabular}{C{1.7cm}|C{2.6cm}C{1.1cm}|C{1.8cm}C{1.1cm}||C{2.6cm}C{1.1cm}}%
\begin{tabular}{c|cr|cr|cr}%
  \hline
\multicolumn{1}{C{1.7cm}|}{\textrm{perturbative order}} & $\sigma_{\rm{no}\Pb}$
[ab]   &  
$\frac{\sigma_{\rm{no}\Pb}}{\sigma_{\rm{no}\Pb,\,\rm{LO}_{1}}}$ & $\sigma_{\Pb}$ [ab]   &  $\frac{\sigma_{\Pb}}{\sigma_{\rm{no}\Pb,\,\rm{LO}_{1}}}$& $\sigma$ [ab]   &  $\frac{\sigma}{\sigma_{\rm{LO}_{1}}}$  \\ 
\hline
$\rm LO_{1}$  &  107.246(5)$^{+35.0\%}_{-24.0\%}$  &  1.0000 & 0.31378(9)& $+$0.0029 &107.560(5)$^{+34.9\%}_{-23.9\%}$& 1.0000\\[0.9ex]
$\rm LO_{2}$  &  0.7522(2)$^{+11.1\%}_{-9.0\%}$ & $+$0.0070  & $-$0.6305(2)& $-$0.0059 &0.1217(3)& $+$0.0011\\[0.9ex]
$\rm LO_{3}$  &  0.2862(1)$^{+3.4\%}_{-3.4\%}$  &
$+$0.0027&0.7879(2)&$+$0.0073 & 1.0742(3)$^{+12.1\%}_{-14.9\%}$& $+$0.0100 \\[0.9ex]
  \hline
${\rm NLO_1}$  & $-$11.4(1)    & $-$0.1072  & 0.518(3)    & $+$0.0048 &$-$10.9(1)& $-$0.1016\\ 
  %${\rm NLO_2}$  & $-$0.89(1)    & $-$0.0083  & 0.051(3)    & $+$0.0005 & $-$0.84(1) & $-$0.0078\\
  ${\rm NLO_2}$  & $-$0.89(1)    & $-$0.0083  & 0.109(3)  & $+$0.0010 & $-$0.78(1) & $-$0.0072\\
${\rm NLO_3}$  & 1.126(4)      & $+$0.0105  & $-$0.089(4) & $-$0.0008 & 1.037(6)   & $+$0.0096\\
${\rm NLO_4}$  &  $-$0.0340(9) & $-$0.0003  & $-$0.0180(9)& $-$0.0002 & $-$0.052(1) & $-$0.0005\\
  \hline
$\rm LO_{1}$+${\rm NLO_1}$ & 95.8(1)$^{+0.4\%}_{-11.2\%}$ &$+$0.8933&0.832(3)& $+$0.0078&96.6(1)$^{+0.4\%}_{-10.7\%}$& $+$0.8984\\[0.9ex]
  \hline
  \hline
    LO & 108.285(5)$^{+34.7\%}_{-23.8\%}$  &$+$1.0097  &0.4713(3) & $+$0.0044 & 108.756(5)$^{+34.5\%}_{-23.7\%}$& $+$1.0111\\[0.9ex]
    %LO+NLO & 97.0(1)$^{+0.5\%}_{-11.2\%}$  & $+$0.9052  &0.932(6) &$+$0.0087 & 98.0(1)$^{+0.4\%}_{-10.7\%}$& $+$0.9114\\[0.9ex]
    LO+NLO & 97.0(1)$^{+0.5\%}_{-11.2\%}$  & $+$0.9052  &0.991(6) &$+$0.0092 & 98.0(1)$^{+0.4\%}_{-10.6\%}$& $+$0.9114\\[0.9ex]
  \hline
\end{tabular}
}
\caption{
  LO cross-sections and NLO corrections (in ab) in the fiducial setup. In the second column all partonic channels are included in $\sigma_{\rm{no}\Pb}$
  except the ones having at least one bottom quark in the initial
  state, while  $\sigma_{\Pb}$ includes all these channels.
  The sum of the two ($\sigma$) is shown in the sixth column. Ratios with respect to the cross-section $\sigma_{\rm{no}\Pb}$ at
  $\loqcd$ accuracy are reported in the third and fifth column. In the seventh column ratios are shown with respect to the full
  $\loqcd$ cross-section including the bottom channels, as
  well. Integration errors are given in parentheses and percentage 7-point scale
  variations as super- and sub-scripts. 
}\label{table:sigmainclNLO_combined}
\end{table*}
Results for the integrated cross-sections, which have been separately shown for the different
partonic channels in \reftas{table:sigmainclLO_channels} and~\ref{table:sigmainclNLO_channels}, are collected in
\refta{table:sigmainclNLO_combined}. Theoretical uncertainties in the results are estimated with
7-point scale variations, as described in \refse{sec:input}. We refrain from showing
scale uncertainties for $\loint$ results in all cases where interference contributions from
bottom-induced diagrams are present. For such contributions, the
QCD-scale (renormalisation- and factorisation-scale) uncertainties do not have a
clear interpretation.
However, the scale uncertainties are shown for the $\loint$ contribution from the $\gamma\Pg$ process.
Uncertainty bands for $\loew$ results come uniquely from factorisation-scale variations,
owing to the EW nature of the contribution, which is also why they turn out to be smaller than the $\loqcd$ ones.

In the second column of \refta{table:sigmainclNLO_combined} we present  results
for the integrated cross-section $\sigma_{\rm{no}\Pb}$ at different perturbative accuracies: all partonic channels are included, except the ones
involving at least one bottom quark in the initial state. The contribution $\sigma_{\Pb}$ to the integrated cross-section,
namely the sum of all bottom-induced contributions, is presented in the fourth column individually for each
perturbative order. The results accounting for all partonic channels ($\sigma=\sigma_{\rm{no}\Pb}+\sigma_{\Pb}$) are reported in the sixth
column of the table. For the bottom contributions alone, we also decided not to report scale uncertainties:
since the running of the bottom PDFs that enters when computing the scale
envelope includes also contributions from other quark flavours, the interpretation
of the scale-uncertainty bands for these comparably small contributions is unclear and not particularly revealing. The scale variations
for bottom-induced contributions are instead properly taken into account for the complete result $\sigma$.

%%%%%%%%%%%%%%%%%%%%%%%%

From \refta{table:sigmainclNLO_combined}, one can notice that
the LO contribution to $\sigma_{\rm{no}\Pb}$ is dominated by the $\loqcd$ term. The largest
NLO correction is represented by the $\nloone$ term: with our scale choice of \refeq{eq:scaleB}
it roughly amounts to a $-10\%$ correction of the $\loqcd$ result and its inclusion significantly
reduces the size of QCD-scale uncertainties.
The $\nlothree$ turns out to be the largest NLO contribution amongst the subleading ones,
providing a positive $1\%$ correction to $\loqcd$. 
Since $\loew$ is a purely EW contribution, no improvement in QCD-scale uncertainty is
obtained when adding the $\nlothree$ correction.
The $\nlotwo$ term is slightly smaller than the $\nlothree$ one, but being negative, largely cancels
the effect of the $\nlothree$ contribution,
so that the sum $\nlotwo+\nlothree$ only corrects the $\loqcd$ by $0.2\%$.
Finally, the $\nlofour$ just affects the NLO result for $\sigma_{\rm{no}\Pb}$
at the sub-per-mille level. Due to their EW nature (with the exception of the small
contribution arising from the QCD corrections to $\loint$) both $\nlotwo$ and $\nlofour$
do not further reduce the scale uncertainties.

The $\lobqcd$ contribution for the bottom channel is comparable in
size to the subleading $\lobint$ and $\lobew$ terms. The sum of these three LO perturbative orders
only modifies the complete LO result for $\sigma$ at the per-mille level,
also as a consequence of large cancellations occurring amongst them. A similar
impact on the full NLO result for $\sigma$ is observed for the set of NLO corrections in bottom-induced
channels. The largest contribution still arises from the $\nlobone$ perturbative order, which changes the
$\nloone$ corrections computed without including the bottom contributions by $-4\%$. 
The $\nlobtwo$ and $\nlobthree$ corrections 
are again comparable in size but, due to their opposite signs, compensate each other to a large extent.
The $\nlofour$ corrections to $\sigma$, $35\%$ of which comes from the bottom-induced channels, still remain
negligible compared to the other NLO terms.
Therefore, the LO$+$NLO result for $\sigma$ only receives roughly a $+1\%$ correction after including
the bottom channels at all LO and NLO perturbative orders.

As a general remark, we stress that the cancellations of contributions
between different partonic channels should not be viewed as a general
feature but might be related to our specific event selection. The sizes of
the individual channels are more characteristic than sums of different
contributions.

\subsection{Differential cross-sections}\label{sec:differential}

In this section we study the impact of including the NLO corrections
at differential level, by examining some relevant distributions.
Since $\nlofour$ corrections have already been shown in the previous section
to be fully negligible and out of reach in any foreseen LHC experimental measurement,
we refrain from including them in our differential results. Unless otherwise stated,
all predictions include bottom-initiated contributions. Whenever presented, the uncertainty
bands are obtained by means of 7-point scale variations performed at the bin-by-bin level.

All plots are characterised by the following structure. In a main panel the three LO contributions
are presented with dashed lines, namely the $\loqcd$ (blue curve), the $\loint$ (magenta curve) and
$\loew$ (turquoise curve). The last two contributions are shown after scaling them up by a factor of 10.
In the same panel, the genuine NLO QCD result, obtained by correcting the $\loqcd$ result by $\nloone$, is shown in red,
together with the full NLO prediction LO$+$NLO, shown in brown. All plots are supplemented with at least
two ratio panels. The first one shows again the $\loew$ curve but also the three NLO corrections, namely
$\nloone$ (in red), $\nlotwo$ (in green), and $\nlothree$ (in goldenrod), normalised to the $\loqcd$ result.
Since the $\loint$ contribution is typically negligible or at most similar in size to the $\loew$ one, we refrain from including its
curve in any ratio panel, with just one exception for the distribution in the invariant mass of the $\tau$-lepton pair
(as justified below).
A second ratio panel compares the $\loqcd+\nloone$ result with the complete LO$+$NLO prediction, both normalised with respect to
the former. For a selected set of distributions two additional ratio panels are included. The first additional panel
reports three curves, normalised to the $\loqcd$ result: the $\nlotwo$ and $\nlothree$ corrections are included again
as a reference together with an additional purple curve, which includes photon-induced channels (namely all channels having
at least one photon as an initial-state parton). The second additional panel is added to illustrate the impact of the
bottom-induced contributions. Our full LO$+$NLO prediction is reported together with the $(\rm{LO}+\rm{NLO})_{\rm no b}$ result (dash-dotted
orange curve), where all channels with at least one initial-state bottom have been excluded: the latter prediction
is also used for normalisation.

We consider observables that are directly measurable at the LHC as well as observables that rely on Monte Carlo truth.
The latter refer to kinematic variables of the top and antitop quark
or of jets initiated by bottom/antibottom quarks arising from the top/antitop decays. As already mentioned,
some partonic channels may include up to three final-state bottom quarks. Moreover, since our jet-clustering
algorithm does not distinguish bottom and antibottom, bottom jets
are identified as originating from a top or antitop quark by maximising the
likelihood function $\mathcal{L}$, defined as a product of two
Breit--Wigner distributions corresponding to the top-quark and
antitop-quark propagators,
\begin{equation}\label{eq:likelihood}
 \mathcal{L}(p_{\Pb_i},\,p_{\Pb_j}) =  
 \frac{1}{\left(p^2_{\Pe^+\nu_\Pe\Pb_i} - m_\Pt^2\right)^2+\left(m_\Pt \Gamma_\Pt\right)^2} \;
 \frac{1}{\left(p^2_{\mu^-\bar\nu_\mu\Pb_j} -
     m_\Pt^2\right)^2+\left(m_\Pt \Gamma_\Pt\right)^2},
\end{equation}
where the momenta $p_{a b c}$ are defined as $p_{a b c} = p_{a} +p_{b} + p_{c}$.
Notice that at the Monte Carlo-truth level the neutrino flavour is accessible.
The combination of momenta $\{p_{\Pb_i}, p_{\Pb_j} \}$ (at most nine for channels with three final-state
bottom quarks) that maximises this function defines the two bottom jets originating
from top quarks: the first momentum of the pair is identified with a bottom quark
and the second with an antibottom quark. The top- and antitop-quark momenta are then computed
as 
\beq
p_\Pt = p_{\Pe^+\nu_\Pe\Pb_i}\qquad\text{and}\quad \qquad p_{\bar\Pt} = p_{\mu^-\bar\nu_\mu\Pb_j}\,,
\eeq
respectively. Owing to the final state involving three different lepton flavours, 
all considered observables refer to charged leptons via their flavours
with no need to introduce a transverse-momentum ordering.
In the dominant $\Pt\bar{\Pt}$-resonant topologies, the $\tau^+\tau^-$ pair is unambiguously
associated to a potential $\PZ/\gamma^*$~boson.
All leptons are understood as dressed leptons (see \refse{sec:input}).

\begin{figure}
  \centering
\subfigure[Azimuthal-angle separation between the positron and the muon in units of $\pi$\label{fig:phi}]{  \includegraphics[scale=0.32]{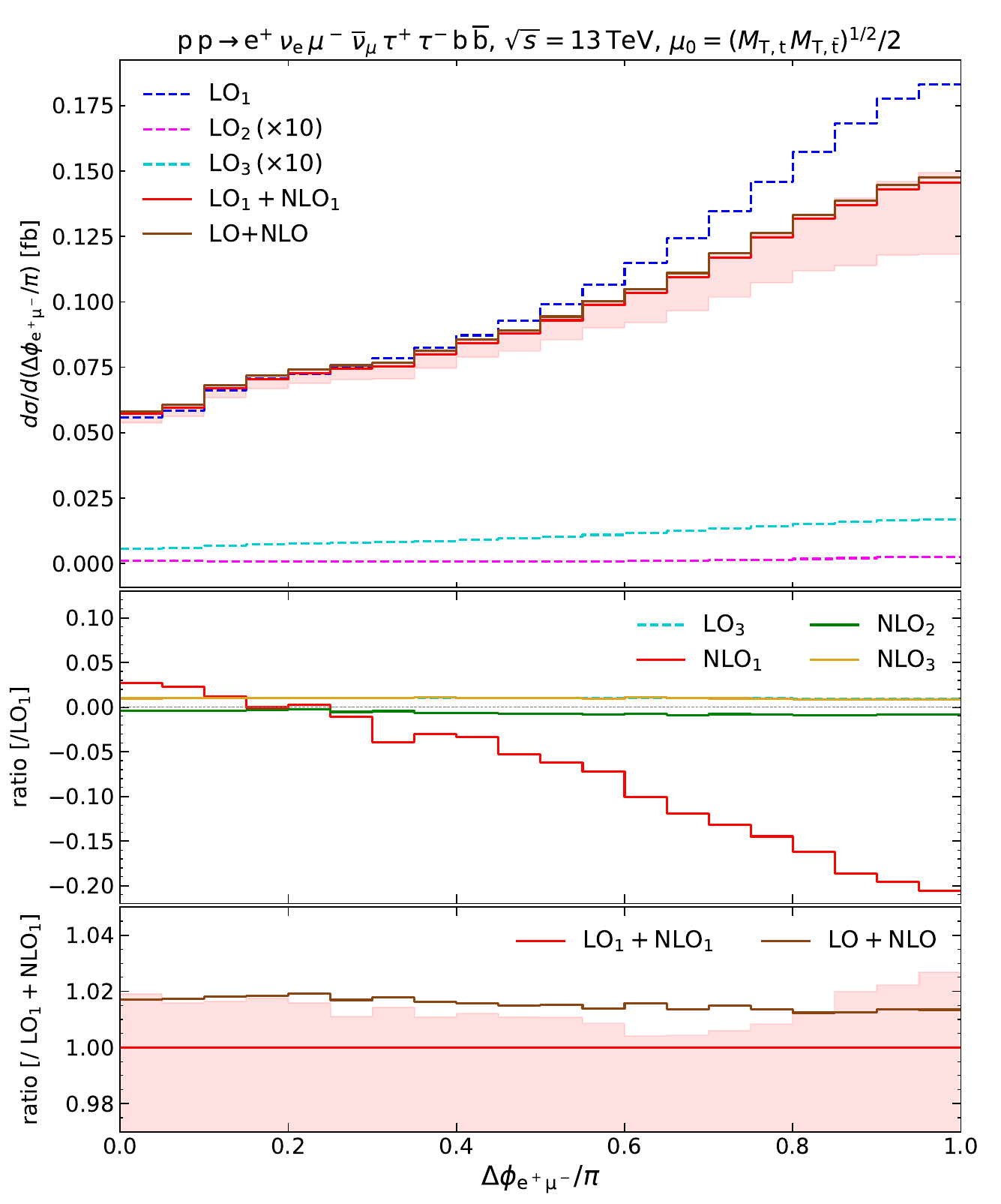}}
\subfigure[Cosine of the angle between $\tau^+$ and $\tau^-$\label{fig:cos}]{  \includegraphics[scale=0.32]{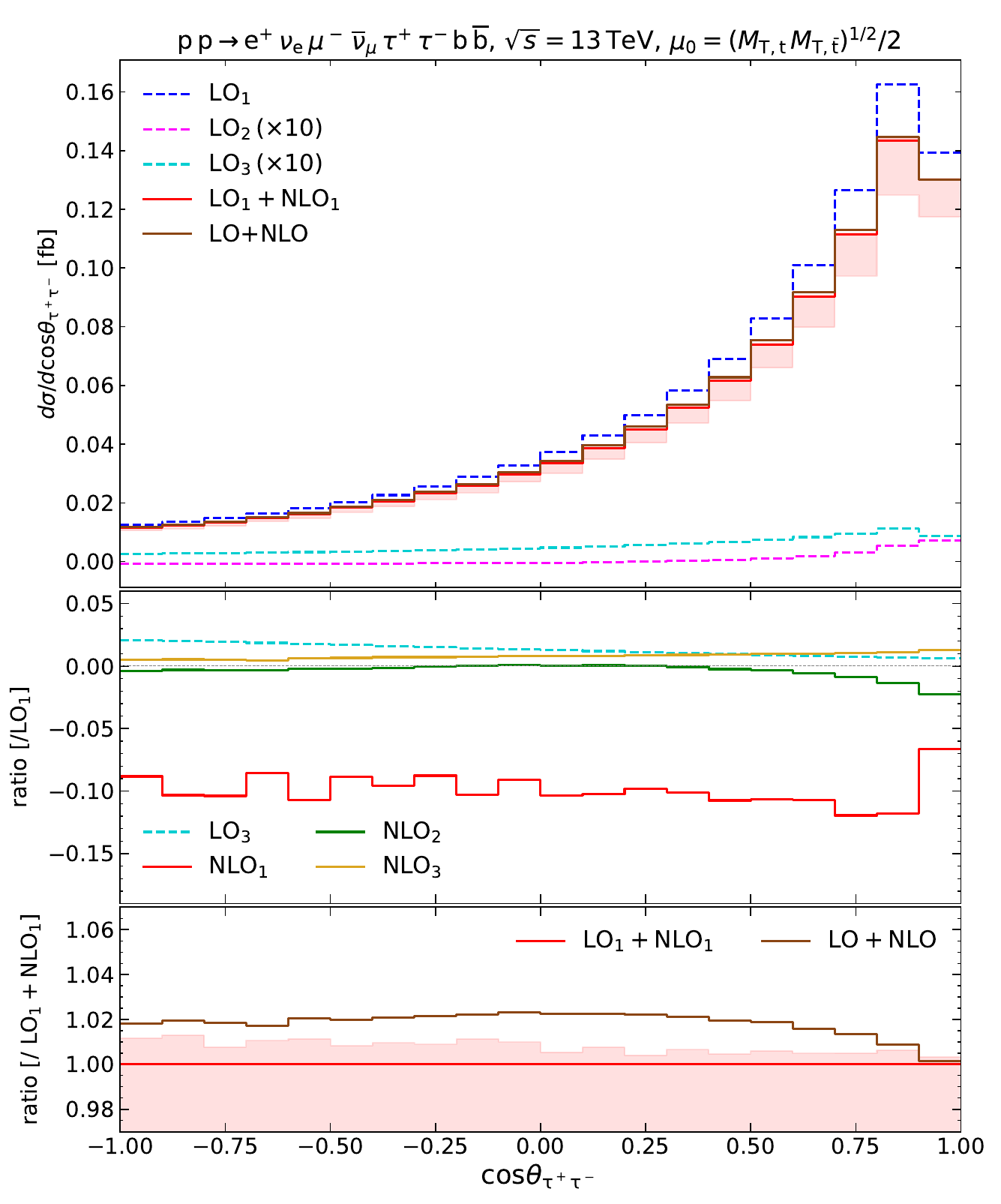}}
  \caption{Distributions in the azimuthal-angle separation between the positron and the muon in units of $\pi$ (left) and in the cosine
    of the angle between the two $\tau$ leptons (right). The different NLO corrections for the observables are compared
  separately (first ratio panels) and at the level of the full prediction (second ratio panel).}\label{fig:diff1}
\end{figure}

We start the discussion by examining the two angular distributions in \reffi{fig:diff1}. 
%%%%%%%%%%%%%%%%%%%%%%%%%%%%%%%%%%
%%%%%%%%%%%%%%%%%%%%%%%%%%%%%%%%%%
In \reffi{fig:phi} the distribution
in the azimuthal-angle separation between the positron and the muon $\Delta\phi_{\Pe^+\mu^-}$ is shown.
This observable is particularly relevant
to constrain some BSM theories and to study the spin correlations of top--antitop pairs. The cross-section shows a minimum 
at small azimuthal separation and monotonically increases towards larger values of $\Delta\phi_{\Pe^+\mu^-}$.
This trend, which is particularly pronounced at $\loqcd$, is partially mitigated by adding
$\nloone$ corrections. Indeed, the $\nloone$ corrections are positive at $\Delta\phi_{\Pe^+\mu^-}\approx 0$, where they amount to $2$--$3\%$, while they become
negative towards large azimuthal separation, reaching more than
$-20\%$ at $\Delta\phi_{\Pe^+\mu^-}\approx \pi$. The other
contributions, \ie $\loint$, $\loew$, $\nlotwo$, and $\nlothree$, are flat
and basically reproduce the relative  corrections to the fiducial
cross-section. Adding them to the
dominant QCD ones ($\loqcd$ and $\nloone$) amounts to
a shift in normalisation of about $+1\%$, giving a combined LO+NLO
curve sitting at the edge of the scale uncertainty band of the
$\loqcd+\nloone$ result for our scale choice \refeqf{eq:scaleB}. 

In \reffi{fig:cos} the distribution in the cosine of the angle between the two $\tau$ leptons is presented. It shows a minimum for
$\theta_{\tau^+\tau^-}=\pi$, where the two leptons are produced in a back-to-back configuration.
Then, the cross-section steeply increases reaching a maximum around
$\cos\theta_{\tau^+\tau^-}\approx 0.85$, where 
the two leptons are almost collinear. 
The right-most bin is lower than
the previous one, owing to the $\Delta R$ distance cut of 0.4 between charged
leptons [see \refeq{eq:setup2}].
$\nloone$ corrections essentially amount to a negative shift in the normalisation of the distribution
varying from $-9\%$ to $-11\%$ throughout the allowed kinematic range, reaching $-7\%$ in collinear configurations.
The $\loint$ contribution weakly increases towards the collinear region until the
very last bin of the distribution.
The $\loew$ contribution reflects the peaked behaviour of the
$\loqcd$ term, even though its size is much smaller: it ranges from $2\%$ to about $0.5\%$ of the $\loqcd$ when moving from the anticollinear
to the collinear region. The impact of subleading NLO corrections on this observable is quite moderate. $\nlotwo$ corrections
are negative and amount to roughly $-0.5\%$ in the anticollinear region, almost vanishing in the central emission range of
$-0.25<\cos\theta_{\tau^+\tau^-}<0.25$ and then decreasing again to $-2\%$ in the collinear-emission region. On the other hand,
$\nlothree$ corrections are positive and grow moderately from $0.5\%$ to $1\%$ at $\theta_{\tau^+\tau^-}=0$.
Due to the partial compensation of the different subleading corrections, their overall effect on the normalisation of
the $\loqcd+\nloone$ result vanishes in the collinear region, while it provides a $2\%$ correction in the anticollinear
and especially central-emission region, mainly due to the $\loew$ and $\nlothree$ terms.
%by roughly $2\%$, which can become as large as $3\%$ in the region $0<\cos\theta_{\tau^+\tau^-}<0.5$
%due to $\nlothree$ corrections.

\begin{figure}
  \centering
  \subfigure[Rapidity of the muon\label{fig:rapm}]{  \includegraphics[scale=0.32]{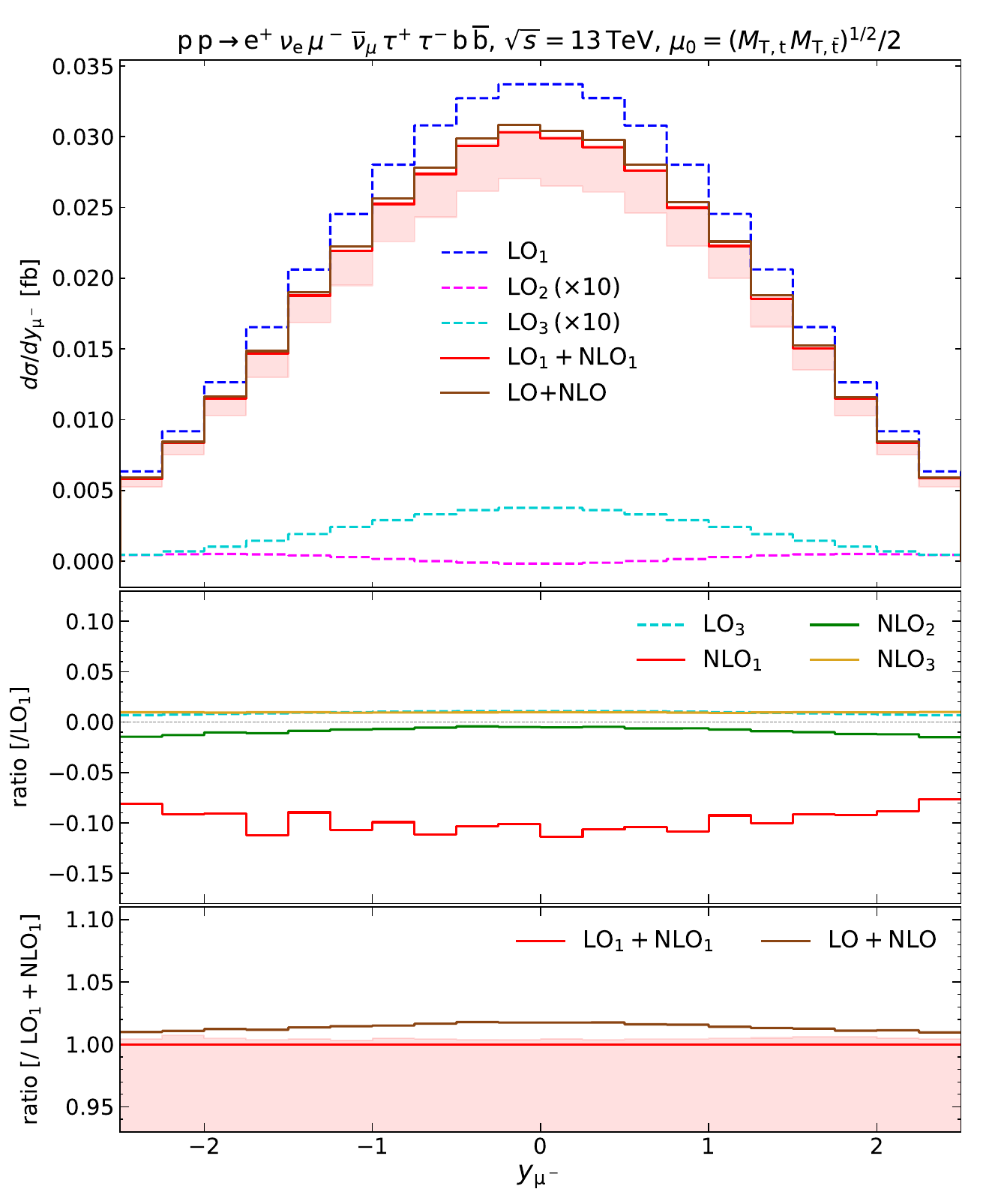}}
  \subfigure[Rapidity of the antitop quark\label{fig:rapat}]{\includegraphics[scale=0.32]{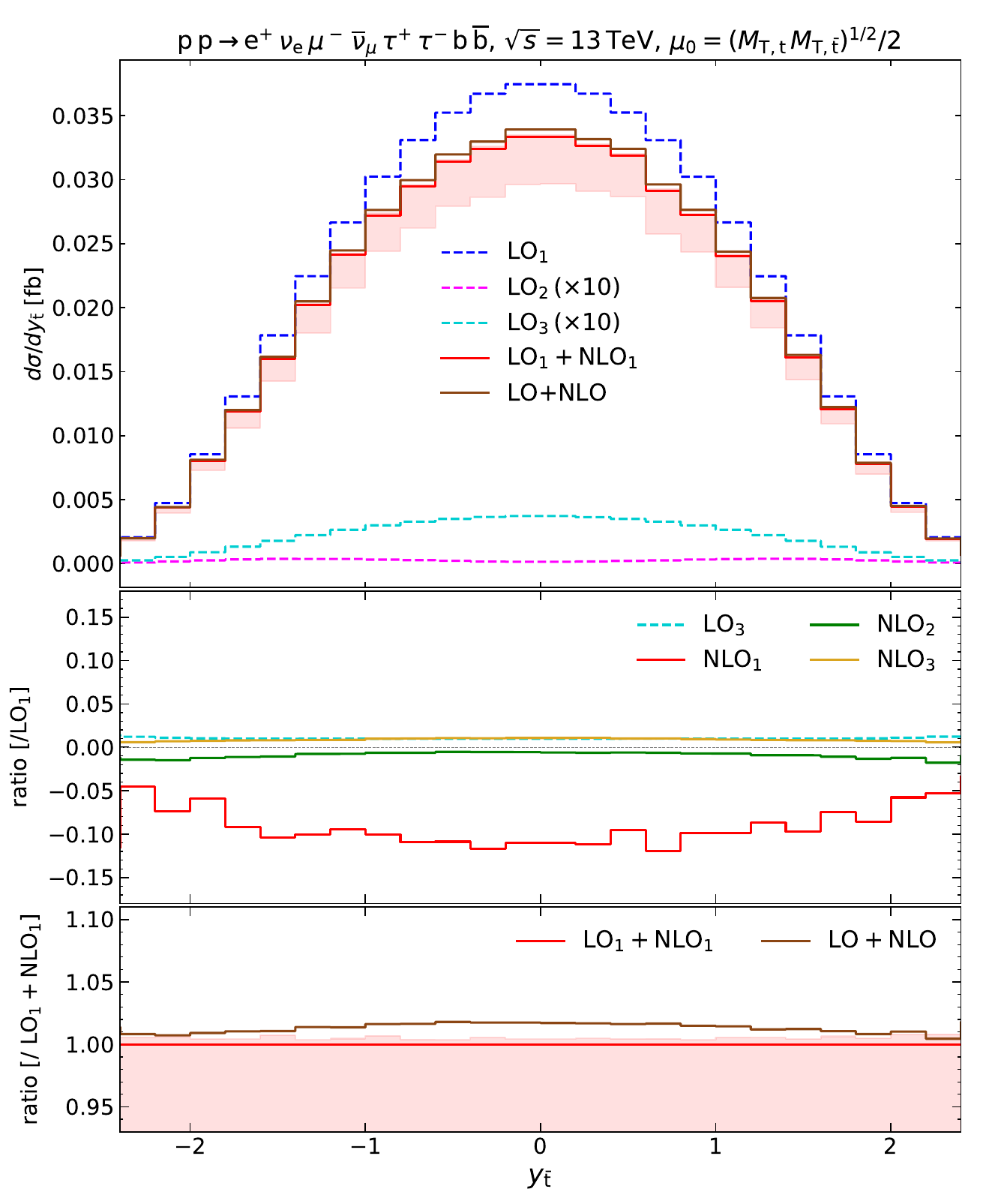}}
  \caption{Distributions in the rapidity of the muon (left) and of the antitop quark (right). The different NLO corrections for the observables are compared
  separately (first ratio panels) and at the level of the full prediction (second ratio panel).}\label{fig:diff2}
\end{figure}

We turn to the distributions in the rapidity of the muon and of the antitop quark 
in \reffis{fig:rapm} and~\ref{fig:rapat}, respectively. Similar considerations hold for both
observables, since the two quantities
are largely correlated (in the dominant doubly-resonant topologies the muon is a decay product of the antitop quark).
This allows to consider to a good approximation the muon rapidity as a proxy for the antitop one, which cannot be
directly measured at the LHC. Both the muon and the antitop quark are
preferably produced at central rapidity.
The muon-rapidity distribution is cut at $|y_{\mu^-}|<2.5$ [see \refeq{eq:setup2}], while the antitop one is highly suppressed
by the rapidity cuts applied on its visible decay products. $\nloone$ corrections are negative and
roughly $-10\%$ throughout the rapidity range, just slowly decreasing in absolute value towards the edges of the distribution
(especially for the antitop case, where they amount to $-5\%$ around
$|y_{\bar{\Pt}}|\approx 2.4$).
The $\loew$ term mimics the shape behaviour of the $\loqcd$ one to a
large extent, even though it represents a small fraction of it (just a
$1\%$). This behaviour was observed for all
distributions that we analysed in phase-space regions dominated by
on-shell top quarks.
The even smaller $\loint$ term
exhibits a dip at central rapidity: this is due to the fact that the $\loint$ result
is the sum of a $\gamma\Pg$ contribution,
which shows the expected behaviour at central rapidity, and bottom-induced interference
terms, which are negative and entirely cancel in our setup the enhancement at zero rapidity of the former contribution.
These large cancellations between the two channels at differential level
are already manifest for the fiducial cross-sections, as visible in the third column
of~\refta{table:sigmainclLO_channels}.
Subleading NLO corrections range between $-2\%$ and  $+1\%$
of the $\loqcd$ result. Owing to the opposite sign of the $\nlotwo$ and $\nlothree$ corrections, their overall impact on the combined NLO results
is moderate, giving a $+1\%$ normalisation factor (reaching a $4\%$  for the antitop quark in very suppressed forward regions).

\begin{figure} 
  \centering
\subfigure[Transverse momentum of the $\tau^+\tau^-$ pair\label{fig:pttt}]{  \includegraphics[scale=0.322]{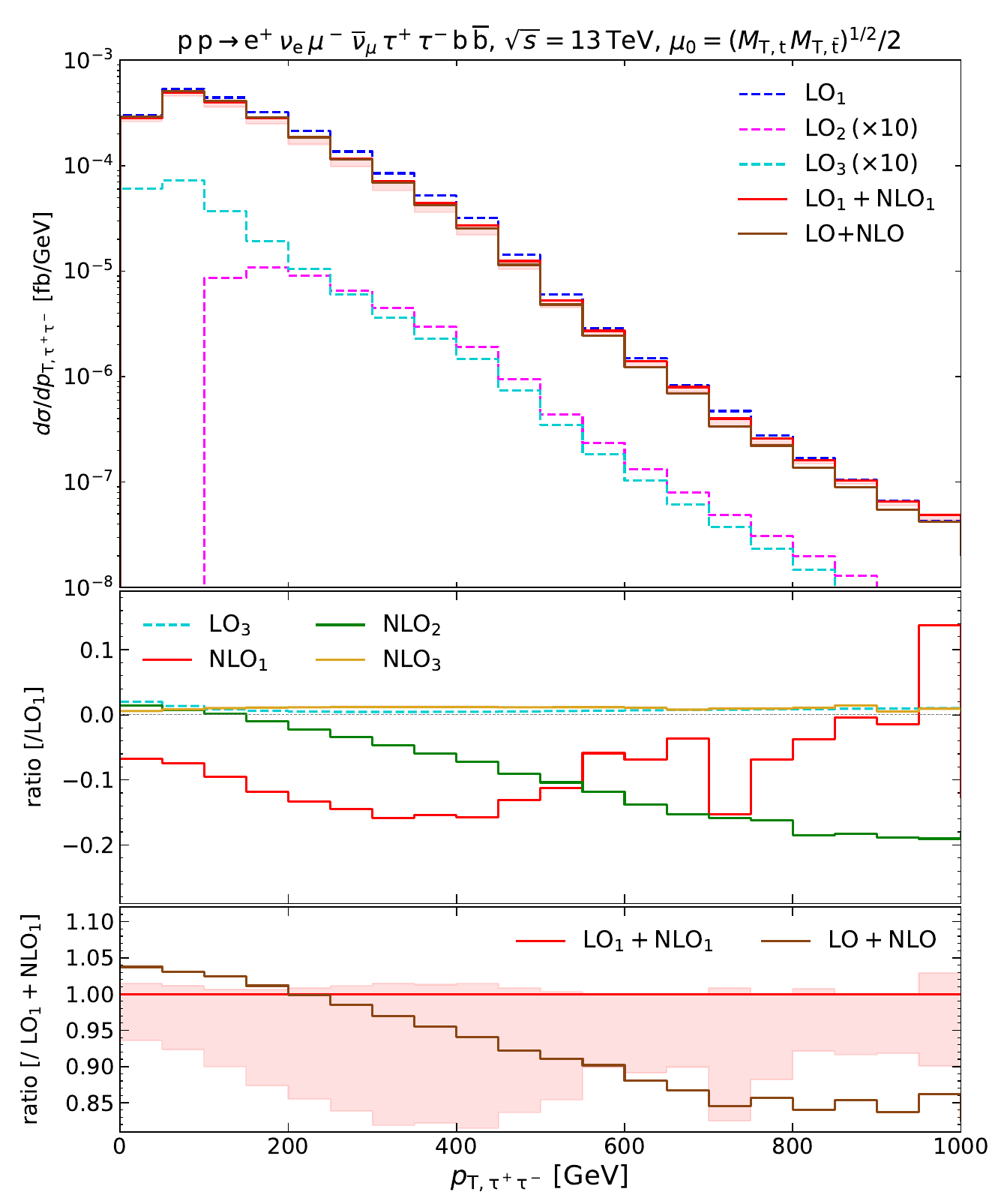}}
\subfigure[Transverse momentum of the antitop quark\label{fig:ptat}]{  \includegraphics[scale=0.32]{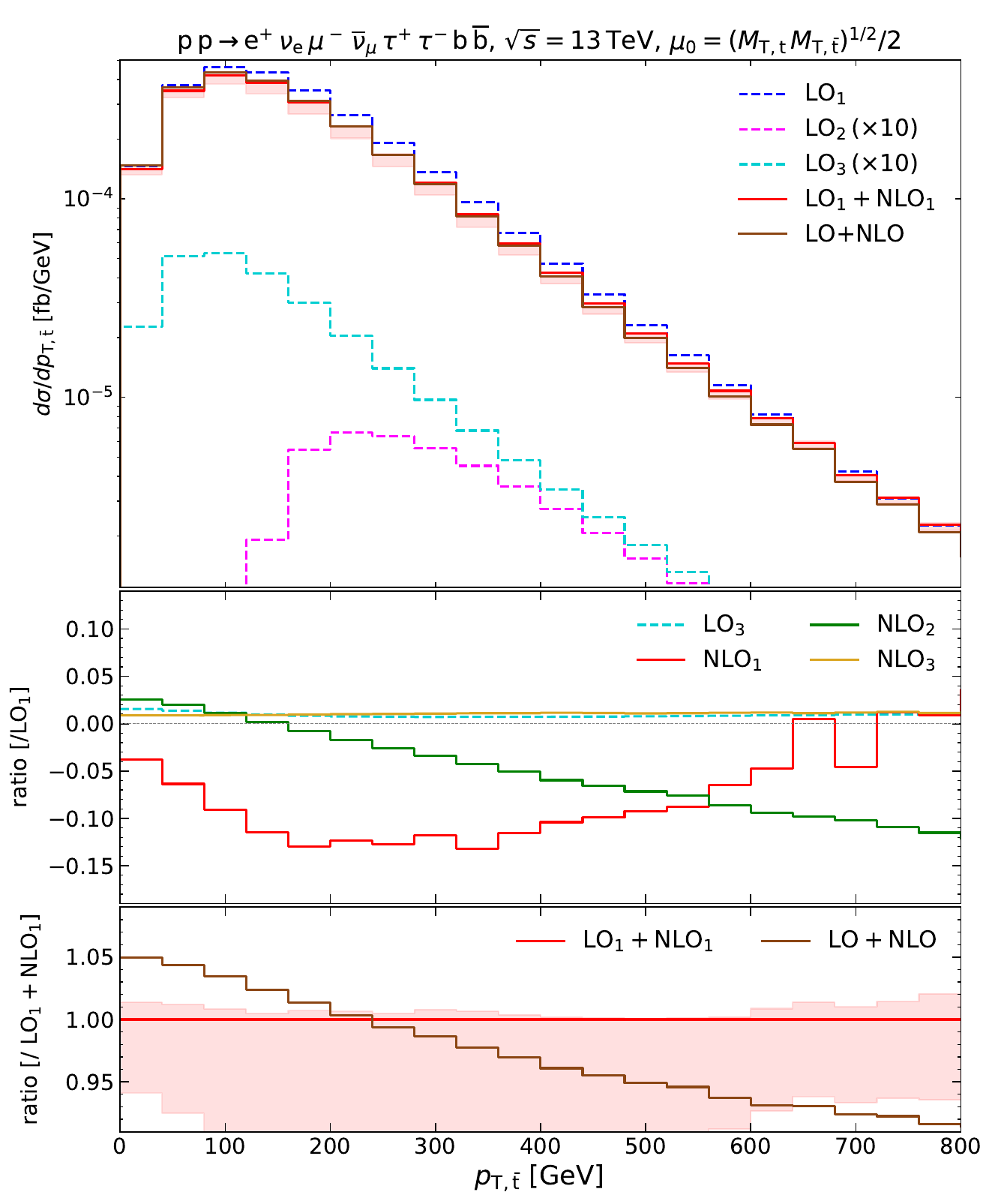}}
  \caption{Distributions in the transverse momentum of the $\tau^+\tau^-$ pair (left) and the antitop quark (right). The different NLO corrections for the observables are compared
  separately (first ratio panels) and at the level of the full prediction (second ratio panel).}\label{fig:diff3}
\end{figure}

We move on presenting some transverse-momentum distributions, which are known to be quite sensitive to EW corrections. In \reffis{fig:pttt} and~\ref{fig:ptat}
we display the distributions in the transverse momentum of the $\tau^+\tau^-$ pair and the antitop quark, the latter being accessible only with Monte Carlo truth.
In both cases the maximum of the cross-section is around $100\,$GeV. The $\nloone$ corrections are negatively increasing in the most populated region, reaching  $-15\%$
around $300\GeV$, 
then they decrease and turn positive
at very high transverse momenta (as visible for
the transverse momentum of the $\tau^+\tau^-$ pair).
While the position of the distribution maximum is unchanged when computing the $\loew$ contributions, for $\loint$  it is
shifted to higher values as a result of the negative interference contribution in bottom-induced channels
at low transverse momentum. Indeed, in this region the bottom-interference terms are as large as the $\gamma\Pg$ ones but of opposite sign. On the other hand,
towards the tails of the distribution the $\loint$
result is dominated by the positive $\gamma\Pg$ cross-section, which even
exceeds the $\loew$ one for  $p_{\rm T,t\bar{t}}$. A similar behaviour of the $\loint$ terms
was found in all transverse-momentum distributions that we
have studied, \ie large cancellations for small transverse momenta 
but dominance of photon-induced contributions at larger $p_{\rT}$.
Moving to NLO subleading corrections, $\nlotwo$ are by far the dominant ones for these
observables. The correction with respect to the $\loqcd$ result is positive at low transverse momentum, but constantly decreases  showing the expected negative  enhancement
at high transverse momentum due to the effect of EW Sudakov logarithms \cite{Denner:2000jv}: for the $\tau^+\tau^-$ pair the corrections reach 
 $-20\%$ at $p_{\rm T,\tau^+\tau^-}=1\,$TeV, while for the antitop distribution we observe a $-12\%$ at $p_{\rm T,\bar{t}}=800\,$GeV.
$\nlothree$ terms only marginally correct the $\loqcd$ result:
the corrections are roughly $1\%$ for both
$p_{\rm T,\tau^+\tau^-}$ and $p_{\rm T,\bar{t}}$
and essentially constant over the whole range.
Clearly, the $\loqcd+\nloone$ distributions are strongly distorted by the inclusion of subleading contributions, with a dominant effect given by $\nlotwo$ corrections.
Both distributions exhibit a positive correction of $4$--$5\%$ in the bulk region, which slowly decreases becoming negative around $250\,$GeV.

  \begin{figure} 
  \centering
\subfigure[Transverse momentum of the bottom-jet pair\label{fig:ptbb}]{  \includegraphics[scale=0.32]{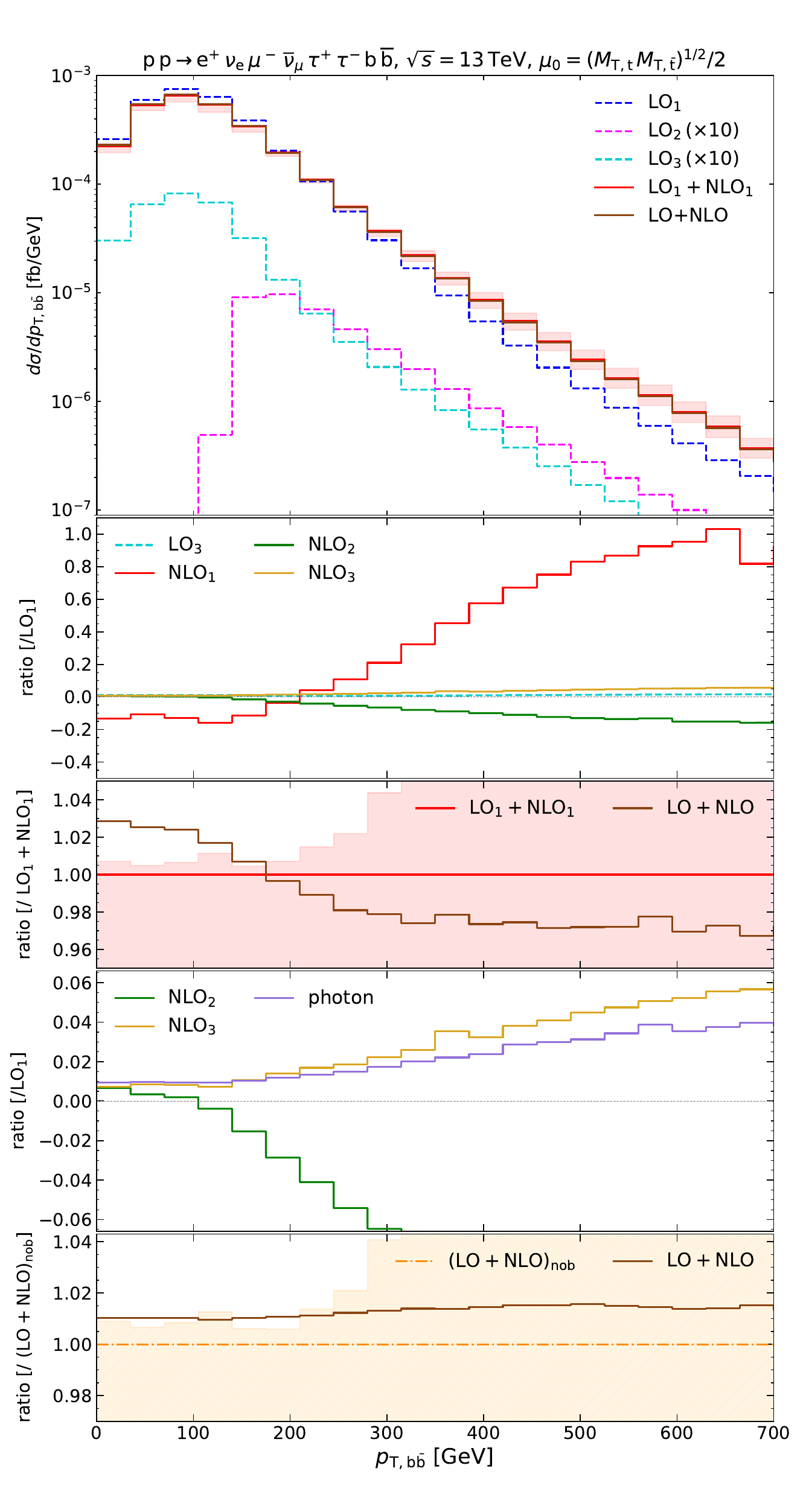}}
\subfigure[$H^{\rm vis}_{\rm T}$ variable\label{fig:ht}]{  \includegraphics[scale=0.32]{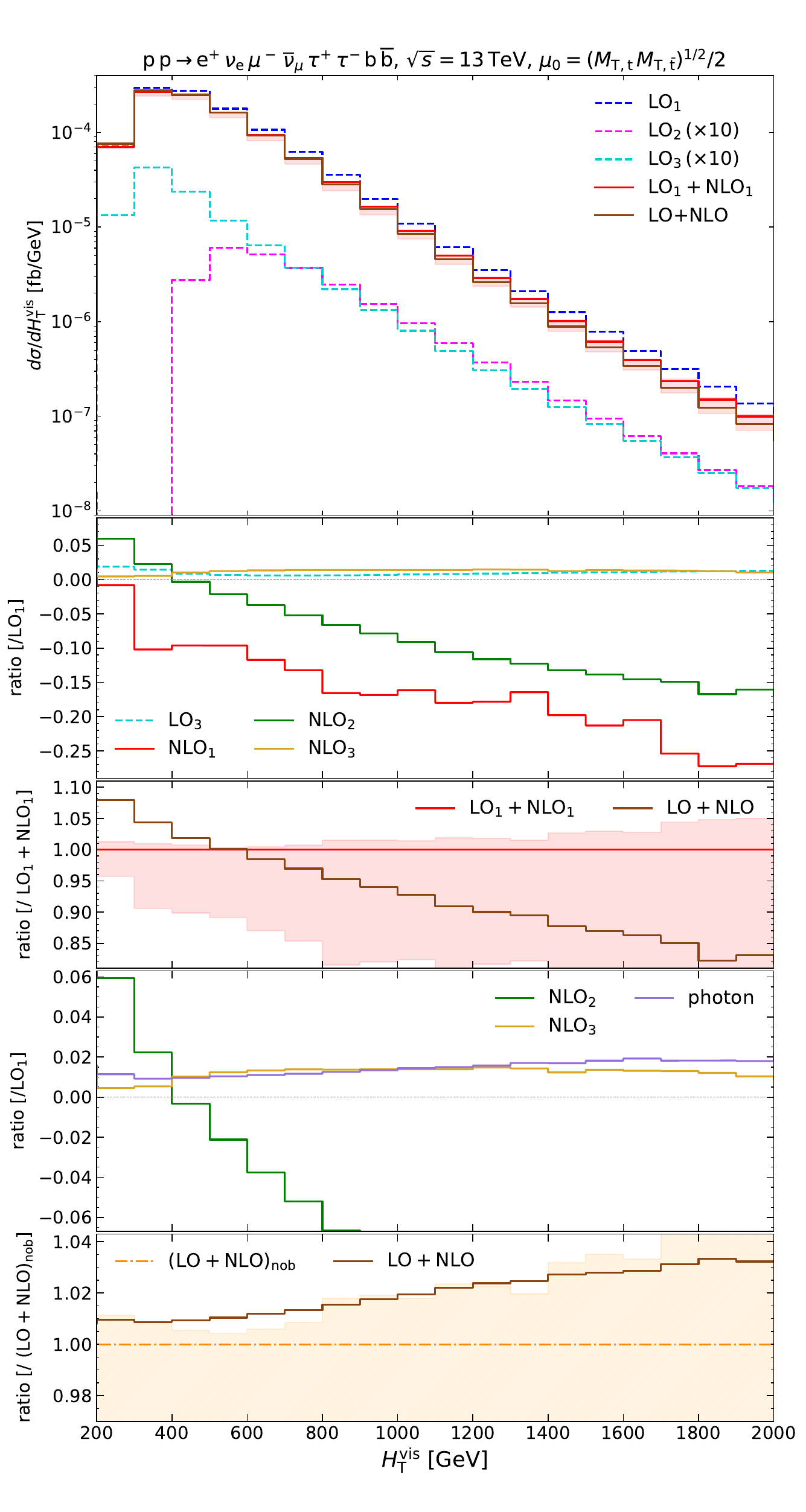}}
  \caption{Distributions in the transverse momentum of the bottom-jet pair (left) and in the $H^{\rm vis}_{\rm T}$ variable (right).
    The different NLO corrections for the observables are compared  separately (first ratio panels) and at the level of the full prediction (second ratio panel).
    The size of photon-induced channels and bottom contributions are presented in the third and fourth ratio panels, respectively.
  }\label{fig:diff4}
  \end{figure}

  In \reffi{fig:diff4} two more transverse variables are reported. In
  \reffi{fig:ptbb} we present our results for the distribution in the
  transverse momentum
  of the $\Pb\bar\Pb$ pair defined at the Monte Carlo-truth level as described at the beginning of this section and  in \refeq{eq:likelihood}.
  As for the transverse momentum of the $\Pt\bar{\Pt}$ pair (not shown here), the transverse momentum of the $\Pb\bar\Pb$ pair is particularly
  sensitive to QCD corrections to $\loqcd$. The QCD corrections are negative and roughly $-15\%$ in the bulk of the distribution,
  then, after changing sign around $200\,$GeV, they steeply increase up to $+100\%$ at $p_{\rm T,\Pb\bar\Pb}\approx 600\,$GeV. %At larger values, it reaches a plateau.
  These huge effects are not simply explained by the presence of hard
  QCD real radiation, but are known as giant QCD $K$-factor \cite{Rubin:2010xp}.
  Indeed, at $\nloone$ topologies of a $\Pt\bar{\Pt}\Pj$ event
  open up where the emission of a soft and/or collinear $\PZ$~boson causes the double-logarithmic enhancement of the cross-section
  in the high transverse-momentum regions of the jet. A similar effect
  was found also in $\ttw$ production
  \cite{Denner:2020hgg}. As a consequence of the very large QCD
  corrections for large $p_{\rT,\Pb\bar\Pb}$, the scale dependence of
  the full NLO QCD result is LO like in these regions of phase space.
 The $\nlotwo$ corrections are once again dominated by large and negative EW Sudakov logarithms.
 They start from small positive values of roughly $0.5\%$
  (as visible in the third ratio panel) to reach $-15\%$ at $p_{\rm T,\Pb\bar\Pb}=700\,$GeV. An opposite behaviour is found for $\nlothree$
  corrections, which are ruled by real gluon-induced contributions: being positive and reaching
$+5\%$ at high transverse momenta,
  they partially balance the effect of $\nlotwo$ corrections in the high-energy tails. Indeed, when considering all subleading contributions,
  the impact of EW Sudakov logarithms is reduced, and subleading corrections never exceed $-3.5\%$.
  In the soft region of the spectrum, \ie for $p_{\rm T,\Pb\bar\Pb}<250\,$GeV, subleading corrections reach at most $+3\%$ of the NLO QCD
  cross-section. Transverse-momentum distributions are known to be more sensitive to the effects of photon-induced
  contributions in the tails. In the third ratio panel we show these contributions at NLO accuracy. All partonic channels 
  involving at least one initial-state photon and their NLO corrections are included. For the process at hand, photon-induced contributions receive both $\nlotwo$
  and $\nlothree$ corrections, the latter being essentially EW corrections to the $\gamma\Pg$ $\loint$ term (QCD corrections to the $\gamma\gamma$
  channel are fully negligible). In agreement with similar studies
  for other processes (see for instance \citere{Denner:2016jyo}), photon contributions slowly increase reaching $4\%$ of the $\loqcd$ contribution at high $p_{\rm T}$
  values. This behaviour is explained by the fact that the photon PDF grows faster than the quark and gluon ones in this phase-space region \cite{Pagani:2016caq}.
  Finally, in a fourth ratio panel we present the impact of the bottom-induced channels together with their NLO corrections.
  As already observed in \citere{Stremmer:2021bnk}, the observables which are expected to be more affected by bottom contributions are the hadronic
  ones whose definition requires at least one $\Pb$~jet. We see that the inclusion of leading and subleading bottom corrections has a moderate impact
  on the result, which is corrected by roughly $+1\%$ in the bulk of
  the distribution and up to $+1.5\%$ in the tails. Nevertheless, these corrections are entirely
  covered by the theory uncertainty bands of the NLO result not including the bottom channels.

In \reffi{fig:ht} we present another variable of interest for LHC
searches, namely $H^{\rm vis}_{\rm T}$, defined as
\beq\label{eq:htvis}
  H^{\rm vis}_{\rm T}=\pt{\Pb_1}+\pt{\Pb_2}+\pt{\tau^+}+\pt{\tau^-}+\pt{ \mu^-}+\pt{\Pe^+}\,,
\eeq
similarly to $H_{\rm T}$ in \refeq{eq:scaleA} but without including the
missing-energy contribution. Since in our calculation up to three $\Pb$~jets can be produced, the sum of the transverse momenta 
is restricted to the leading and subleading $\Pb$ jets (here $\Pb_1$ and $\Pb_2$, respectively) defined according to a $p_{\rm T}$ ordering.
We see that also for this observable $\nloone$ QCD
corrections are large, reaching more than $-25\%$ at $2\,$TeV. $\nlotwo$ corrections amount to
$6\%$ at low $H^{\rm vis}_{\rm T}$ values, steadily decreasing and becoming negative towards higher values. The impact of EW Sudakov logarithms is less
pronounced than in the $p_{\rm T,\Pb\bar\Pb}$ distribution, with a correction relative to $\loqcd$ of around $-15\%$ at $2\,$TeV.
The  $\nlothree$ contributions
account for at most $+1\%$ and are essentially flat up to very high $H^{\rm vis}_{\rm T}$ values.
Overall, the subleading NLO corrections are dominated by the $\nlotwo$ corrections in the high-energy regime.
The impact of photon-induced contributions is slightly milder than for $p_{\rm T,\Pb\bar\Pb}$, 
amounting to roughly $1\%$ of $\loqcd$ at low $H^{\rm vis}_{\rm T}$ values and increasing to almost a $2\%$ towards the tail. 
Since $\Pb$~jets are involved in the definition  of the observable in \refeq{eq:htvis}, we might expect that bottom-induced channels
play a role. Indeed, in the fourth ratio panel in \reffi{fig:ht} some shape effects are found, owing to the inclusion of the bottom
contributions, which correct the $\sigma_{\rm{nob}}$ result
from $+1\%$ in the very first bin up to $+3.5\%$ at $H^{\rm vis}_{\rm T}=2\,$TeV.

\begin{figure*}
  \centering
 \subfigure[Invariant mass of the antitop quark\label{fig:mat}]{\includegraphics[scale=0.32]{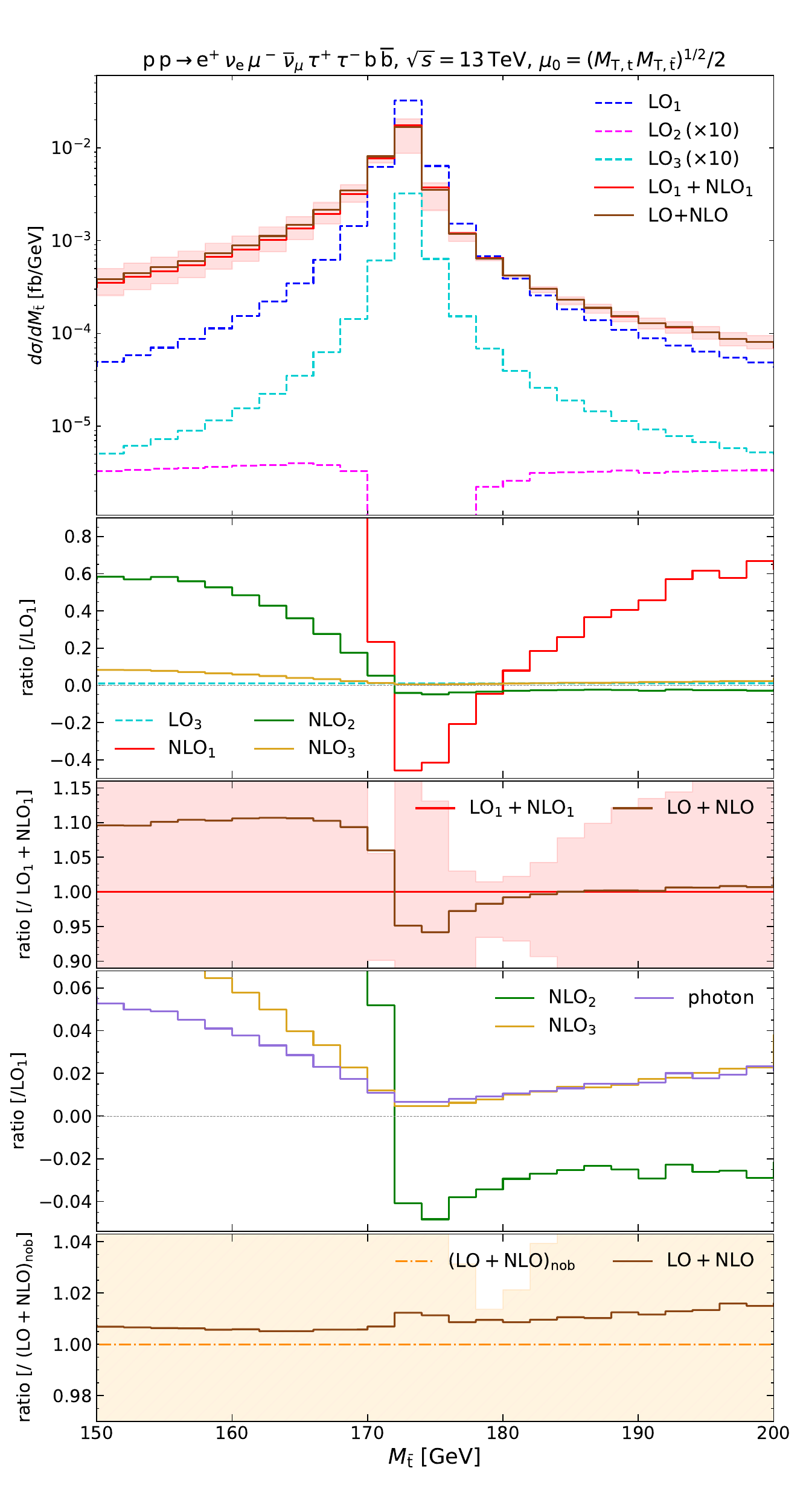}}
\subfigure[Invariant mass of the $\tau^+\tau^-$ pair\label{fig:mtt}]{  \includegraphics[scale=0.32]{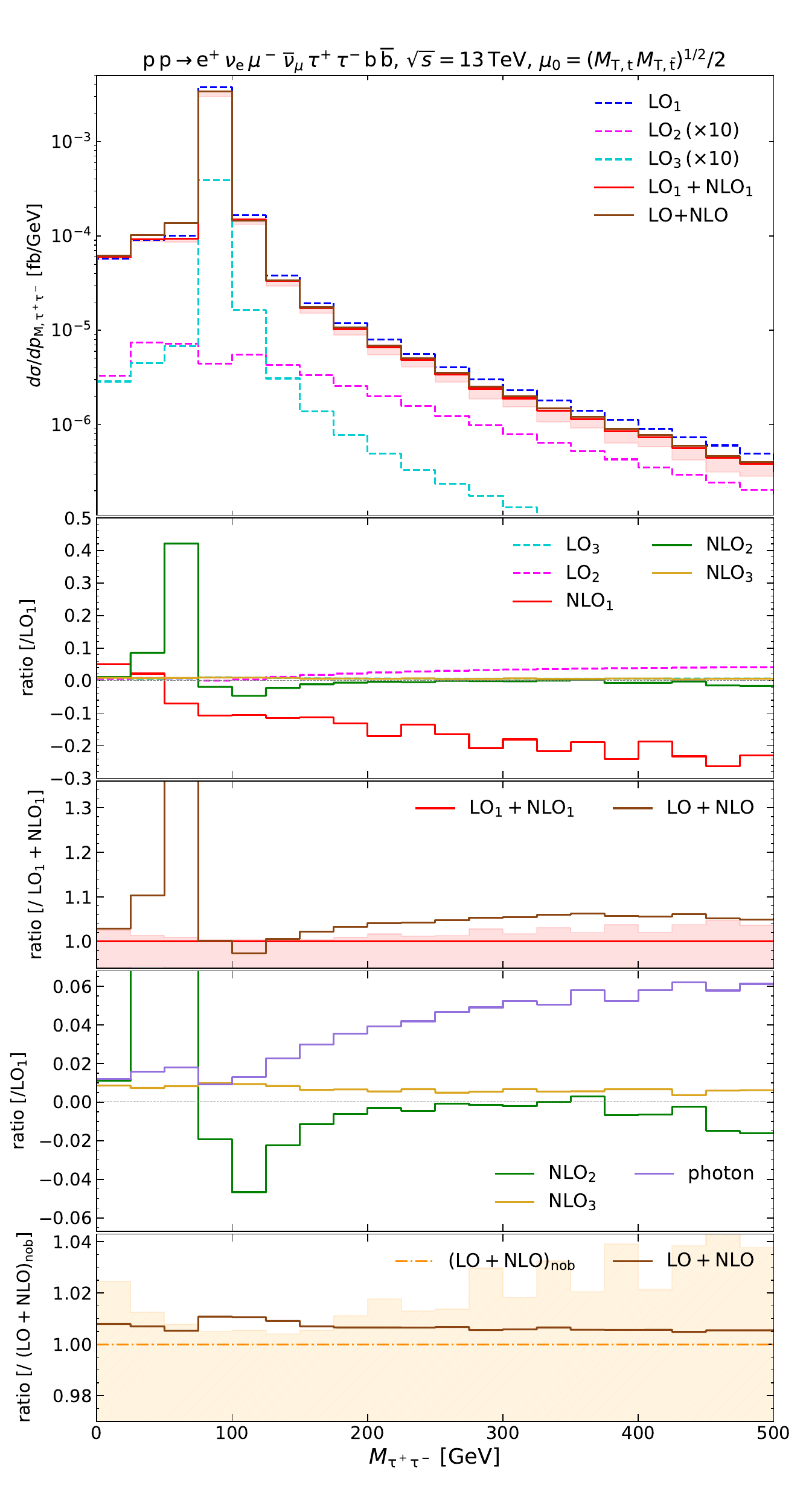}}
  \caption{Distributions in the invariant mass of the antitop quark (left) and of the $\tau^+\tau^-$ pair (right).
    The different NLO corrections for the observables are compared  separately (first ratio panels) and at the level of the full prediction (second ratio panel).
  The size of photon-induced channels and bottom contributions are presented in the third and fourth ratio panels, respectively.}\label{fig:diff5}
\end{figure*}  

In \reffi{fig:diff5} we consider two invariant-mass variables. We start with the distribution in the invariant mass of
the antitop quark displayed in \reffi{fig:mat}. The characteristic invariant-mass peak at $\Mt=173\,$GeV is found
in the $\loqcd$ and in the $\loew$ results.
Conversely, the $\loint$ shows a dip in that region, as a consequence of the negative bottom-interference
contributions, which overcompensate the resonance peak of the $\loint$
$\gamma\Pg$ term leading to a slightly negative $\loint$ contribution
near the nominal top mass. Our results for the $\nloone$ corrections are in
line with the results for this observable found in $\Pt\bar{\Pt}$
\cite{Denner:2012yc} and $\ttw$ \cite{Denner:2020hgg} production.
The QCD corrections are negative (around $-40\%$) on the peak, while
they become positive in regions where the antitop quark is off shell:
for regions of a reconstructed antitop mass above the peak, \ie $M_{\bar{\Pt}}>\Mt$, the corrections progressively increase reaching up to
$+60\%$ for $M_{\bar{\Pt}}\approx\Mt+25\,$GeV; for $M_{\bar{\Pt}}<\Mt$ the $\loqcd+\nloone$ result is one order of magnitude larger than 
its $\loqcd$ one. The enhancement below the peak is well-known and due to QCD radiation that is
not clustered with the $\Pb$~jet arising from the antitop-quark decay.
Radiative-tail effects are also present in subleading NLO corrections, even though to a lesser extent.
Indeed, in $\nlotwo$ corrections real-photon radiation can also be emitted
from the muon arising from the antitop quark. Owing to the suppression
of the EW versus the QCD coupling, this gives rise to smaller
corrections with respect to $\loqcd$, reaching $+60\%$
for $M_{\bar{\Pt}}\approx\Mt-20\,$GeV. The $\nlotwo$ corrections are negative at the peak (around $-4\%$) as well as in the off-shell region above the pole mass.
Yet another behaviour is found for the $\nlothree$
contribution, which remains positive throughout the considered spectrum. It
reaches a minimum of $0.5\%$ around the top resonance,
while increasing in the off-shell regions: for $M_{\bar{\Pt}}\approx\Mt+25\,$GeV a $2\%$ correction is found, and for $M_{\bar{\Pt}}\approx\Mt-20\,$GeV
the radiative tail generates a $10\%$ correction. The behaviour of $\nlothree$ corrections is dominated by quark--gluon partonic channels,
which involve a quark in the final state that cannot arise from the radiative decay of the top or the
antitop quark, differently from a final-state gluon or photon (as observed in \citere{Denner:2021hqi}).
The complete set of subleading corrections to the NLO QCD result just
reinforces the behaviour of the QCD corrections: it amounts to an essentially constant $+10\%$
effect below the peak, to a negative correction of roughly $-5\%$ around $M_{\bar{\Pt}}\approx\Mt$,
and is almost vanishing for $M_{\bar{\Pt}}>\Mt$. 
As already observed for $\Pt\bar{\Pt}$ production \cite{Denner:2016jyo}, the role of photon-initiated contributions in invariant-mass
distributions can be quite significant. In our case we see that the full set of photon-induced channels mimics the $\nlothree$ behaviour,
being always positive and growing towards the off-shell regions. The radiative-tail enhancement mostly comes from the $\gamma\Pg$ channel,
reaching $+5.5\%$ at $M_{\bar{\Pt}}\approx\Mt-20\,$GeV. Since the definition of this variable
involves just one bottom quark for the reconstruction of the antitop-quark mass at Monte Carlo-truth level, the role of the bottom contributions is slightly
smaller than for other distributions analysed previously. They essentially give a contribution to the combined NLO cross-section ranging from $0.5\%$ to $1.5\%$.

In \reffi{fig:mtt} we present the distribution in the invariant mass of the $\tau^+\tau^-$ pair. The peak at the $\PZ$-boson
mass is observed for  $\loqcd$ and $\loew$, but appears to be
completely absent for $\loint$. In fact there is even a slight dip in
the bin containing the $\PZ$-boson mass. As discussed for \reffi{fig:mat},
this results once again from large cancellations between the
$\gamma\Pg$ and the bottom-induced interference terms in the resonant region.
The harder tail of the $\loint$ curve as compared to $\loew$ is instead entirely due to the $\gamma\Pg$ channel,
which is by far the largest subleading contribution in the tail,
accounting for a $+4\%$ correction with respect to $\loqcd$.
The $\nloone$ corrections constantly decrease from $+4\%$ in the very first bin of the distribution to roughly $-20\%$ in the far off-shell
region. Since the observable at hand is fully leptonic, large radiative effects are generated by $\nlotwo$ corrections, owing to real
photons that are not clustered with $\tau$ leptons: the $\nlotwo$ corrections
with respect to $\loqcd$ are indeed
more than $+40\%$ right below the peak and $-2\%$ for $M_{\tau^+\tau^-}\approx \MZ$. Above this region they increase, almost
vanish around $300\,$GeV, and then mildly decrease again. Since the $\nlothree$ contributions are effectively dominated by QCD corrections, they do not
affect the position of the resonance peak, like $\nloone$, but they provide an essentially constant correction to the $\loqcd$ cross-section ranging from $+0.5\%$ to $+1\%$.
Therefore, subleading contributions to $M_{\tau^+\tau^-}$ mainly arise from $\loint$
contributions in the region $M_{\tau^+\tau^-}\gtrapprox 150\,$GeV, while the $\nlotwo$ corrections dominate around and below the $\PZ$-boson mass.
The photon-induced contributions are quite relevant, increasing from roughly $1\%$ at $M_{\tau^+\tau^-}\approx \MZ$ up to $6\%$ at $500\,$GeV,
due to the leptonic nature of the observable.
The impact of bottom-initiated contributions is instead negligible:
they only mildly distort the shape of the distribution by a positive correction of at most $1\%$,
which is entirely contained within the theory uncertainty of the result without bottom contributions.

\section{Conclusions}\label{sec:conclusions}
We have presented the first calculation
of the off-shell production of a top--antitop pair in association with a $\PZ$ boson
that is accurate both at NLO QCD and NLO EW for $13\,$TeV proton--proton collisions.
To be more precise, we have computed the entire tower of LO contributions and NLO corrections to a final state
that involves four different charged leptons, two bottom jets, and missing transverse momentum. All off-shell effects have been retained
in order to provide a realistic description of the process including decay effects at NLO accuracy.

Owing to the very high multiplicity of the final state, this calculation proved to be extremely intensive from the computation-time
and book-keeping point of view.
In this sense, the off-shell $\ttz$ process lies at the frontier of LHC signatures that can be simulated with optimised Monte Carlo
integrators and one-loop amplitude providers that are currently available.
Given the high complexity of the simulations, we have performed a careful
validation of the QCD corrections with existing results
\cite{Bevilacqua:2022nrm}, finding very good agreement both in integrated and in differential cross-sections.

Even if bottom-induced partonic processes were already considered
in the context of NLO QCD predictions for $\Pt\bar{\Pt}$-associated processes~\cite{Stremmer:2021bnk},
our calculation is the first one that accounts
for all  LO and NLO contributions to these channels, \ie we provide predictions in a 
truly five-flavour scheme. Also photon-induced processes have been
taken into account, in spite of the expected minor impact on the final results.
Throughout our calculation full spin correlations have been kept, while including both resonant and non-resonant diagrams.
The mixing of QCD and EW radiative corrections as well as all interference effects have been taken into account.

We provide integrated and differential results in a realistic fiducial setup, using
a resonance-aware definition for the renormalisation and factorisation scales,
which is known to improve the convergence of the perturbative expansion in $\alphas$~\cite{Denner:2021hqi}.
With such a choice, the QCD corrections to the LO QCD ($\loqcd$) cross-section, namely $\nloone$, are sizeable and negative ($-10\%$) at integrated level,
and decrease the QCD-scale uncertainties from $30\%$ down to $10\%$ when going from $\loqcd$ to $\loqcd+\nloone$. 
The $\nlotwo$ corrections (EW corrections to LO QCD and QCD
corrections to LO interference) account for less than a percent of
the LO fiducial cross section and are comparable in size with the
$\nlothree$ contributions (EW corrections to LO interference and QCD
corrections to LO EW), where the expected $\alpha$ suppression
is balanced by the enhancement owing to $\Pt\PZ$-scattering topologies in the real corrections.
On the other hand, our calculation confirms the expectation that $\nlofour$ corrections  (EW corrections to LO EW) are very strongly suppressed,
giving a sub-per-mille effect that is definitely out of reach at the LHC even after the planned luminosity upgrades.
The contributions of bottom-induced and photon-induced channels are each
about $1\%$ of the complete NLO cross-section.
In particular, we have observed in our fiducial setup strong cancellations between the bottom-induced contributions and photon--gluon-induced ones,
occurring already at the integrated level for the LO$_2$ cross-section.
This interplay amongst the various LO and NLO corrections gives an even more intricate structure at differential level.
All NLO corrections do not just cause a change of normalisation but also a distortion of the distribution shapes, even for some
more inclusive observables like angles and rapidities.
Remarkable shape changes at $\nloone$ are motivated by an overall strong scale dependence, but also by the
presence of hard real radiation, which gives LO-like scale bands in
tails of some distributions. 
After including $\nlotwo$ and $\nlothree$ corrections, the
differential cross-sections change sizeably w.r.t.\ the pure QCD result
($\loqcd+\nloone$) especially in transverse-momentum distributions,
pointing out that the $\nloone$, $\nlotwo$, and $\nlothree$
corrections are unavoidable for a description of the $\ttz$ process when
aiming at precise predictions over the full phase space.
In particular, a sizeable enhancement from EW Sudakov logarithms gives $\nlotwo$ corrections that reach up to $-20\%$
of the $\loqcd$ result at moderate-to-high transverse momenta. The differential $\nlothree$ corrections are usually flatter
than the other two. As a general comment, in phase-space regions where the cross-section is
sizeable, the subleading NLO corrections are below 5\%.

We conclude stressing that, although an on-shell calculation (or one including decay effects via a production$\times$decay approximation)
is expected to capture most of the $\ttz$ features in the bulk regions of the phase space, it is essential to simulate
this process including all kinds of off-shell effects, especially in the case where exclusive fiducial selections are applied
and differential measurements are performed.

\section*{Acknowledgements}
The authors are indebted to Sandro Uccirati for fixing the \recola code in the case of processes with four external (anti)bottom quarks and intermediate
top-quark propagators, to Jean-Nicolas Lang for testing the fixed \recola code, and to Christopher Schwan for useful discussions.
 This work is supported by the German Federal Ministry for
Education and Research (BMBF) under contract no.~05H21WWCAA.

\bibliographystyle{JHEPmod}
\bibliography{ttv}

%%%%%%%%%%%%%%%%%%%%%%%%%%%%%%%%%
\end{document}